\documentclass[aps,prd,onecolumn,superscriptaddress,11pt,nofootinbib,preprintnumbers]{revtex4-1}
\usepackage{amsmath,amssymb,hyperref}
\usepackage{tikz}
\usetikzlibrary{decorations.pathmorphing,decorations.pathreplacing}
\tikzset{gluon/.style={decorate
	,decoration={coil,amplitude=5pt, segment length=7pt, post length=0mm}}}
\tikzset{photon/.style={decorate
	,decoration={snake,amplitude=1mm, segment length=3mm, post length=0mm}}}

\usepackage{braket}	

\newcommand{\be}{\begin{equation}}
\newcommand{\ee}{\end{equation}}

\newcommand{\LQ}{\Lambda_{\rm QCD}}
\newcommand{\tL}{\tilde{\Lambda}}
\newcommand{\lt}{\left}
\newcommand{\rt}{\right}
\newcommand{\bz}{\beta_0}
\newcommand{\non}{\nonumber \\}
\newcommand{\mf}{\mu_f}
\newcommand{\IR}{\rm IR}
\newcommand{\UV}{\rm UV}
\newcommand{\Li}{{\rm Li}}

\def\sla#1{\rlap/#1}

\begin{document} 

\preprint{TU--1037, TTP16-056}

\title{
Subtracting IR Renormalons from Wilson Coefficients:\\
Uniqueness and power dependences on $\Lambda_\mathrm{QCD}$
}

\author{Go Mishima}
\email{go.mishima@kit.edu}
\affiliation{
Department of Physics, Tohoku University, Sendai, 980-8578 Japan}
\affiliation{
  Institute for Theoretical Particle Physics (TTP), 
  Karlsruhe Institute of Technology, Wolfgang-Gaede-Stra{\ss}e 1, 
  76128 Karlsruhe, Germany
}
\affiliation{
  Institute for Nuclear Physics (IKP), Karlsruhe Institute of
  Technology, Hermann-von-Helmholtz-Platz 1, 76344
  Eggenstein-Leopoldshafen, Germany
}
\author{Yukinari Sumino}
\email{sumino@tuhep.phys.tohoku.ac.jp}
\affiliation{
Department of Physics, Tohoku University, Sendai, 980-8578 Japan}
\author{Hiromasa Takaura} 
\email{t\_hiromasa@tuhep.phys.tohoku.ac.jp} 
\affiliation{
Department of Physics, Tohoku University, Sendai, 980-8578 Japan}


\begin{abstract}
In the context of OPE and
using the large-$\bz$ approximation,
we propose a 
method to define Wilson coefficients free from uncertainties
due to IR renormalons.
We first introduce
a general observable $X(Q^2)$
with an explicit IR cutoff, and then we extract
a genuine UV contribution $X_{\UV}$ as a cutoff-independent part.
$X_{\UV}$ includes power corrections $\sim (\LQ^2/Q^2)^n$
which are independent of renormalons.
Using the integration-by-regions method, we observe that
$X_{\UV}$ coincides with the leading Wilson coefficient
in OPE and also 
clarify that
the power corrections originate from UV region.
We examine scheme dependence of $X_{\UV}$ and
single out a specific scheme favorable
in terms of analytical properties.
Our method
would be optimal 
with respect to systematicity, analyticity and stability.
We test our formulation with the
examples of the Adler function, QCD
force between $Q \bar{Q}$, and $R$-ratio in $e^{+} e^{-}$ collision.
\end{abstract}

\maketitle

\newpage

\section{Introduction}
In perturbative quantum field theory, 
perturbative series are considered to be asymptotic and divergent.
It suggests that we have to truncate the series at finite order, 
and thus perturbative calculation cannot reach arbitrary precision. 
The idea of renormalon is a powerful tool to discuss an inevitable uncertainty of perturbative calculation \cite{Beneke:1998ui}. 
It is related to divergent behaviors of perturbative series, 
and it provides an estimate of the size of uncertainty 
in an optimal prediction. 
In perturbative QCD, 
infrared (IR) renormalons give essential uncertainties of order $(\LQ/Q)^n$ in the prediction, 
where $Q$ is a typical energy scale of an observable $X$. 
IR renormalons stem from low-energy region of loop momenta 
in Feynman integrals.  
Such uncertainties cannot be removed even by a resummation or Borel summation.  
This indicates that another framework is needed to overcome perturbative uncertainties induced by IR renormalons. 

Operator product expansion (OPE) is a
framework, in which the perturbative uncertainties can be eliminated  systematically. 
An OPE of an observable $X(Q^2)$ consists of two components: 
Wilson coefficients and non-perturbative matrix elements. 
In the Wilsonian picture, Wilson coefficients are calculated from ultraviolet (UV) modes, which are higher than a factorization scale $\mf$, 
whereas non-perturbative matrix elements are described by a low-energy effective theory valid below the scale $\mf$. 
As a result, Wilson coefficients are free from uncertainties induced by IR renormalons and can be calculated unambiguously in perturbation theory (in principle). 
Non-perturbative matrix elements are determined from IR dynamics and show the same power dependence on $\LQ/Q$ as the uncertainties due to IR renormalons in the
original perturbative series of $X$. 
Note, however, that each non-perturbative matrix element
is no longer an uncertainty but a definite quantity,   at least conceptually. 
Therefore, one can go beyond perturbation theory in the OPE framework. 

In OPE an observable $X(Q^2)$ is evaluated by expansion in $1/Q^2$.
To realize the concept of the Wilsonian picture, it is  natural to
introduce a hard cutoff ($\mf$) in momentum space
for factorizing UV and IR dynamics.%
\footnote{
In conventional analyses of renormalons, 
a UV scale is assumed to be much larger than 
any scale involved in the calculation. 
In this paper, however, we use the terminology ``UV'' for scales
above the factorization scale $\mu_f$ in 
the context of OPE.
In particular $Q$ is regarded as a UV scale.
} 
Then the IR renormalons are clearly eliminated from 
perturbative calculation of Wilson coefficients,
and the $1/Q^2$-expansion
(derivative expansion) in the low-energy effective theory is
well justified since the active modes satisfy
$k/Q\leq \mf/Q \ll 1$.
It is, however, disadvantageous in practical computations to introduce
a hard cutoff due to the following reasons:
(1)~One should include an additional scale $\mf$ in computations, 
which complicates the
computations considerably.
(2)~Generally it generates
apparent power-like strong dependences on $\mf$ of Wilson coefficients. 
Although they should eventually cancel in physical predictions,
they can be sources of strong instability of the 
predictions in practice \cite{Kiyo:2015ooa}.
(3)~If we adopt a too naive cutoff regularization scheme,
it may violate gauge invariance.
For these reasons today it is customary to compute perturbative
series of Wilson coefficients in dimensional regularization.
This regularization circumvents the above difficulties.
Nevertheless, as a trade-off, the perturbative series contain IR renormalon
uncertainties since each integral region extends from
$k\sim 0$ to infinity.
Hence, several ways to subtract
the contributions of IR renormalons
have been explored
\cite{Beneke:1998rk,Pineda:2001zq,Sumino:2005cq,Hoang:2009yr,Hoang:2014wka,Scimemi:2016ffw}.

In this paper we investigate the Wilson coefficient of the
leading operator in OPE (equals to the identity operator in our 
explicit examples)
and aim at removing a factorization scale dependent part, 
which destabilizes the prediction. 
Our basic tool is perturbation theory  
in the so-called large-$\bz$ approximation 
\cite{Beneke:1994qe, Neubert:1994vb,Broadhurst:1993ru}.
We proceed in the following steps:
(i)~We consider an observable $X(Q^2)$ with an IR cutoff $\mf$.
(ii)~We extract a $\mf$-independent part $X_{\UV}$
systematically, which can be regarded
as a genuine UV contribution.
(iii)~We examine the scheme dependence of $X_{\UV}$.
(iv)~We single out a favorable scheme in terms of
analyticity of $X_{\UV}$.

It turns out that
$X_{\UV}$ includes power corrections $\sim (\LQ^2/Q^2)^n$ 
which stem from UV physics
and are totally different from renormalon ambiguities.
We will see that (1)~the power corrections are 
consistent with the framework of OPE, and
(2)~the power corrections are crucial for understanding
the short-distance behavior of $X(Q^2)$.
This is one of the main focuses of our discussion. 
Our method would also be useful 
in extracting 
non-perturbative matrix elements numerically,
since the leading Wilson coefficient which we construct does not contain 
intrinsic uncertainties of the order of the matrix elements.

An analytical evaluation of a resummed perturbative series in the large-$\beta_0$ approximation was first performed in Ref.~\cite{Ball:1995ni}. 
In fact 
many building blocks in our method are taken from their analysis.
Their analysis starts from a regularized Borel integral, which
removes IR renormalons by contour deformation.
In their method a physical quantity can be 
separated into the real part and imaginary part. 
The real part is predicted reliably 
within perturbation theory, whereas the imaginary part is regarded as a perturbative uncertainty. 
The real part in their method and the cutoff independent part in our method have 
the same expanded form in $1/Q$. 
We also use an idea in their analysis
related with the pseudo gluon mass 
to extract a cutoff independent part in our method. 

Characteristic features of our method can be stated as follows.  
By starting from a well-defined integral with an explicit cutoff, we give
a solid basis to our method, thereby the relation to 
OPE in the Wilsonian picture is made clear.
We also reinforce our argument
using the integration-by-regions (expansion-by-regions)
method or comparison with the perturbative series up to
large orders.
Furthermore, we compare the perturbative series in
the large-$\bz$ approximation
with the known
exact perturbative series and confirm consistency or
validity of the approximation we use.
These analyses utilize theoretical developments
which took place after the analysis \cite{Ball:1995ni},
and it is worthwhile to examine their impact.

Related subjects have also been studied in 
Refs.~\cite{Mueller:1984vh,Beneke:1994qe,Neubert:1994vb}
(see also Refs.~\cite{Narison:2001ix,Chetyrkin:1998yr}).
In particular, existence of power corrections in the UV contribution
to observables
has been discussed, e.g., using a resummation of the perturbative 
series \cite{Ball:1995ni}, and in certain model calculations \cite{Narison:2001ix,Chetyrkin:1998yr}.
Our work can be regarded as an extension of 
the analyses in
Refs.~\cite{Beneke:1994qe,Neubert:1994vb,Ball:1995ni}
and more directly of
the formulation used in the analysis of the static QCD potential
\cite{Sumino:2003yp,Sumino:2004ht,Sumino:2005cq,Sumino:2014qpa}.
Part of the analysis presented in this paper,
in particular its application to the Adler function,
have been reported in the letter article \cite{Mishima:2016xuj}.

The important and new points provided in this paper can be summarized as follows.
First, we propose a systematic method to construct a Wilson coefficient ($X_{\UV}$)
which is free from both renormalon ambiguities and cutoff dependence. 
Such a quantity is important for precise predictions
as it gives a foundation for order-by-order predictions in the OPE framework,
and in connection with this, it enables us to determine non-perturbative matrix elements numerically.
Secondly, we support existence of power corrections $\sim (\LQ^2/Q^2)^n$ in $X_{\UV}$
which are independent of renormalons.
Although such a term is found in the literature,
the existence was vague 
since it (in particular, the coefficient of such a term) is 
generally dependent on a resummation prescription. 
Nevertheless, we clarify that a specific prescription
is uniquely favored from the viewpoint of analyticity in our framework.
Hence, a natural coefficient is specified
and it supports, for instance, the existence of a $\LQ^2/Q^2$-term in the Adler function.

The outline of this paper is as follows. 
In Sec.~\ref{sec.BM}, 
we explain our method to extract a cutoff independent part 
from a general observable defined with an IR cutoff. 
We also test our method with 
the Adler function and the force between $Q \bar{Q}$. 
In Sec.~\ref{sec.EbR}, we investigate the relation between our method and OPE 
using the method of integration by regions 
and also clarify which region gives  each power correction. 
In Sec.~\ref{sec.RB}, we show that the power corrections in $X_{\UV}$ is 
included in the large-order perturbative series.
We also compare our results with known exact perturbative
series. 
Through Secs.~\ref{sec.BM}--\ref{sec.RB} only Euclidean quantities
are examined.
In Sec.~\ref{sec.ET}, we study the $R$-ratio in $e^{+} e^{-}$ collision 
as an example of a timelike quantity, 
and how our method can be applied. 
Conclusions and discussion are given in Sec.~\ref{sec.C}. 
Details of our analyses are collected
in Appendixes.

\section{Extraction of cutoff-independent part from
UV contributions}
\label{sec.BM}

In this section
we present a method to extract a
cutoff-independent part from
UV contributions to physical quantities.
In Sec.~\ref{ss:defs},
basic notions are reviewed.
In Sec.~\ref{sec.GC},
the method to extract a cutoff-independent part is explained.
As examples, we investigate the Adler function in Sec.~\ref{ss:adler} and the force between static quark and antiquark in Sec.~\ref{ss:force}.
In Sec.~\ref{sec.SD}, 
we examine a scheme dependence inherent 
in our method.
In Sec.~\ref{sec.FM}, we show that a specific scheme
is favored 
from analytical properties of the extracted UV part. 
In Sec.~\ref{sec.FM2},
some detailed features of this specific scheme
are analyzed.

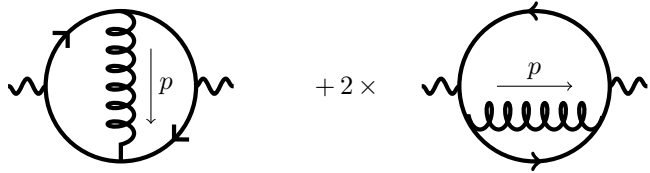
\begin{figure}
\centering
\begin{tikzpicture}[baseline={(0,0)},scale=1]
\draw [ultra thick,photon] (0,0)--(-0.5,0);
\draw [ultra thick] (1,0) circle [radius=1];
\draw [ultra thick,photon] (2,0)--(2.5,0);
\draw [ultra thick,gluon] (1,1)--(1,-1);
\draw [->] (1.4,0.5)--(1.4,-0.5);
\draw (1.6,0) node {$p$};
\draw [ultra thick] (-1/1.414+1,1/1.414+0)--(-1/1.414+1,1/1.41-0.2);
\draw [ultra thick] (-1/1.414+1,1/1.414+0)--(-1/1.414+1-0.2,1/1.41+0);
\draw [ultra thick] (1/1.414+1,-1/1.414+0)--(1/1.414+1,-1/1.41+0.2);
\draw [ultra thick] (1/1.414+1,-1/1.414+0)--(1/1.414+1+0.2,-1/1.41+0);
\draw (4,0) node {$+\, 2\, \times$};
\draw [ultra thick,photon] (0+5.5,0)--(-0.5+5.5,0);
\draw [ultra thick] (1+5.5,0) circle [radius=1];
\draw [ultra thick,photon] (2+5.5,0)--(2.5+5.5,0);
\draw [ultra thick,gluon] (1.9+5.5,-0.4)--(0.1+5.5,-0.4);
\draw [->, ultra thick] (1+5.5,-1)--(1.1+5.5,-1);
\draw [->, ultra thick] (1+5.5,1)--(0.9+5.5,1);
\draw [->] (6,0)--(7,0);
\draw (6.5,0.2) node {$p$};
\end{tikzpicture}
\caption{\small 
Leading-order diagrams which contribute to the reduced Adler function.
The spiral line represents a gluon (with momentum $p$),
and the solid line represents a massless quark.
The external wavy line represents
an insertion of the electromagnetic current.
}
\label{Fig.Adlerdiagram}
\end{figure}

\subsection{Definitions and basics (review)}
\label{ss:defs}
We consider a dimensionless spacelike observable $X(Q^2)$ 
whose leading order contribution
is given by one-gluon-exchange diagrams, 
such as the ones shown in Fig.~\ref{Fig.Adlerdiagram}.
For simplicity we focus on a quantity which depends on
a single scale $Q^2>0$ in perturbative QCD.
All the external and loop
momenta are taken to be in the Euclidean region and
we use the Euclidean metric through Secs.~\ref{sec.BM}--\ref{sec.RB},
except where stated otherwise.
Explicitly we consider the case where the leading order (LO) contribution
to $X(Q^2)$ in perturbation theory
can be written in the form 
\begin{align}
X_\mathrm{LO}(Q^2)
= \alpha_s(\mu)
\int_0^{\infty} \frac{d \tau}{2 \pi \tau} \, w_X \! \lt(\frac{\tau}{Q^2} \rt) \, .
\label{oneloop}
\end{align}
$\tau$ represents the modulus-squared of the Euclidean gluon momentum $p$
$(\tau=p^2)$,
and integrations over 
all the other loop momentum variables are included in $w_X$.
We call $w_X$ as 
``weight function,'' or simply ``weight.'' 
In this form $w_X$
reduces to a function of the single variable $\tau/Q^2$.
We assume that the integral is finite both in IR ($\tau\to 0$)
and UV ($\tau \to \infty$) regions.
The strong coupling constant 
$\alpha _s(\mu)$ is factored out,
where $\mu$ is the renormalization scale.
We adopt the modified minimal-subtraction
($\overline{\rm MS}$) renormalization scheme,
in which $\alpha _s(\mu)$ 
at the one-loop level
is given by
\begin{align}
\alpha_s(\mu)=\frac{4 \pi}{\bz} \frac{1}{\log{(\mu^2/\LQ^2)}}\, .
\end{align}
$\beta_0=11-2n_f/3$ denotes
the leading-order coefficient of the beta function for
$n_f$ active quark flavors.

We evaluate $X(Q^2)$
in the large-$\beta_0$ approximation, which can be obtained as follows.
We consider insertions of a chain of fermion bubbles into the gluon propagator
of $X_{\rm LO}$.
Each bubble diagram produces a factor proportional to 
$\alpha_s(\mu ) n_f \log(\mu^2 e^{-C}/p^2 )$, 
where $C$ is a scheme dependent constant and
$C=-5/3$ in the $\overline{\rm MS}$ scheme.
Taking the infinite sum of the chains and replacing $n_f \to n_f-33/2=-3 \bz/2$, 
we obtain the all-order perturbative series in the large-$\beta_0$ 
approximation \cite{Beneke:1994qe, Neubert:1994vb,Broadhurst:1992si}
\begin{align}
X_{\beta_0}(Q^2)
=\alpha_s(\mu )
\sum_{n=0}^{\infty} \int_0^{\infty} \frac{d \tau}{2 \pi \tau}\,
w_X \! \lt(\frac{\tau}{Q^2} \rt)
 \left[
\frac{\beta_0 \alpha_s(\mu )}{4 \pi} 
\log \left(\frac{\mu^2  e^{5/3} }{\tau}\right)
\right] ^n
.\label{ExLB}
\end{align}
After resummation of the infinite series in Eq.~\eqref{ExLB},
the expression reduces to the same form as Eq.~\eqref{oneloop}
with the strong coupling replaced by an effective
coupling $\alpha_{\beta_0} (\tau)$: 
\begin{align}
X_{\beta_0}^{\rm resum}(Q^2)
=\int_0^{\infty} \frac{d \tau}{2 \pi \tau}\,
w_X \!\lt(\frac{\tau}{Q^2} \rt)
\alpha_{\beta_0}(\tau)\,,
\label{ReLB}
\end{align}
where
\begin{align}
\alpha_{\beta_0}(\tau)
=\frac{4 \pi}{\beta_0} \frac{1}{\log(\tau e^{-5/3}/\LQ^2)} \, .
\label{a0}
\end{align}
The effective coupling $\alpha _{\beta_0}(\tau)$
has a pole at $\tau =e^{5/3} \LQ^2$,
and the existence of this pole on the integral path
makes the integral ill-defined.
The uncertainty which arises from this
pole in this approach is attributed to IR renormalons. 

We can make use of the Borel transformation
to understand properties of the series in Eq.~\eqref{ExLB}.
The Borel transform of $X_{\beta_0} (Q^2)$ is defined as 
\begin{align}
\hat{B}_X(u)
&\equiv
\sum_{n=0}^{\infty} 
\frac{u^n}{n!}
\int_0^{\infty} \frac{d \tau}{2 \pi \tau} \, w_X \! \lt(\frac{\tau}{Q^2} \rt)
\left[
\log \left(\frac{\mu^2  e^{5/3} }{\tau}\right)
\right] ^n
\nonumber\\
&=
\int_0^{\infty} \frac{d \tau}{2 \pi \tau} \, w_X\! \lt(\frac{\tau}{Q^2} \rt)
\left(\frac{\mu^2  e^{5/3} }{\tau}\right) ^u.
\label{a1}
\end{align}
$\hat{B}_X(u)$
plays the role of a generating function for 
the coefficients of the original series $d_n$
after accelerating convergence by $1/n!$:
\begin{align}
\hat{B}_X(u)
&=\sum_{n=0}^{\infty} \frac{d_n}{n!} u^n \,,
\label{a2}\\
X_{\beta_0} (Q^2)
&=\alpha _s(\mu )
\sum_{n=0}^{\infty} 
d_n
\left[
\frac{\beta_0 \alpha_s(\mu )}{4 \pi} 
\right] ^n \label{a3} \, .
\end{align}
In general, singularities of $\hat{B}_X(u)$
characterize diverging behaviors of the original series. 
Singularities of $\hat{B}_X(u)$ located on the positive real axis 
are called IR renormalons and 
those on the negative real axis are called 
UV renormalons.
Due to assumed finiteness of the integral Eq.~\eqref{oneloop},
$\hat{B}_X(u)$ is regular at
$u=0$ [since Eq.~\eqref{a1} reduces to Eq.~\eqref{oneloop}].
The first IR renormalon at
$u=u_{\IR}>0$, closest to the origin, is known to 
give an inevitable uncertainty of $\mathcal{O}((\LQ^2/Q^2)^{u_{\IR}})$ in perturbative prediction.  

Since the renormalization scale dependence
of $\hat{B}_X(u)$ is factorized in Eq.~\eqref{a1},
we further define 
\begin{align}
B_X(u)
\equiv 
\lt(\frac{Q^2 e^{-5/3}}{\mu^2} \rt)^{u} 
\hat{B}_X(u) \, .
\label{a4}
\end{align}
The weight $w_X(x)$
and the Borel transform $B_X(u)$
are related by \cite{Neubert:1994vb}
\begin{align}
B_X(u)
&=\int_0^{\infty} \frac{d x}{2 \pi} \,
w_X(x) \, x^{-u-1} \,,
\label{Borel} \\
w_X(x)
&=\frac{1}{i} \int_{u_0-i \infty}^{u_0+i \infty} du \, B_X(u) \, x^{u} \, ,
\label{invB}
\end{align}
where $u_0$ is located between 
the first IR renormalon and the first UV renormalon. 
In particular, the small-$x$ behavior of the
weight $w_X(x)$ is detected from the singularities of $B_X(u)$ explicitly as\footnote{
\label{fn:loginc}
In the case that $B_X(u)$ has a multiple pole
in $u$, the corresponding residue
includes a polynomial of $\log x$.
For simplicity we explain in the case
where $w_X$ is expanded as a Taylor series in $x$.
}
\begin{align}
w_X(x)=\sum_{n \in U_{\IR}} b_n x^n \label{invBexp} 
= -2 \pi \sum_{n \in U_{\IR}}  {\rm Res}_{u=n}[B_X(u) x^u] \, ,
\end{align}
where $U_{\IR}$ denotes the set of IR renormalons ($U_{\IR}=\{u_{\IR},\dots\}$).

As mentioned below Eq.~\eqref{a0},
the expression Eq.~\eqref{ReLB}
has an ambiguity because of
the pole of $\alpha _{\beta_0}(\tau )$.
In order to avoid this ambiguity
we introduce an IR cutoff scale $\mu _f$
to the gluon momentum
and eliminate contributions
whose momentum scales are smaller than 
$\mu_f$~\cite{Neubert:1994vb}:
\begin{align}
X_{\bz}(Q^2; \mf)\equiv 
\int_{\mf^2}^{\infty} \frac{d \tau}{2 \pi \tau} \,
w_X\!\lt(\frac{\tau}{Q^2} \rt) \alpha_{\bz}(\tau) 
\label{start} \, .
\end{align}
The factorization scale is chosen to satisfy 
$e^{5/3} \LQ^2 \ll \mf^2 \ll Q^2$. 
Now that the integral path does not 
contain the pole,  
the integral is well defined. 
We choose this well-defined
quantity as the starting point of our discussion.
We will see in explicit examples that 
$X_{\bz}(Q^2; \mf)$ corresponds to the Wilson coefficient
of the leading operator in OPE (see Sec.~\ref{sec.EbR}).

The subtraction of IR contributions
also removes the IR renormalons of
$X_{\bz}(Q^2)$
since they stem from the divergence of the 
integral \eqref{Borel} around $x=0$ for some positive $u$.
One can verify this by restarting from Eq.~\eqref{ExLB} with 
the IR cutoff $\mu_f$ and
tracing the above discussion.

\subsection{Extraction of cutoff-independent part:
General case}
\label{sec.GC}

The IR-subtracted quantity \eqref{start}
is free from the ambiguity caused by IR renormalons.
However, it has a cutoff dependence.
This dependence makes the prediction of Eq.~\eqref{start} unstable 
under the change of the artificial cutoff scale $\mu _f$
(which should eventually be canceled in a physical prediction).  
In this subsection we explain a method to extract 
a cutoff-independent part from this quantity.

Our method consists of two steps:
(i)~Rewrite the weight $w_X(x)$ by a new function $W_X(z)$
which is analytic in the upper half-plane and
 is related to $w_X(x)$ by
\be
2\, {\rm Im} \, W_X(x)=w_X(x) \qquad
(x \in \mathbb{R}~ \text{and}~ x>0) \label{rel} \, .
\ee
 We call $W_X$ as ``pre-weight.'' 
(We will shortly present a construction of $W_X$.)
(ii)~Deform the integral path in the complex $\tau$-plane.
The original integral path is decomposed as follows:
\begin{equation*}
\begin{tikzpicture}[baseline={(0,0.4)},scale=0.5]
\draw [->] (-1,0)--(7,0);
\draw [->] (0,-1)--(0,3);
\draw (6,3)--(6,2)--(7,2);
\draw (6.5,2.5) node {$\tau$};
\draw[-](3,-0.25)--(3,0.25);
\draw [->,ultra thick,red] (3,0)--(7,0);
\draw (1.7,0) node {$\times$};
\draw (1.65,-0.8) node {$e^{5/3} \LQ^2$};
\draw (3.1,0.7) node {$\mf^2$};
\end{tikzpicture}
\quad=\quad
\begin{tikzpicture}[baseline={(0,0.4)},scale=0.5]
\draw [->] (-1,0)--(7,0);
\draw [->] (0,-1)--(0,3);
\draw[-](3,-0.25)--(3,0.25);
\draw (1.7,0) node {$\times$};
\draw [->, ultra thick,red] (1.7,1.3)--(1.8,1.3);
\draw [ultra thick,red] (0,0)--(0.4,0);
\draw [ultra thick,red] (3,0) arc (0:180:1.3);
\draw [ultra thick,red] (3,0)--(7,0);
\draw [->,ultra thick,red] (4,0)--(6,0);
\draw (3,2) node {$C_a$};
\end{tikzpicture}
\quad - \quad
\begin{tikzpicture}[baseline={(0,0.4)},scale=0.5]
\draw [->] (-1,0)--(7,0);
\draw [->] (0,-1)--(0,3);
\draw[-](3,-0.25)--(3,0.25);
\draw (1.7,0) node {$\times$};
\draw [->, ultra thick,red] (1.7,1.3)--(1.8,1.3);
\draw [ultra thick,red] (0,0)--(0.4,0);
\draw [ultra thick,red] (3,0) arc (0:180:1.3);
\draw (3,2) node {$C_b$};
\end{tikzpicture}
\end{equation*}
Then Eq.~\eqref{start} is rewritten as
\begin{align}
X_{\beta_0}(Q^2; \mf)
=&{\rm Im} \int_{\mf^2}^{\infty} \frac{d \tau}{\pi \tau} \,
W_X \! \lt(\frac{\tau}{Q^2} \rt) \alpha_{\bz}(\tau)
\label{g0}\\
=&{\rm Im} \int_{C_a} \frac{d \tau}{\pi \tau}  \,
W_X\! \lt(\frac{\tau}{Q^2} \rt) \alpha_{\bz}(\tau)-
{\rm Im} \int_{C_b} \frac{d \tau}{\pi \tau}  \,
W_X\! \lt(\frac{\tau}{Q^2} \rt) \alpha_{\bz}(\tau) \, .
\label{g1}
\end{align}
The first term of Eq.~\eqref{g1}
(integral along $C_a$)
is clearly independent of $\mf$.
Although the second term
(integral along $C_b$)
is apparently $\mu_f$-dependent,
we can show that it also includes a
$\mf$-independent part.

Since $\mu_f^2\ll Q^2$ it would be
justified to expand $W_X(\tau/Q^2)$ about $\tau=0$ along $C_b$.
In this way
the second term of Eq.~\eqref{g1} 
is expressed in the large-$Q^2$ expansion: 
\be
{\rm Im} \int_{C_b} \frac{d \tau}{\pi \tau}  \,
W_X\! \lt(\frac{\tau}{Q^2} \rt) \alpha_{\bz}(\tau)
={\rm Im}\, \sum_{n \geq 0} 
c_n  \int_{C_b} \frac{d \tau}{\pi \tau}  \lt(\frac{\tau}{Q^2} \rt)^n \alpha_{\bz}(\tau) \label{sum_n_int} \, ,
\ee
with\footnote{
\label{fn:smallzexp}
We assume that the small-$z$ expansion of $W_X(z)$
exists, where 
the expansion can include half-integer powers of $z$ or 
powers of $\log z$.
For simplicity we explain in 
the case where $W_X$ is expanded as a
Taylor series in $z$.
In other cases, it only matters whether the integrand 
satisfies the relation $\{f(z)\}^*=f(z^*)$ or not
in classifying the Cases (I) and (II) in the following discussion.
}
\be 
W_X(z)=\sum_{n\geq 0} c_n z^n \, .
\label{eq_Ws}
\ee 
The $\mf$-dependence of the integral of
each term of Eq.~\eqref{sum_n_int}
can be classified into two cases.
\vspace*{2mm}
\\
{\bf Case (I)}: If the coefficient $c_n$ is real, 
the complex conjugate of the integral along $C_b$ 
becomes the integral along $C_b^*$
since the integrand satisfies the relation $\{f(z)\}^*=f(z^*)$.
Hence, we obtain
\begin{align}
{\rm Im} \int_{C_b} \frac{d \tau}{\pi \tau} \,
c_n \lt(\frac{\tau}{Q^2} \rt)^n \alpha_{\bz}(\tau)
&=\frac{1}{2 i} \lt(\int_{C_b}-\int_{C_b^*} \rt) \frac{d \tau}{\pi \tau} \,
c_n \lt(\frac{\tau}{Q^2} \rt)^n \alpha_{\bz}(\tau) \non
&=\frac{1}{2 \pi i} \int_{C_{\LQ} } \frac{d \tau}{\tau}  \,
c_n \lt(\frac{\tau}{Q^2} \rt)^n \alpha_{\bz}(\tau) \non
&=-\frac{4 \pi c_n}{\bz} \left(\frac{e^{5/3} \LQ^2}{Q^2} \right)^n \, ,
\label{Cauchy}
\end{align}
where the integration contours
$C_b^*$ and $C_{\LQ}$
are defined as below.
\begin{equation*}
\begin{tikzpicture}[baseline={(0,0.4)},scale=0.5]
\draw [->] (-1,0)--(5,0);
\draw [->] (0,-1)--(0,3);
\draw[-](3,-0.25)--(3,0.25);
\draw (1.7,0) node {$\times$};
\draw [->, ultra thick,red] (1.7,1.3)--(1.8,1.3);
\draw [ultra thick,red] (0,0)--(0.4,0);
\draw [ultra thick,red] (3,0) arc (0:180:1.3);
\draw (3,2) node {$C_b$};
\end{tikzpicture}
\quad - \quad
\begin{tikzpicture}[baseline={(0,0.4)},scale=0.5]
\draw [->] (-1,0)--(5,0);
\draw [->] (0,-1)--(0,3);
\draw[-](3,-0.25)--(3,0.25);
\draw (1.7,0) node {$\times$};
\draw [->, ultra thick,red] (1.7,-1.3)--(1.8,-1.3);
\draw [ultra thick,red] (0,0)--(0.4,0);
\draw [ultra thick,red] (0.4,0) arc (180:360:1.3);
\draw (3,-1.3) node {$C_b^*$};
\end{tikzpicture}
\quad = \quad
\begin{tikzpicture}[baseline={(0,0.4)},scale=0.5]
\draw [->] (-1,0)--(5,0);
\draw [->] (0,-1)--(0,3);
\draw[-](3,-0.25)--(3,0.25);
\draw (1.7,0) node {$\times$};
\draw [ultra thick,red] (2.2,0) arc (0:360:0.5);
\draw (3,0.7) node {$\mf^2$};
\draw [->,ultra thick,red] (1.7,0.5)--(1.8,0.5);
\draw (1.7,1.3) node {$C_{\LQ}$};
\end{tikzpicture}
\end{equation*}
Here we use the fact that $C_b-C_b^*$ becomes a closed contour
surrounding the pole at $e^{5/3} \LQ^2$. 
Therefore the result is $\mf$-independent 
and can be calculated analytically 
by the Cauchy theorem. 
We see that positive powers of $\LQ$ appear.
\vspace*{2mm}
\\
{\bf Case (II)}: If the coefficient $c_n$ has a non-zero imaginary part,
the above argument does not hold
since the integrand does not satisfy the relation $\{f(z)\}^*=f(z^*)$.
In this case $\mf$-dependence remains in the result:
\be
{\rm Im} \int_{C_b} \frac{d \tau}{\pi \tau} 
c_n \lt(\frac{\tau}{Q^2} \rt)^n \alpha_{\bz}(\tau)
=\mathcal{O}((\mf^2/Q^2)^n) \, .
\label{nonCauchy}
\ee
Thus, $\mf$-independent part appears 
not only from the integral along $C_a$ 
but also from the integral along $C_b$ 
depending on whether the expansion coefficient $c_n$ is real or complex.

We can find
whether the coefficient $c_n$ in Eq.~\eqref{eq_Ws}
is real or complex 
without knowing the concrete form of $W_X$.
The insight is obtained using
the expansions of $w_X$ [Eq.~\eqref{invBexp}] and $W_X$ 
[Eq.~\eqref{eq_Ws}] and the relation between them [Eq.~\eqref{rel}].
Schematically the relation can be understood as follows:
\begin{eqnarray}
\begin{array}{l}
n \notin U_{\IR} 
~~ \longleftrightarrow 
~~2 \, {\rm Im}\, c_n=b_n=0
~~ \longleftrightarrow
~~c_n\in \mathbb{R} ~\longleftrightarrow \text{case\,~(I)} 
\\
n \in  U_{\IR} 
~~ \longleftrightarrow 
~~2 \, {\rm Im}\, c_n=b_n \neq 0
~~ \longleftrightarrow
~~c_n\notin \mathbb{R} ~\longleftrightarrow \text{case (II)}
\end{array}
\label{exprel}
\end{eqnarray}
Namely, the knowledge on the IR renormalons of $X_{\beta_0}(Q^2)$ is 
sufficient to judge $\mu_f$-independence of
each term of Eq.~\eqref{sum_n_int}.

From the above discussion, 
by taking the terms for $0\le n<u_{\IR}$ of Eq.~\eqref{sum_n_int} 
and the first term of Eq.~\eqref{g1},
we obtain the general result for $X_{\bz}(Q^2;\mf)$, where the $\mf$-independent part is separated: 
\be
X_{\bz}(Q^2;\mf)
=X_{\UV}(Q^2)
+\mathcal{O}
\left(
\lt(\mf^2/Q^2 \rt)^{u_{\IR}}
\right) \, . \label{genres}
\ee 
We have extracted the $\mf$-independent part $X_{\UV}$ given by 
\be
X_{\UV}(Q^2)
={\rm Im} \int_{C_a} \frac{d \tau}{\pi \tau} \,
W_X \!\lt(\frac{\tau}{Q^2} \rt) \alpha_{\bz}(\tau)
+\sum_{0 \leq n<u_{\IR}} \frac{4 \pi c_n}{\bz} 
\lt( \frac{e^{5/3}\LQ^2}{Q^2} \rt)^n 
\label{SUV} \, .
\ee
This is one of the main results in this paper. 
$X_{\UV}(Q^2)$ is insensitive to IR physics and can be 
regarded as a genuine UV contribution.
 
We rewrite $X_{\UV}$ as
\be
X_{\UV}(Q^2)=X_0(Q^2)
+\sum_{0 < n<u_{\IR}} \frac{4 \pi c_n}{\bz} 
\lt( \frac{e^{5/3}\LQ^2}{Q^2} \rt)^n \label{SUV2} \, ,
\ee 
with
\be
X_0(Q^2)
={\rm Im} \int_{C_a} \frac{d \tau}{\pi \tau} \,
W_X \! \lt(\frac{\tau}{Q^2} \rt) \alpha_{\bz}(\tau)+\frac{4 \pi c_0}{\bz} 
\label{S0} \, .
\ee
The asymptotic form of $X_0$ 
as $Q^2 \to \infty$ is given by
\be
X_0(Q^2) \to d_0 \alpha_s(Q)
=\frac{4 \pi d_0}{\bz} \frac{1}{\log({Q^2/\LQ^2})} 
\label{asym}  \, .
\ee
This is the leading term of the asymptotic expansion of $X_0$
that will be derived in Eq.~\eqref{asymD0} below;
it is also a consequence of
the renormalization-group (RG) equation.\footnote{
Since the leading logarithmic  terms 
are proportional to $\alpha_s(\mu)[\bz \alpha_s(\mu) \log(Q/\mu)]^n$, they
are incorporated correctly by 
the large-$\bz$ approximation.
The modification of the perturbative series by the IR cutoff
is power-suppressed $\sim (\mf^2/Q^2)^{k}$,
hence the leading large-$Q^2$ behavior is determined by
the one-loop RG equation.
}
This gives a more dominant contribution 
than power behaviors for large $Q^2$. 
Therefore $X_{\UV}(Q^2)$ indeed 
gives a leading behavior of $X_{\bz}(Q^2;\mf)$ for large $Q^2$.
In explicit examples in Secs.~\ref{ss:adler} and \ref{ss:force}, we will see that
Eq.~\eqref{SUV2} represents a separation of $X_{\UV}(Q^2)$ into
a logarithmic term\footnote{
By a ``logarithmic term'' we mean a term 
which is closest
to $(Q^2/\LQ^2)^P$ with $P=0$
in the entire range $0<Q^2<\infty$, if it is
compared with a single power dependence on $Q^2$
(for an integer $P$); see Figs.~\ref{Fig.Adler1},\,\ref{Fig.Force1} and
Sec.~\ref{sec.FM2}.
}
 (non-power correction term) $X_0$ and 
power correction terms $\sim (\LQ^{2}/Q^{2})^n$.

Up to this point we have considered a general pre-weight $W_X(z)$, 
which is analytic in the upper half-plane and satisfies 
the relation \eqref{rel}.
A pre-weight which satisfies these conditions
can be constructed systematically 
as
\be
W_X(z)=\int_{0}^{\infty} \frac{d x}{2 \pi} \frac{w_X(x)}{x-z-i0}  
\label{massglu}  \, ,
\ee
due to the relation 
${\rm Im} \{(x-z-i0)^{-1} \}=\pi \delta(x-z)$ for $z\in \mathbb{R}$. 
The integral in Eq.~\eqref{massglu} always converges 
according to our assumption on the convergence of $X_{\rm LO}$.
Note that there are potentially an 
infinite number of candidates for the pre-weight $W_X$
since Eq.~\eqref{rel} does not restrict its real part on the 
positive real axis. 
Thus, $W_X$ defined by Eq.~\eqref{massglu} 
represents just one possibility and   
we refer to the choice Eq.~\eqref{massglu} 
as ``massive gluon scheme.'' 
This is because this construction
is equivalent to replacing the gluon propagator to that with 
a tachyonic mass $m^2=-\tau$
in the leading order contribution Eq.~\eqref{oneloop}:
\footnote{
There exist many studies on 
low-energy QCD phenomena
(especially chiral symmetry breaking
and confinement)
in terms of massive gluons 
\cite{Maskawa:1974vs,Cornwall:2010ap,RodriguezQuintero:2010wy}.
We stress, however, that we study perturbative (UV) contributions
using $W_X$.
}
\begin{align}
\int_0^{\infty} \frac{d (p^2)}{2 \pi} \frac{w_X(p^2/Q^2)}{p^2}  
~~\to ~~
& \int_0^{\infty} \frac{d (p^2)}{2 \pi} \frac{ w_X(p^2/Q^2) }{p^2-\tau-i0}
= W^{(m)}_X(\tau/Q^2)
\, , \label{replace}
\end{align}
where $W^{(m)}_X$ denotes the pre-weight in the massive gluon
scheme.

We note that one does not have to start from $w_X$ to obtain $W_X$
in the massive gluon scheme.
It is sufficient to use the gluon propagator with a 
tachyonic mass in the usual loop calculation, i.e.,
starting from the expression retaining all the loop momentum integrals,  
since it coincides with the right-hand side of Eq.~\eqref{replace}.
If we take this route, we rather obtain the weight
$w_X$ via the relation \eqref{rel}
{\it after} calculating the pre-weight $W_X$.

For later convenience, we introduce $W_{X+}$ from the pre-weight in
the massive gluon scheme as
\be
W^{(m)}_{X+}(z)\equiv
W^{(m)}_X(-z)=\int_{0}^{\infty} \frac{d x}{2 \pi} 
\frac{w_X(x)}{x+z-i0}  
\label{W+}  \,  .
\ee
This function is real for $z>0$ since 
$w_X(x)$ is real
and $x+z>0$.
Using this function, Eq.~\eqref{S0} can be expressed as%
\footnote{
A quantity similar to $X_{\UV}$ with this $X_0$ 
is derived in Ref.~\cite{Ball:1995ni}
using a regularized Borel integral.
Our derivation is different from theirs
in that our result does not contain renormalon uncertainties 
since we subtract IR modes in Eq.~\eqref{start}.
}
\be
X_0(Q^2)=\int_{0}^{\infty} \frac{d \tau}{\pi \tau} \,
W^{(m)}_{X+} \!\lt(\frac{\tau}{Q^2} \rt) 
{\rm Im} \, \alpha_{\bz}(-\tau+i0)
+\frac{4 \pi c_0}{\bz} \, ,
\label{S0rot} 
\ee
\be
{\rm Im} \, \alpha_{\bz}(-\tau+i0)=\frac{4 \pi}{\bz} \frac{-\pi}{\log^2{(\tau e^{-5/3}/\LQ^2)}+\pi^2} \, ,
\ee
in the case that it is justified to deform the integral path $C_a$ 
to the straight line connecting $\tau=0$ to $-\infty$. 
This expression has a good analytical property as we will see later
(end of Sec.~\ref{ss:adler} and Sec.~\ref{sec.FM}).
In calculating the asymptotic form 
of $X_0(Q^2)$ as $Q^2 \to \infty$ or $Q^2 \to 0$,
the following expression, obtained by partial integration, is useful:
\begin{align}
X_0(Q^2)=&-\int_0^{\infty} \frac{d x}{\pi} \,
W^{(m)\,\prime}_{X+} (x) \, \frac{4 \pi}{\bz}\, {\rm Im} \log{\log{(Q^2/(e^{5/3} \LQ^2))}}  
\non
&-\int_0^{\infty}  \frac{d x}{\pi}  \,
W^{(m)\,\prime}_{X+} (x) \, \frac{4 \pi}{\bz} \tan^{-1} 
\lt[ \frac{\pi}{\log{(Q^2/(e^{5/3} \LQ^2))}+\log{x}} \rt] 
\label{S0forAsymp} \, .
\end{align}

\subsection{Example 1: Adler function}
\label{ss:adler}

As an application of the general framework
presented in the previous subsection,
we examine large-$Q^2$ behavior 
of the Adler function~\cite{Mishima:2016xuj}.
This observable is suited to test our method, in particular since 
OPE can be performed. 
The first IR renormalon is located at $u_{\IR}=2$ \cite{Mueller:1984vh}, 
and thus the renormalon uncertainty is fairly suppressed. 

We study the reduced Adler function $D(Q^2)$
with one massless quark,  defined as
\be
D(Q^2)=4 \pi^2 Q^2 \frac{d \Pi(Q^2)}{d Q^2}-1\, ,
\ee
where $\Pi(Q^2)$ is a correlator%
\footnote{
Eq.~\eqref{defPi} uses the Minkowski metric,
where $q$ denotes the four-momentum of the vacuum polarization.
In our letter~\cite{Mishima:2016xuj}
the sign of the corresponding equation [Eq.~(2)] was incorrect and
should be reversed.
}
of 
the quark current $J^{\mu}(x)= \bar{q}(x) \gamma^{\mu} q(x)$,
\begin{eqnarray}
(q^{\mu} q^{\nu} -g^{\mu \nu} q^2)\Pi(Q^2)
=-i \int d^4 x \,  e^{-i q \cdot x}  \braket{0|
\mathrm{T}
J^{\mu}(x) J^{\nu}(x)|0}
\,,
~~~
Q^2=-q^2>0 \,.
\non 
\label{defPi}
\end{eqnarray}

We define
the reduced Adler function 
in the large-$\bz$ approximation 
with an IR cutoff  as
\begin{align}
D_{\bz}(Q^2;\mf)
=\int_{\mf^2}^{\infty}
\frac{d \tau}{2 \pi \tau}\, w_D \!\lt(\frac{\tau}{Q^2} \rt) \alpha_{\bz}(\tau) \, .
\label{ad}
\end{align}
The weight $w_D(x)$ is given by \cite{Neubert:1994vb}
\begin{align}
&w_D(x)=
\frac{N_C C_F}{3} \times 
\nonumber\\
&
\left\{
\begin{array}{ll}
 (7-4 \log{x}) x^2+4 x(1+x) \{\Li_2(-x)+\log{x} \log{(1+x)}\} 
&;x<1\\
3+2 \log{x}+4x(1+\log{x})
+4x(1+x) \{ \Li_2(-x^{-1})-\log{x} \log{(1+x^{-1})} \} 
&;x>1
\end{array}
\right.
\, ,
\label{wD}
\end{align}
where $N_C=3$ is the number of colors and $C_F=4/3$ is 
the Casimir operator of the fundamental representation.
The first IR renormalon is located at $u_{\IR}=2$, as 
can be seen from the expansion of $w_D(x)$ and Eq.~\eqref{invBexp}:
\be
w_D(x)=N_C C_F x^2+\dots \, .
\ee
The pre-weight
$W^{(m)}_D$ and $W^{(m)}_{D+}$ in
the massive gluon scheme, obtained via Eq.~\eqref{massglu} or by calculating 
the two-loop integral,
are given by
\begin{align}
 W&^{(m)}_D(z)
 =\frac{N_C C_F}{12\pi} 
 \Bigl[ 3+16z(z+1) H(z)
-14 z^2 \log{(-z)} \nonumber \\ 
&+8 z(z+1) \{ -\log (-z)
{\rm Li}_2(-z)+{\rm Li}_3(z)+{\rm Li}_3(-z)\}\nonumber \\
&+4\{2 z^2+2z+1-4z(z+1)\log{(1+z)}\}{\rm Li}_2(z)  \nonumber \\
& +2(7 z^2-4 z-3)\log{(1-z)}-8\zeta_2z(z+1) \log{(1+z)}
\nonumber \\ 
&+4\{z^2-z(z+1)\log({1+z})\} \log^2{(-z)}
\nonumber \\
&+2(4\zeta_2-7 \zeta_3 )z^2+2(11-7 \zeta_3)z 
\, \Bigr] 
\label{WD} \, 
\end{align}
and
\be
W^{(m)}_{D+}(z) \equiv W^{(m)}_D(-z) \, .
\ee
Here, we define $H(z)=\int_{z}^{1} dx\, x^{-1}\log{(1+x)} \log{(1-x)}$;
${\rm Li}_n(z)=\sum_{k=1}^\infty\frac{z^k}{k^n}$ denotes the
polylogarithm;
$\zeta_k=\zeta(k)$ denotes the Riemann zeta function.\footnote{
$H(z)$ can be expressed using the harmonic polylogarithms.
}
We present another expression of $W^{(m)}_D$ in App.~\ref{app:adler},
which is lengthier but exhibits the structure of the singularities 
more clearly. 
The first few terms
of the small-$z$ expansion of $W^{(m)}_D$ are given by%
\footnote{
This series expansion was obtained in Ref.~\cite{Ball:1995ni}.
}
\be
W^{(m)}_D(z)
=N_C C_F \lt[ \frac{1}{4 \pi}
+\frac{2(4-3 \zeta_3)}{3 \pi} z
+\frac{10-12 \zeta_3-3 \log{z}+3 i \pi}{6 \pi} z^2
+\dots \rt] \,  \label{WDexp} .
\ee
Following the discussion in the general case, we can extract 
the $\mf$-independent part $D_{\UV}$: 
\begin{align}
D_{\beta_0}(Q^2;\mf)=D_{\UV} (Q^2)
+\mathcal{O}(\mf^4/Q^4)
  \label{a20}
\end{align}
with 
\begin{align}
&D_{\UV}(Q^2)=D_0(Q^2)+\frac{8(4-3 \zeta_3) e^{5/3} N_C C_F}{3 \bz} \frac{\LQ^2}{Q^2} \, ,
\label{defDUV} \\
&D_0(Q^2)=\int_{0}^{\infty} \frac{d \tau}{\pi \tau} \,
W^{(m)}_{D+}\! \lt(\frac{\tau}{Q^2} \rt) {\rm Im} \, \alpha_{\bz}(-\tau+i0)
+\frac{N_C C_F}{\bz} \, .
\label{defD0}
\end{align}
The $\LQ^2/Q^2$-term arises from 
the $z^1$-term
of the pre-weight $W_D(z)$; see Eq.~\eqref{WDexp}.
The large-$z$ behavior of $W_D(z)$ allows rotation
of the integration contour and
we write $D_0$ as in Eq.~\eqref{S0rot}.
The asymptotic behaviors of $D_0(Q^2)$
are obtained as
\be
D_0(Q^2) \to 
\begin{cases}
\frac{N_C C_F}{\bz} \frac{1}{\log{(Q^2/\LQ^2)}} ~~ \text{as}~ Q^2 \to \infty \\
\frac{N_C C_F}{\bz}         ~~~~~~~~~\,\,~~~~~~~~  \text{as}~  Q^2 \to 0
\end{cases} \, ,
\ee
and these asymptotic forms are interpolated smoothly in the intermediate
region.
Hence, qualitatively $D_0$ behaves as a constant term with a logarithmic
correction at large $Q^2$.

 \begin{figure}[t]
 \centering
 \begin{tabular}{ll}
 \begin{minipage}{0.5\hsize}
 \centering
 \includegraphics[width=7cm]{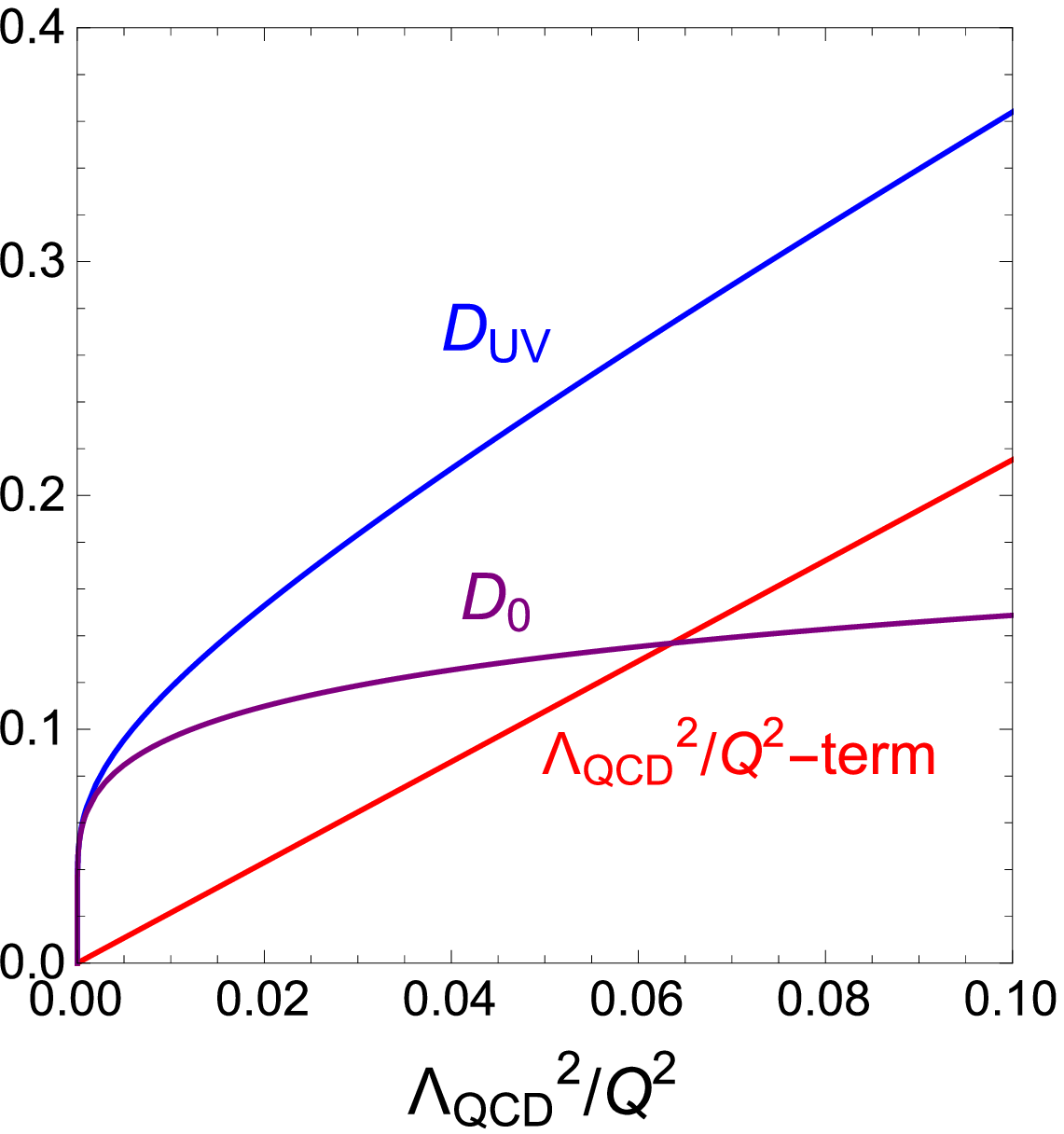}
 \end{minipage}
 \begin{minipage}{0.5\hsize}
 \centering
 \includegraphics[width=7cm]{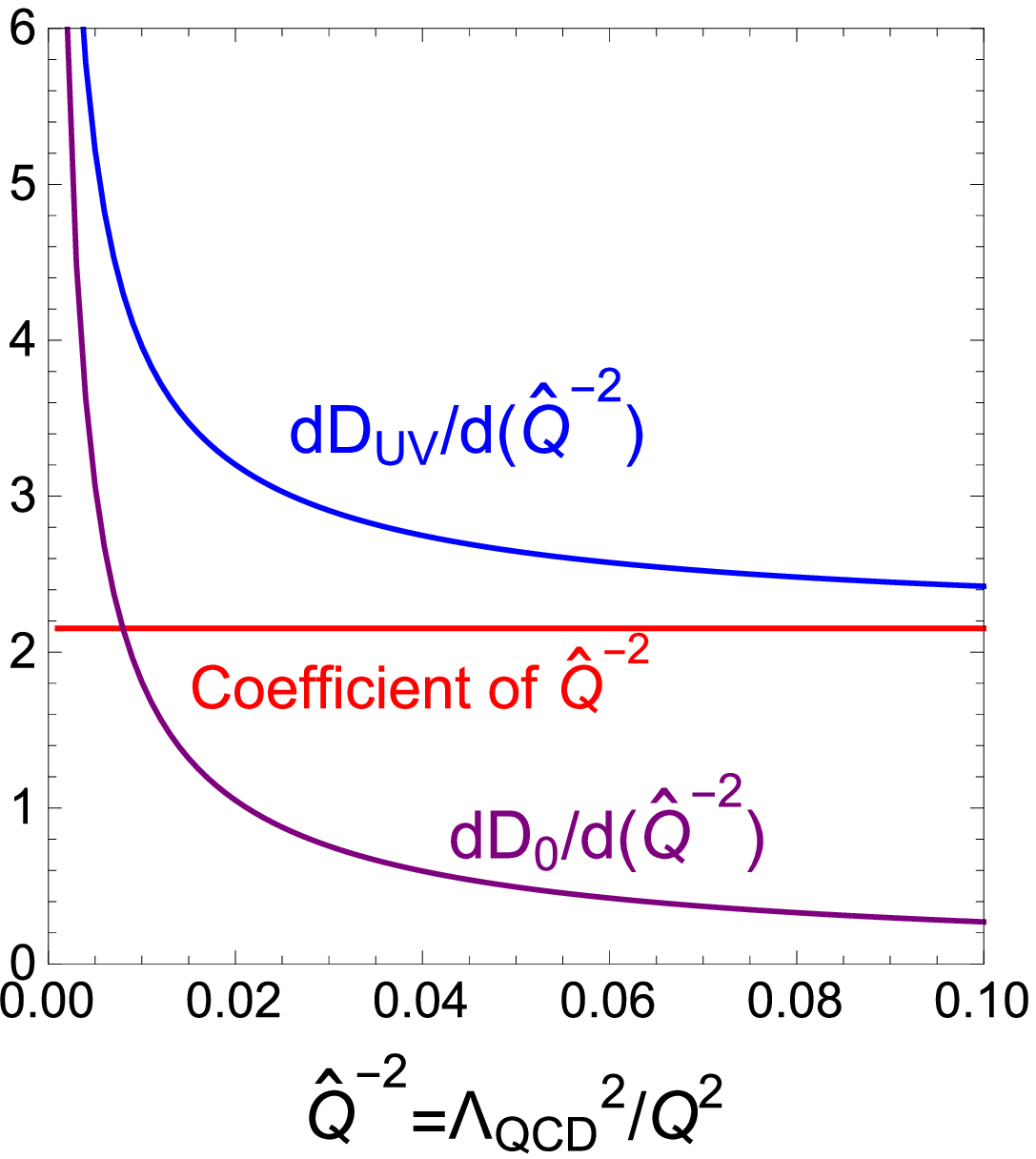}
 \end{minipage}
 \end{tabular}
 \caption{\small [Left]
 $D_{\rm UV}$ [Eq.~\eqref{defDUV}], 
 $D_0$ [Eq.~\eqref{defD0}] 
 and the $\LQ^2/Q^2$-term 
 [Eq.(\ref{defDUV})] as functions of $\LQ^2/Q^2$.
 [Right]
 Derivatives of 
 $D_{\UV}$, $D_0$ and the $\LQ^2/Q^2$-term with respect to 
$\hat{Q}^{-2} \equiv \LQ^2/Q^2$.}
 \label{Fig.Adler1}
 \end{figure}
 
In Fig.~\ref{Fig.Adler1},
$D_\mathrm{UV}$, $D_0$
and the $\LQ^2/Q^2$-term of Eq.~\eqref{defDUV}
are plotted as functions of $\LQ^2/Q^2$.
The $\LQ^2/Q^2$-term naturally explains the power-like behavior of $D_{\UV}$, which looks linear 
in this figure. 
In fact, the derivative of $D_{\UV}$ is given by 
the $\LQ^2/Q^2$-term dominantly in the range $\LQ^2/Q^2 \gtrsim 0.01$. 
In Sec.~\ref{sec.RB} we will compare $D_{\rm UV}$ with the large-order perturbative
prediction in the large-$\beta_0$ approximation as well as
with the known exact perturbative series, where
we will find good agreement.

The $\mu_f$-dependence of 
the $1/Q^4$-term in Eq.~\eqref{a20}
shows a sensitivity to IR dynamics
and can be interpreted in the context of OPE.
In OPE, 
the reduced Adler function is expressed in terms of 
vacuum expectation values (VEVs) 
of operators which are invariant under
Lorentz and gauge symmetries:
\be
D(Q^2)
=C_1+C_{GG} 
\frac{\braket{0|G^{a \mu \nu} G^{a}_{\mu \nu}|0}}{Q^4}
+\dots \, , \label{OPE.Adler}
\ee  
where $C_1$ and $C_{GG}$ represent the Wilson coefficients of the
operators $\bold{1}$ and $G^{a \mu \nu} G^{a}_{\mu \nu}$, respectively.
The VEV of $G^{a \mu \nu} G^{a}_{\mu \nu}$,
known as the local gluon condensate,
has mass-dimension four and
hence it is accompanied by the factor $1/Q^4$.
The gluon condensate is determined by IR dynamics and 
it would have a dependence on the UV cutoff scale $\mf$ 
of the low energy effective theory.   
We can interpret that the IR cutoff dependence of $D_{\bz}(Q^2;\mf)$ 
at the order $1/Q^4$ in Eq.~\eqref{a20} is a counterpart of 
the UV cutoff dependence of the gluon condensate.
In other words, if we include the gluon condensate 
as determined by IR dynamics, the leading $\mf$-dependence of
$D_{\bz}(Q^2;\mf)$ would be canceled and the
$1/Q^4$-term is expected to be reduced to order $\LQ^4/Q^4$.

In the OPE framework,
$D_{\UV}$ including the $\LQ^2/Q^2$-term 
is identified with $C_1$ in Eq.~\eqref{OPE.Adler}
as we will clarify in Sec.~\ref{sec.EbR}.
In this sense, the $\mf$-independent $\LQ^2/Q^2$-term does not conflict 
with the structure of OPE, and
what we have found in this subsection is
a non-trivial behavior 
of the Wilson coefficient $C_1$ of the reduced Adler function.
Due to this power correction,
we conclude that the Adler function has the leading power dependence
as $\LQ^2/Q^2$
rather than $\LQ^4/Q^4$ at large $Q^2$ 
as long as the large-$\bz$ approximation is valid.
(We discuss subtle issues on the $\LQ^2/Q^2$-term further
in Secs.~\ref{sec.SD} and \ref{sec.FM}.)

Finally we comment on the analytic structure of the Adler function.
 It is known that the Adler function in perturbative QCD
is an analytic function in the complex $Q^2$-plane, 
with a cut along the negative axis from $Q^2=0$
corresponding to the   threshold of
massless partons,
and with the $1/(\bz \log Q^2)$ singularity at
$Q^2=\infty$ dictated by the RG equation.
One can see that the expression of $D_{\UV}$ of Eq.~\eqref{defDUV} with Eq.~\eqref{defD0}
indeed satisfies these requirements. 
The cut arises from the property of $W^{(m)}_{D+}$
that it has an imaginary part when the argument becomes negative 
due to the relation \eqref{rel}. 
However, if we represent $D_0$ as in Eq.~\eqref{S0}, it cannot
be regarded as an analytic function of $Q^2$
since it is given by the imaginary part of a function. 
The representation \eqref{S0} is defined only for 
real positive $Q^2$,
whereas the representation \eqref{S0rot} is defined in the entire complex $Q^2$ plane.
They are equivalent only if we limit $Q^2$ to a real positive parameter.
Thus, from the viewpoint of  analyticity, the latter representation 
turns out to be superior
to the former.

\subsection{Example 2: Force between static quark-antiquark pair}
\label{ss:force}
As another application of
the method presented in Sec.~\ref{sec.GC},
we consider the short-distance
behavior of the force between a static quark-antiquark pair,
which is obtained from the derivative of
the static QCD potential.
The static QCD potential has been studied extensively to understand
the nature of the force between the quark and antiquark.
At short distances 
perturbative QCD prediction is accurate, whereas at large distances lattice
QCD predictions are accurate.
There is a significant overlap region 
at intermediate distances, where both predictions agree well.
Presently the exact perturbative series are known 
up to NNNLO~\cite{Anzai:2009tm, Smirnov:2009fh,Lee:2016cgz}. 
In addition, the low energy effective theory ``potential non-relativistic QCD 
(pNRQCD)''
is known, in which OPE can be performed, and
there is a good theoretical understanding 
of the connection between UV and IR contributions.
Therefore the QCD potential (or the force)
is an optimal observable
to examine our formulation.

The potential energy between 
the static quark $Q$ and antiquark $\bar{Q}$
(QCD potential)
in the large-$\bz$ approximation and with an IR cutoff
is given by
\be
V_{\bz}(r;\mf)
=-
{\hbox to 18pt{
\hbox to -5pt{$\displaystyle \int$} 
\raise-16pt\hbox{$\scriptstyle p>\mf$} 
}}
\frac{d^3 \vec{p}}{(2 \pi)^3} \, e^{i \vec{p} \cdot \vec{r}} \, 
\frac{4 \pi  C_F}{p^2} \, \alpha_{\bz}( p^2)
=-\frac{1}{r} \int _{\mf^2}^{\infty} \frac{d \tau}{2 \pi \tau} 
2C_F \sin(\sqrt{\tau}r) \alpha_{\bz}(\tau)
\, .
\label{f1}
\ee
Here, the typical (energy) scale is $r^{-1}$, 
the inverse of the distance between $Q \bar{Q}$.
Comparing Eq.~\eqref{f1} with Eq.~\eqref{start},
the weight of the (dimensionless)
QCD potential $r V_{\bz}(r)$ is 
given by
\begin{align}
w_V(x)=
-2C_F\sin (\sqrt{x}) \, .
  \label{wV}
\end{align}
Comparing its expansion and Eq.~\eqref{invBexp}
we find that the first IR renormalon
of the QCD potential is located
at $u=1/2$.
The pre-weight in the massive gluon scheme
is obtained using Eq.~\eqref{massglu} as
\begin{align}
W^{(m)}_V(z)=-C_F e^{i \sqrt{z}} \, .
\label{pre_weight_of_V}
\end{align}
This function is utilized throughout the analyses in 
Refs.~\cite{Sumino:2004ht,Sumino:2005cq,Sumino:2014qpa} and 
turns out to correspond to the massive gluon scheme.
By using this pre-weight, $V_{\bz}$ is expressed as
\begin{align}
V_{\bz}(r;\mf)=\frac{1}{r} \, {\rm Im} \int_{\mf^2}^{\infty} \frac{d \tau}{\pi \tau} W_V^{(m)} (\tau r^2) \alpha_{\bz}(\tau) \, . \label{VwithW}
\end{align}
Continuing the discussion given in Sec.~\ref{sec.GC}, 
one obtains the result for the QCD potential in the large-$\bz$ approximation 
\cite{Sumino:2004ht,Sumino:2005cq,Sumino:2014qpa}. 
However, in order to circumvent the first IR renormalon at $u=1/2$,
which is relatively close to the origin, 
we analyze the force between $Q \bar{Q}$, 
$F_{\bz} (r^2)=-dV_{\bz}(r)/dr$ \cite{Sumino:2001eh}.\footnote{ 
The first IR renormalon only gives an uncertainty to the constant 
($r$-independent) part of the potential.}

The force between $Q \bar{Q}$ with an IR cutoff is
obtained by differentiating Eq.~\eqref{f1} and Eq.~\eqref{VwithW} with respect to $r$:
\begin{align}
F_{\bz} (r^2 ;\mf) 
&\equiv -C_F \frac{\alpha_{F , \bz}(1/r^2;\mf)}{r^2} \label{f2} \\
&=-\frac{C_F }{r^2} \int_{\mf^2}^{\infty} 
\frac{d \tau}{2 \pi \tau}\, w_F(\tau r^2)\, \alpha_{\bz}(\tau) \\
&=-\frac{C_F }{r^2} \, {\rm Im} \int_{\mf^2}^{\infty}  \frac{d \tau}{\pi \tau}\, W_F^{(m)}(\tau r^2)\, \alpha_{\bz}(\tau)
\, ,
\end{align}
where the weight $w_F(x)$ and the pre-weight $W_F^{(m)}(z)$
are given by\footnote{
The convergence of $w_F$ in UV region
is not always sufficient to derive some of the relations
discussed in this paper.
In such a case,
generally convergence in UV region is better with $w_V$,
and we can differentiate by $r$ after the $\tau$-integral.
The same result can be obtained with $w_F$ directly,
if we regularize the integral measure first as $d \tau \to d \tau \, \tau^{-\epsilon}$
and take the limit $\epsilon \to 0$
after the $\tau$-integral 
(we do not encounter divergences).
}
\begin{align}
w_F(x)=
2  (\sin{\sqrt{x}}-\sqrt{x} \cos{\sqrt{x}}) 
\label{f3}
\end{align}
\begin{align}
W_F^{(m)}(z)=e^{i \sqrt{z}} (1-i \sqrt{z}) \, .
\end{align}
In the following we deal with the dimensionless force 
(or the $F$-scheme coupling)
$\alpha_{F , \bz}$, defined by Eq.~\eqref{f2}. 
The expansion of the weight reads
\begin{align}
w_F (x)
=\frac{2}{3} x^{3/2}-\frac{1}{15} x^{5/2}+\cdots \, ,
\label{f4}
\end{align}
hence, the first IR renormalon of 
$\alpha_{F,\bz}$ is 
indeed shifted to $u_{\IR}=3/2$.
The expansion of the pre-weight is given by
\be
W^{(m)}_F(z)=1+\frac{z}{2}+\frac{i}{3} z^{3/2}+\cdots . \label{Walphaexp}
\ee
From the general discussion we can extract the
$\mf$-independent part $\alpha_{F , \UV}(1/r^2)$ from
$\alpha_{F , \bz}(1/r^2;\mf)$ as 
\be
\alpha_{F , \bz}(1/r^2;\mf)
=\alpha_{F,\UV}(1/r^2)+\mathcal{O}(\mf^3 r^3)
\label{f7}
\ee
with
\be
\alpha_{F , \UV}(1/r^2)
=\alpha_{F ,0}(1/r^2)+\frac{2 \pi}{\bz} \LQ^2 e^{5/3} r^2  \,,
\label{defalphaFUV}
\ee
\be
\alpha_{F,0}(1/r^2)
=\int_0^{\infty} \frac{d \tau}{\pi \tau} \,
W^{(m)}_{{F +}} \! \lt(\tau r^2 \rt) {\rm Im} \, 
\alpha_{\bz}(-\tau+i0)+\frac{4 \pi}{\bz}
\, ,
\label{f8}
\ee
where $W_{F+}^{(m)}(z)=e^{-\sqrt{z}}(1+\sqrt{z})$.
The $\LQ^2 r^2$-term arises from the $z^1$-term of 
the pre-weight $W^{(m)}_F(z)$ [Eq.~\eqref{Walphaexp}].
This power behavior corresponds to a linear potential 
in the QCD potential.
The asymptotic behaviors of $\alpha_{F, 0}$ are obtained via
Eq.~\eqref{S0forAsymp} as
\be
\alpha_{F,0}(1/r^2) \to 
\begin{cases}
\frac{4 \pi}{\bz} \frac{1}{|\log{(r^2 \LQ^2)}|} ~~~ \text{as}~ r^2 \to 0 \\
\frac{4 \pi}{\bz}         ~~~~~~~~~\,\,~~~~~~~~  \text{as}~  r^2 \to \infty
\end{cases} \, .
\ee

 \begin{figure}[t]
 \centering
 \begin{tabular}{ll}
 \begin{minipage}{0.5\hsize}
 \centering
 \includegraphics[width=7cm]{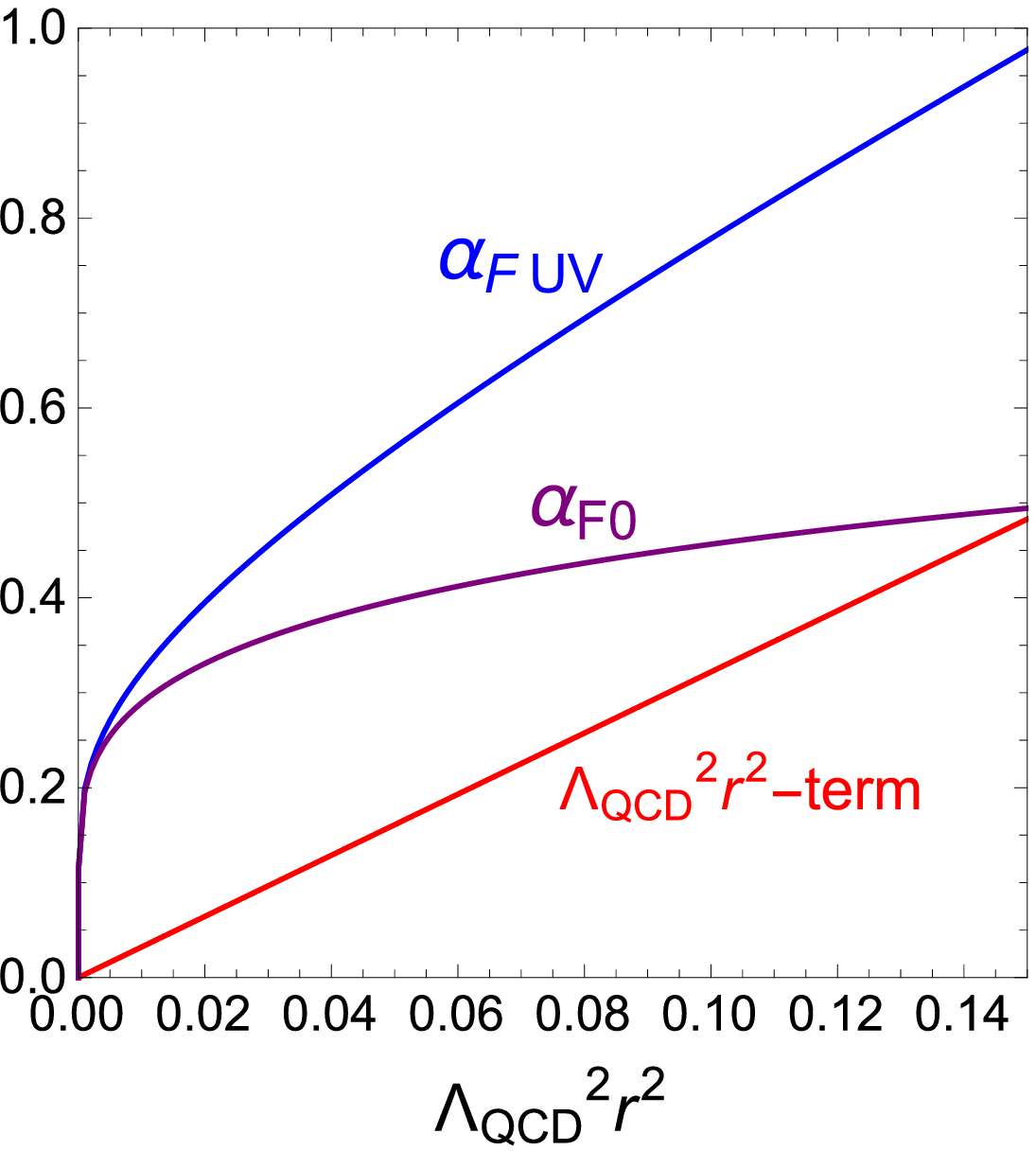}
 \end{minipage}
 \begin{minipage}{0.5\hsize}
 \centering
 \includegraphics[width=7cm]{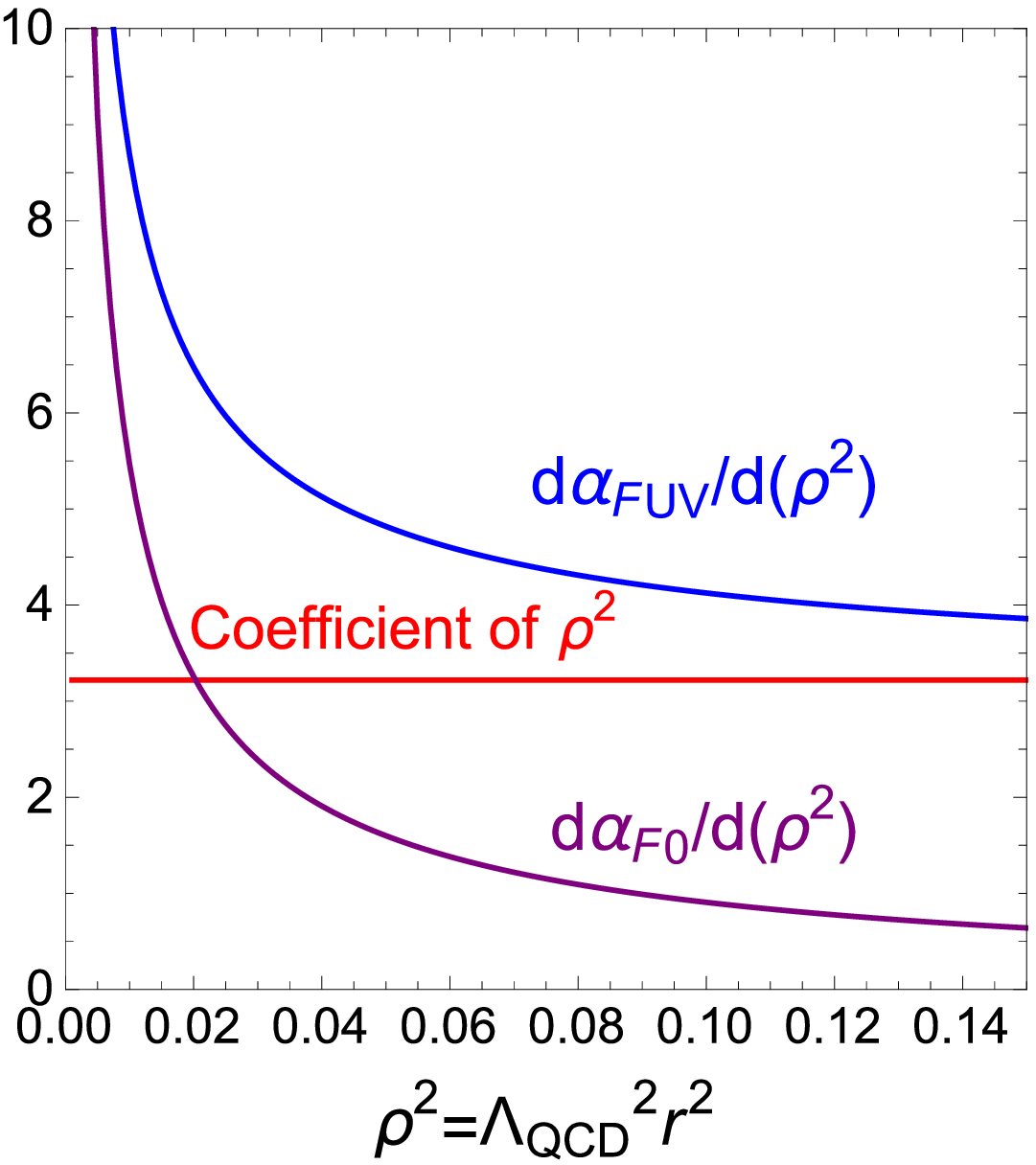}
 \end{minipage}
 \end{tabular}
 \caption{\small [Left]
$\alpha_{F , \UV}$ [Eq.~\eqref{defalphaFUV}], 
$\alpha_{F , 0}$ [Eq.~\eqref{f8}],
and the $\LQ^2 r^2$-term 
 [Eq.~\eqref{defalphaFUV}] 
 as functions of $\LQ^2 r^2$. 
 The number of flavors is set to $n_f=1$.
 [Right]
 Derivatives of 
 $\alpha_{F , \UV}$, $\alpha_{F,0}$ and the $r^2 \LQ^2$-term with respect to 
$\rho^2 \equiv \LQ^2 r^2$.
 }
\label{Fig.Force1}
\end{figure}
 
In Fig.~\ref{Fig.Force1},
$\alpha_{F , \UV}$, 
$\alpha_{F , 0}$,
and the $\LQ^2 r^2$-term 
of Eq.~\eqref{defalphaFUV}
are plotted as functions of $\LQ^2 r^2$.
Qualitatively
they show similar behaviors
to those of the reduced Adler function
(Fig.~\ref{Fig.Adler1}), and the derivative of
$\alpha_{F , \UV}$ is dominated by the $\LQ^2 r^2$-term
especially in the range $\LQ^2 r^2 \gtrsim 0.02$.
Comparisons with the large-order predictions in the large-$\beta_0$
approximation and with the known exact perturbative series
will be presented in Sec.~\ref{sec.RB}.

In Eq.~\eqref{f7}, 
the $\mf$-dependent term starts from order $r^3$. 
Let us discuss this $\mu_f$-dependence in the context of OPE.
The relevant low-energy effective theory is known 
as pNRQCD, in which
the QCD potential is expressed in expansion in $\vec{r}$
(multipole expansion) as \cite{Brambilla:1999xf}
\be
V_{\rm QCD}(r) \approx
V_S(r)
-\frac{2\pi i \alpha_s}{N_C} \int_0^{\infty} dt \,
e^{-i t \Delta V(r)} 
\braket{\vec{r} \cdot \vec{E}^a(t) \, \vec{r} \cdot  \vec{E}^a(0)}
+\mathcal{O}(r^3)  
\label{opeQCD} \, .
\ee
Here, $V_S(r)$ represents the Wilson coefficient for the (leading)
identity operator and
has the meaning of the energy of the $Q\bar{Q}$ singlet state; 
$\Delta V$ denotes the energy difference between the octet and singlet states; 
$\vec{E}^a$ denotes the color electric field. 
If we compute $V_S(r)$ in the large-$\bz$ approximation and
with an explicit cutoff in the gluon momentum,
it  is identified with  $V_{\bz}(r;\mf)$.\footnote{
See also the discussion at the end of Sec.~\ref{ss:force2}.
}
It has been confirmed that
the $\mf$-dependence of $V_{\bz}(r;\mf)$ 
at order $\mf^3 r^2$, originating from the $u=3/2$ renormalon, 
is canceled against the $\mf$-dependence of the 
non-perturbative matrix element in the second term 
of Eq.~\eqref{opeQCD}~\cite{Brambilla:1999xf,Sumino:2014qpa}. 
Differentiating with respect to $r$,
the leading $\mf$-dependence of $\alpha_{F,\bz}(1/r^2;\mf)$ at order $\mf^3 r^3$ 
is also canceled by that of the non-perturbative matrix element.
We expect that a similar cancellation between $D_{\bz}(Q^2;\mf)$ and
the local gluon condensate would
hold for the Adler function, 
although the relevant low energy effective theory is as yet unknown.

\subsection[Scheme dependence by choice of pre-weight $W_X$]
{\boldmath Scheme dependence by choice of pre-weight $W_X$}
\label{sec.SD}

As we already pointed out, the pre-weight $W_X$
introduced in Sec.~\ref{sec.GC}
is not unique, and
we clarify its effect in this subsection. 
We first show that the dependences of $X_0$ and the power corrections
in $X_\mathrm{UV}$ on the choice of $W_X$
almost cancel in the sum ($X_{\UV}$).
We then discuss its relevance in determination 
of a non-perturbative matrix element in OPE.
Finally we discuss why the power corrections in $X_{\UV}$ 
can vary from the viewpoint of the asymptotic property 
of the perturbative series and how it is related to 
variation of $W_X$.

The pre-weight $W_X$ which satisfies Eq.~\eqref{rel} 
is not unique 
since its real part on the positive real axis 
is not restricted.
Although the original $\mu_f$-dependent integral \eqref{g0}
is independent of the choice of $W_X$,
the $\mu_f$-independent part $X_\mathrm{UV}$
generally depends on the choice of $W_X$. 
Namely, $X_{\UV}$ is a functional of $W_X$. 
We can regard that $X_{\UV}$ determined by different $W_X$
correspond to different scheme choices. 
We first discuss the scheme dependence of $X_{\UV}$.

Consider two different pre-weights
$W_{X}^{(i)}$ ($i$=1,2) both
satisfying the relation \eqref{rel}.
Correspondingly we obtain $X_{\UV}$ in different schemes 
via Eqs.~\eqref{SUV2} and \eqref{S0}:
\begin{align}
X_\mathrm{UV}^{(i)}
=X_0^{(i)}(Q^2)
+\sum_{0<n<u_{\IR}} 
\frac{4 \pi c_n^{(i)}}{\bz} 
\lt(\frac{e^{5/3} \LQ^2}{Q^2} \rt)^n
\label{s1}
\end{align}
with
\begin{align}
X_0^{(i)}(Q^2)
={\rm Im} \int_{C_a} 
\frac{d \tau}{\pi \tau}\, W_X^{(i)}\! \lt(\frac{\tau}{Q^2} \rt)
\alpha_{\bz}(\tau)+\frac{4 \pi c_0^{(i)}}{\bz} \, ,
  \label{s2}
\end{align}
where $W_X^{(i)}(z)
=\sum_{n \geq 0} c_n^{(i)} z^n$.
The difference between $X_0^{(1)}$ and $X_0^{(2)}$ 
is given by 
\begin{align}
X_0^{(2)}
-X_0^{(1)}
=&{\rm Im} \int_{C_a} 
\frac{d \tau}{\pi \tau} 
\left\{W_X^{(2)} \left(\frac{\tau}{Q^2} \right)-
W_X^{(1)} \left(\frac{\tau}{Q^2} \right) \right\} 
\alpha_{\bz}(\tau)
+\frac{4 \pi (c_0^{(2)}-c_0^{(1)})}{\bz}  
\label{dif} \, .
\end{align}
In the integral along $C_b$
we assumed that it is justified to
expand $W_X(z)$ for sufficiently small $|z|$.
Accordingly, we assume that
$\delta W(z)\equiv W_X^{(2)}(z)-W_X^{(1)}(z)$ is regular
at any point 
$z_0 \in \mathbb{R}$ and $0<z_0<\epsilon$ for
$\exists \epsilon>0$ (sufficiently close to the origin).\footnote{
The reason to exclude $z_0=0$ is to cope with possible existence
of $\log z$ or $\sqrt{z}$.
(See footnote \ref{fn:smallzexp}.)
Note that even if the small-$z$ expansion of $W_X(z)$ includes
$\log z$ or $\sqrt{z}$ we expect that
the expansion has a domain of convergence close to the origin;
see the examples in Secs.~\ref{ss:adler} and \ref{ss:force}.
}
Namely, $\delta W(z)$ can be expanded in
Taylor series about $z=z_0$
with a non-zero radius of convergence:
\begin{eqnarray}
\delta W(z) = \sum_{n\ge 0} A_n(z_0)\, (z-z_0)^n \,,
~~~~~~z_0 \in \mathbb{R}~~\text{and}~~0<z_0<\epsilon \, .
\end{eqnarray}
Since $\text{Im}\, \delta W=0$ on the positive real axis,
(i)~the integral along $C_a$ in Eq.~\eqref{dif} is equal
to that along $C_b$, and
(ii)~$A_n(z_0) \in \mathbb{R}$, hence
$\{ \delta W(z) \}^*=\delta W(z^*)$ is satisfied along the path $C_b$
if $Q^2\gg \mf^2$.
Then, by exploiting the same procedure as in Eq.~\eqref{Cauchy},
the first term of Eq.~\eqref{dif} can be reduced to
\begin{align}
&{\rm Im} \int_{C_a} \frac{d \tau}{\pi \tau} \,
\delta W\! \left(\frac{\tau}{Q^2} \right) \alpha_{\bz}(\tau)
=
{\rm Im} \int_{C_b} \frac{d \tau}{\pi \tau} \,
\delta W\! \left(\frac{\tau}{Q^2} \right) \alpha_{\bz}(\tau)
\non 
&~~~ =-\frac{4 \pi}{\beta_0} \,
\delta W \! \left(\frac{e^{5/3} \Lambda^2_{\rm QCD}}{Q^2} \right)
= -\frac{4 \pi}{\beta_0} \sum_{n  \geq 0} 
(c_n^{(2)}-c_n^{(1)})
\left(\frac{e^{5/3} \Lambda^2_{\rm QCD}}{Q^2} \right)^n  
\label{partDif} \, .
\end{align}
It means that the difference of $X_0^{(i)}(Q^2)$ is given by%
\footnote{Note that the right-hand side of Eq.~\eqref{S0dif} is 
$\mathcal{O}(\LQ^2/Q^2)$ 
and the asymptotic form of $X_0(Q^2)$ 
at $Q^2 \to \infty$ 
shown in Eq.~\eqref{asym} is not modified.}
\begin{align}
X_0^{(2)}(Q^2)
-X_0^{(1)}(Q^2) 
= -\frac{4 \pi}{\beta_0} \sum_{n>0} 
(c_n^{(2)}-c_n^{(1)})
\left(\frac{e^{5/3} \Lambda^2_{\rm QCD}}{Q^2} \right)^n  
\label{S0dif} \, .
\end{align}
Furthermore, according to Eq.~\eqref{s1}
we obtain the difference of $X_{\UV}^{(i)}$ as
\begin{align}
X_{\UV}^{(2)}(Q^2)
-X_{\UV}^{(1)}(Q^2)
&=-\sum_{n \geq u_{\IR}} 
\frac{4 \pi (c_n^{(2)}-c_n^{(1)})}{\bz} 
\lt(\frac{e^{5/3} \LQ^2}{Q^2} \rt)^n 
\non
&= \mathcal{O} 
\left( \lt({\LQ^2}/{Q^2} \rt)^{u_{\IR}} \right) \, .
\end{align}
Thus, the differences of the power corrections
$(1/Q^2)^n$ with $0<n<u_{\IR}$ in Eq.~\eqref{s1}
are canceled by the change of $X_0(Q^2)$. 
As a result, the difference of $X_{\UV}$ in different schemes 
is smaller than the last included term of the $(\LQ^2/Q^{2})^n$-terms
in $X_{\UV}(Q^2)$.  
Namely, the $\mf$-independent part $X_{\UV}$
has a minor dependence on the scheme, 
which is the same order 
as an uncertainty induced by the first IR renormalon, 
and we confirm validity of our result of $X_{\UV}$ 
taking into account the scheme dependence.  

It is worth emphasizing 
that the scheme dependence
discussed above
is {\it not} a renormalon uncertainty.
In fact the scheme dependence can be removed 
by including higher orders of the $1/Q^{2}$ expansion. 
Let us clarify this point. 
Suppose we consider $X_{\bz}(Q^2;\mf)$ up to $1/(Q^2)^n$ 
in different schemes:
\be
X^{(i)}_{\bz}(Q^2;\mf)\Bigr|_{1/(Q^2)^n}
=X_0^{(i)}(Q^2)
-\sum_{k=1}^n\,
{\rm Im} 
\int_{C_b} \frac{d \tau}{\pi \tau} \, c_k^{(i)}
\left( \frac{\tau}{Q^2}\right)^k \alpha_{\bz}(\tau) 
\,.
~~~(i=1,2)
  \label{s4}
\ee
We show that $X^{(2)}_{\bz}-X^{(1)}_{\bz}\Bigr|_{1/(Q^2)^n}$
is order $(\LQ^2/Q^2)^{n+1}$.
(The previous argument already proves this for the case 
$n=u_{\rm IR}-1$.)

Note that since ${\rm Im}\,c_k^{(i)}$
is fixed by Eq.~\eqref{rel},
there is no scheme dependence, hence
$c_k^{(2)}-c_k^{(1)}\in \mathbb{R}$.
This enables reducing the difference of the second term of
Eq.~\eqref{s4} as
\be
\sum_{k=1}^n\,
{\rm Im} 
\int_{C_b} \frac{d \tau}{\pi \tau} \, (c_k^{(2)}-c_k^{(1)})
\left( \frac{\tau}{Q^2}\right)^k \alpha_{\bz}(\tau) 
= \frac{4 \pi}{\beta_0} \sum_{k=1}^n
(c_k^{(2)}-c_k^{(1)})
\left(\frac{e^{5/3} \Lambda^2_{\rm QCD}}{Q^2} \right)^k  
\label{schemedif} \, .
\ee
Combining with Eq.~\eqref{S0dif},
we see that 
$X^{(2)}_{\bz}-X^{(1)}_{\bz}\Bigr|_{1/(Q^2)^n}={\cal O}((\LQ^2/Q^2)^{n+1})$.
Such a property follows from the fact that 
the original $\mf$-dependent integral \eqref{g0} 
is independent of the choice of $W_X$. 
Therefore the scheme dependence is gradually eliminated 
by including higher order terms in $1/Q^2$.

In the case of the Adler function, 
this fact 
is important if we want to determine the local gluon condensate using
our formulation, for instance, by comparing with
an evaluation of $D(Q^2)$ by a lattice calculation.
The OPE up to the ${\cal O}(1/Q^4)$ terms (in the large-$\bz$ approximation)
is written as
\be
D(Q^2)=D_{\bz}(Q^2;\mf)\Bigr|_{1/Q^4}
+C_{GG}(\mf) \frac{\braket{0|G^{a \mu \nu} G^{a}_{\mu \nu}|0}(\mf)}{Q^4}
+\mathcal{O}(\LQ^6/Q^6) \, .
\ee 
We expect that $\mf$-dependences up to $1/Q^4$-terms are canceled. 
According to the above discussion, the variation due to the scheme difference 
(choice of $W_X$) satisfies
\be
\Delta_\mathrm{scheme} 
\lt(C_{GG}(\mf) \frac{\braket{0|G^{a \mu \nu} G^{a}_{\mu \nu}|0}(\mf)}{Q^4} \rt)
=\mathcal{O}(\LQ^6/Q^6) \, .
\ee
Thus, the error becomes higher order than the term which we want to determine. 
[Note that $C_{GG}$ would also include power corrections $\sim (\LQ^2/Q^2)^n$.]

Although we have shown that $\delta W $ changes 
$X_{\UV}$ only at subleading order, 
it alters $X_0$ and the power corrections $(\LQ^2/Q^2)^n$
with $n<u_{\IR}$ individually; 
see Eqs.~\eqref{s1} and \eqref{S0dif}. 
In the rest of this subsection, 
we discuss the reason why the coefficients of 
the $(\LQ^2/Q^2)^{n}$-terms can be altered. 

We can show that 
$X_0(Q^2)$ has the same
asymptotic expansion in $\alpha_s$
as the perturbative series of $X_{\bz}(Q^2)$:
\be
X_0(Q^2)-\sum_{k=0}^{n-1} d_k(\mu=Q) \lt( \frac{\bz}{4 \pi} \rt)^k \alpha_s^{k+1}(Q)=\mathcal{O} (\alpha_s(Q)^{n+1}) \label{asymD0} \, ,
\ee
as $\alpha_s(Q)\to 0$.
(We sketch the proof in App.~\ref{AppF}.)
This shows that, 
although $X_0(Q^2)$ is  expansible with respect to $\alpha_s(Q)$, it is {\it not} 
expansible with respect to $\LQ^2/Q^2$ since $\alpha_s(Q) \sim 1/\log{(Q^2/\LQ^2)}$.
Reflecting this fact,
$X_{\bz}(Q^2;\mf)$, 
which is related to $X_0(Q^2)$ by Eqs.~\eqref{genres} and \eqref{SUV2},
is also not expansible with respect to $1/Q^2$.
This is a short answer to the question 
why the $(\LQ^2/Q^2)^n$-terms in $X_{\bz}(Q^2;\mf)$ is not uniquely determined.

Note that
$X_{\bz}(Q^2;\mf)-X_0(Q^2)$ is expansible in $1/Q^2$
and the $(\LQ^2/Q^2)^n$-terms
are regarded as a part of this series expansion.
In this respect
Eq.~\eqref{asymD0} is essential 
since it ensures that the singularities of $X_{\bz}(Q^2;\mf)$
caused by $\alpha_s(Q)^k$
cancel with those of $-X_0(Q^2)$.
Considering the fact that $X_{\bz}(Q^2;\mf)$ 
is a uniquely-defined quantity,
it is deduced that
the non-uniqueness of the  $(\LQ^2/Q^2)^n$-terms in $X_{\bz}(Q^2;\mf)-X_0(Q^2)$
is caused by the non-uniqueness of $X_0(Q^2)$.
In fact there are potentially many candidates of $X_0(Q^2)$ 
satisfying the property \eqref{asymD0}.
A new $X_0$ constructed by adding $(\LQ^2/Q^2)^n$ to the old one 
also
satisfies Eq.~\eqref{asymD0}, 
since all the series coefficients of 
$\LQ^2/Q^2=e^{-4 \pi/(\bz \alpha_s(Q^2))}$ 
in $\alpha_s(Q^2)$
are zero.

The non-uniqueness of $X_0$
stems from the non-uniqueness of $W_X$ in our method.
The variation of $W_X$ indeed changes $X_0$ by powers of $\LQ^2/Q^2$ 
as shown in Eq.~\eqref{S0dif} while keeping the asymptotic expansion
\eqref{asymD0}.
This change of $X_0$ is compensated by the change of
the $(\LQ^2/Q^2)^n$-terms
as shown below Eq.~\eqref{schemedif}. 
Thus, the non-uniqueness of the power corrections
is also attributed to the non-uniqueness of $W_X$.

At this stage, 
it suggests that it would be meaningless to focus 
on the power corrections $(\LQ^2/Q^2)^n$ alone in $X_{\UV}$
since it becomes definite only after we specify $X_0$, 
and only the sum of them ($X_{\UV}$) is a meaningful quantity.  
Nevertheless, it turns out that if we limit schemes to a reasonable class, 
the separation of $X_{\UV}$ into  $X_0$ and $(\LQ^2/Q^2)^n$-terms 
becomes unique by a uniqueness of $W_X$.
We will elaborate on this point in the next subsection.

\subsection{Massive gluon scheme
as the optimal scheme 
}
\label{sec.FM}

We discuss which scheme is favored from 
the analytical properties of $X_{\UV}(Q^2)$ when
we extend it to a function of 
the complex variable $\LQ^2/Q^2$.
Since the power-correction terms in $X_{\UV}$ are obviously analytic 
in the whole $\LQ^2/Q^2$-plane,
we mainly focus on the analytic structure of $X_0(Q^2)$.

$X_0(Q^2)$ in the massive gluon scheme
can be expressed 
as an analytic function of $\LQ^2/Q^2$
by Eq.~\eqref{S0rot},
provided that the integral path can be rotated.
Using this expression we can show that $X_0$
has a cut along the negative real axis starting from
the origin and is regular everywhere else
in the  $\LQ^2/Q^2$-plane.
It follows from the fact that $W^{(m)}_{X+}(z)$ 
in this scheme can have cuts along
the negative real axis starting only from $z=0$ and $z=-1$  
and is regular everywhere else.%
\footnote{
This can be shown using the property that
$W_X^{(m)}(z)$ can have singularities only at $z=0$, 1, $\infty$,
where $z=\tau/Q^2=(-\tau)/q^2=1$ corresponds to the threshold of 
the massive gluon plus
massless partons.
In passing,
since $2 \,{\rm Im} \, W_X^{(m)}(x)=w_X(x)$ holds for $x>0$, 
$w_X(x)$ can have singularities only at $x=0$, 1, $\infty$ along the
integral path of Eq.~\eqref{W+};
c.f., Eqs.~\eqref{wD} and \eqref{f3}.
}
Thus, $X_0(Q^2)$ in this scheme (hence, $X_{\UV}(Q^2)$)
satisfies the required analyticity 
in the complex plane in terms of perturbative QCD,
where the form of the singularity at $\LQ^2/Q^2=0$ is dictated by
the renormalization-group equation.
We have already seen this favorable feature of the massive
gluon scheme for the Adler function
in Sec.~\ref{ss:adler}.

In a general scheme, i.e., for a general pre-weight,
$X_0(Q^2)$ can be expressed 
as an analytic function in the following manner.
We rewrite the pre-weight as the sum of $W_X^{(m)}(z)$, 
which is the pre-weight in the massive gluon scheme, 
and the rest as 
$W_X(z)=W^{(m)}_X(z)+\delta W_X(z)$.
We can follow the same steps which led to Eq.~\eqref{partDif}
in the previous section,
assuming regularity of $\delta W_X(z)$ close to the
origin, and obtain, for 
sufficiently small $|\LQ^2/Q^2|$,
\be
X_0(Q^2)
=X_0^{(m)}(Q^2)+\frac{4 \pi }{\bz} \lt[\delta{W}_X(0)- \delta{W}_X\! \lt(\frac{e^{5/3} \LQ^2}{Q^2} \rt) \rt] \,,
\label{DifX0}
\ee
where $X_0^{(m)}(Q^2)$ represents $X_0$ in the massive gluon scheme.
Then we can enlarge the domain of this function  by
analytic continuation to the entire 
$\LQ^2/Q^2$-plane, except at singular points of $\delta W_X(
{e^{5/3} \LQ^2}/{Q^2})$ and the origin.

In this construction we can regard that
the essential part is determined by
$W^{(m)}_X(z)$, which already gives the required analyticity 
of $X_0(Q^2)$.
$\delta W_X(z)$ is subsidiary in the sense that it 
is not necessary in an essential way and
should not
have addditional singularities (except at $\LQ^2/Q^2=\infty$) in order not
to violate the required analyticity of $X_0(Q^2)$ or $X_{\UV}(Q^2)$.
Thus, we may say that the massive gluon scheme is an optimal
(or minimal) scheme in terms of the analyticity,
according to this construction of $X_0(Q^2)$.

We would like to know how many pre-weights are allowed as a reference 
scheme in the above construction of $X_0(Q^2)$, or in other words,
how many minimal schemes exist.
The integral expression
\eqref{S0rot} is used to define the reference scheme,
and this expression is realized naturally by the following conditions
on the pre-weight:%
\footnote{The condition $(0)$ is already included in the definition
of a general $W_X(z)$.
Note also that $w_X(0)=0$ due to our assumption that $X_{\rm LO}$ is
IR finite [see Eq.~\eqref{oneloop}], hence the conditions
(0) and (1) are mutually consistent at $x=0$.}
\begin{align}
&(0)~~~\text{$W_X(z)$ is analytic in the upper half-plane, and}\non
&~~~~~~~~~~~~
2 \,{\rm Im} \, W_X(x)=w_X(x)  ~~~\text{for}~~x \geq 0 \,. \\
&(1)~~~{\rm Im}\, W_X(x)=0~~~\text{for}~~x \leq 0  \,.  \\
&(2)~~\int_{C_R}\! \frac{d z}{\pi z} \, W_X(z)~ 
\text{is absolutely convergent to 0 as}~R \to \infty,
\non
&~~~~~~~~~~~~~~~~~~~~~~~~~~~~~~~~~~~~~~~~~
~\text{where}~C_R=\{R e^{i \theta}| 0 \leq \theta \leq \pi \} . 
~~~~~~~~~
\label{con2}
\end{align}
The pre-weight in the massive gluon scheme $W^{(m)}_X(z)$
satisfies the conditions (0) and (1).
If it also satisfies 
the condition (2),\footnote{
\label{fn:condrotpath}
We can show that $W_X^{(m)}$ satisfies the condition (2), 
if $|w_X(z)|={\cal O}(|z|^a)$ for $\exists a<0$ for sufficiently 
large $|z|$ in the lower half-plane.
} 
  we can
rotate the integration path to the negative axis, and
the expression \eqref{S0rot} is obtained,
namely, $X_{\UV}$ satisfies the required analyticity.

We now prove that the above conditions (0)--(2) are sufficient
to determine the pre-weight uniquely.
Let us examine the difference of 
the pre-weights satisfying the above conditions:
\be
\delta W_X(z)=W_X^{(2)}(z)-W_X^{(1)}(z) .
\ee
We can translate the conditions $(0)-(2)$ into conditions for $\delta W_X$ as 
\be
{\rm Im} \, \delta W_X(x)=0~\text{for}~x \in \mathbb{R} \label{imzero} \,,
\ee
\be
\int_{C_R} \frac{d z}{\pi z} \delta W_X(z)~ 
\text{is absolutely convergent to 0 as}~R \to \infty. \label{conv}
\ee
Using Eq.~\eqref{conv}, we can show
\be
{\rm Pr.} \int_{-\infty}^{\infty} \frac{d x}{\pi} \frac{\delta W_X(x)}{x-x'}= i \, \delta W_X(x') ,
\ee
where {Pr.} denotes the principal value integral and $x'$ is assumed to be a real parameter.
Taking the imaginary part of this equation and using Eq.~\eqref{imzero}, we obtain
\be
{\rm Re} \, \delta W_X(x) \equiv 0~\text{for}~x \in \mathbb{R}. \label{rezero}
\ee
One can see from Eq.~\eqref{imzero} and Eq.~\eqref{rezero} that
$\delta W_X$ is identically zero in the upper half-plane
including the real axis (by the identity theorem).
Since we do not expect any physical singularity to disturb enlargement
of this analyticity domain,\footnote{
Note that a singularity in $\delta W_X(z)$ 
except for a cut along the negative real axis
generates an additional singularity
in $X^{(2)}_{0}(Q^2)$ compared with $X_0^{(1)}(Q^2)$ as 
one can see from Eq.~\eqref{DifX0}.
}
we can conclude that $\delta W_X=0$
in the entire complex plane:
\be
\delta W_X(z) \equiv 0 ~\text{for}~ z \in \mathbb{C} .
\ee
Hence, if $W^{(m)}_X(z)$ in
the massive gluon scheme satisfies the condition (2),
the allowed scheme is uniquely determined to this one. 
This is the case in 
the Adler function and the QCD potential, 
which are considered as explicit examples in Secs.~\ref{ss:adler} and \ref{ss:force}.

This argument shows that
through the above conditions (0)--(2)
we can realize the correct analytic structure of $X_{\UV}$ minimally 
and single out the pre-weight uniquely simultaneously. 
Therefore, due to the uniqueness of the pre-weight, 
the separation of $X_{\UV}$ into
$X_0$ and power corrections
$\sim (\LQ^2/Q^2)^n$ is fixed,
\footnote{
In the case of the force between $Q \bar{Q}$,
$W_F^{(m)}$ does not satisfy the condition (2),
although the integral path $C_a$ can be rotated 
owing to $\alpha_{\bz}(\tau)$ in $\alpha_{F,\UV}$.
In fact, it would be more general to adopt the condition $(2)'$
instead of the condition (2) to define the minimal schemes:
\begin{align}
(2)' ~~\text{The integral of $|W_X(z)/z|$ along $C_R$ is bounded, i.e.,} \non
~~~~\int_0^{\pi} d \theta |W_X(R e^{i \theta})| <\exists M ~\text{for a sufficiently large}~ R \, ,
\end{align}
since the rotation of the path $C_a$ is assured even in this case.
For simplicity we discussed with a stronger condition [$(2)$] above.
If we adopt the conditions $(0),(1)$ and $(2)'$, 
$\delta W_X=(\text{real const.})$ follows by a similar discussion.
The constant shift of a pre-weight, however, does not change
$X_0$ (and obviously power corrections in $X_{\UV}$)
as seen from Eq.~\eqref{S0dif}. 
Hence, the main result still holds 
that the separation into $X_0$ and $(\LQ^2/Q^2)^n$-terms
is unique under these relaxed conditions.
} 
and in particular the coefficient of the $(\LQ^2/Q^2)^n$-term 
is no longer changeable within the minimal schemes.
It should be regarded as a natural one among all possibilities.

\subsection{Behaviors of $X_{\rm UV}$ in massive gluon scheme}
\label{sec.FM2}

In the previous subsection, we pointed out that 
the massive gluon scheme can be regarded as special
among all the schemes.
We examine some details of  the behaviors of
$X_0(Q^2)$ and the power corrections
in $X_{\UV}$, respectively, in this scheme.
\footnote{In this subsection, we assume that $W_X^{(m)}(z)$ has a good convergence
for large $|z|$ in the upper half-plane including the real axis.}

\subsubsection*{Behavior of $X_0(Q^2)$}
\label{ss:A}

As discussed below Eq.~\eqref{asymD0},
the behavior of $X_0(Q^2)$ close to $1/Q^2=0$
is determined by the fact that $X_0(Q^2)$ has
the same asymptotic expansion as the perturbative series of $X(Q^2)$;
see Eq.~\eqref{asym}.
Namely the behavior of $X_0(Q^2)$ at large $Q^2$
is almost insensitive to the scheme of $W^{(m)}_X$.
In contrast,
the global behavior of $X_0(Q^2)$ generally
depends on the scheme of $W_X$.

Let us examine some details about the massive gluon scheme.
The limit of $X_0$ in this scheme at 
$1/Q^2 \to \infty$ is
calculated from Eq.~\eqref{S0forAsymp} as
\begin{align}
X_0(Q^2) \to \frac{4 \pi}{\bz} W^{(m)}_{X+}(0)
=\frac{4 \pi}{\bz} d_0 \, .
  \label{m1}
\end{align}
Namely $X_0(Q^2)$ approaches a
constant for sufficiently large $1/Q^2$.

In addition, if we regard $X_0(Q^2)$ as a function of $\hat{Q}^{-2}=\LQ^2/Q^2$, 
we can see that $X_0$ and its derivatives have definite signs 
at least for the two examples which we studied:
\be
X_0 \geq 0~~ ;~~ \frac{dX_0}{d (\hat{Q}^{-2})} \geq 0 ~~;~~ \frac{d^2 X_0}{d (\hat{Q}^{-2})^2 } \leq 0  ~~\text{for $X=D, \alpha_F$} \label{mono} \, .
\ee
This property follows from
$W'_{X+}(x)\leq 0$, $W''_{X+}(x)\geq 0 
$
for $x\geq 0$, and
\begin{align}
&\frac{d^nX_0}{d (\hat{Q}^{-2}) ^n}
=\int_0^{\infty} \frac{d \tau}{\pi \tau} 
\left( \frac{\tau}{\LQ^2} \right)^n
\left.
\frac{d^n W^{(m)}_{X+}(x)}{dx^n} 
\right|_{x=\frac{\tau}{Q^2}}
\frac{4 \pi}{\bz} 
\frac{-\pi}{\log^2{(\tau e^{-5/3}}{\LQ^2})+\pi^2}
\, .
\label{m4}
\end{align}
As a result, 
combined with the asymptotic forms at $1/Q^2=0, \infty$, 
the behavior of each $X_0$ is determined globally and the form is simple
(and similar), 
as seen from Figs.~\ref{Fig.Adler1} or \ref{Fig.Force1}. 

 %
\subsubsection*{Power corrections in $X_{\UV}$}
\label{ss:B}

We show that the power corrections in $X_{\UV}$ 
can be detected generally from the Borel transformation. 
Consider an integral
\be
C_X(v) \equiv \int_0^{\infty} \frac{dz}{2 \pi}\, W^{(m)}_X(z)\, z^{-v-1}  \label{Cdef} \, .
\ee
The expansion of $W^{(m)}_X(z)$ for small-$z$ is determined 
by the singularities of $C_X(v)$ as
[c.f., Eqs.~\eqref{Borel} and \eqref{invBexp}] (see footnote \ref{fn:loginc})
\be
W^{(m)}_X(z)=-2 \pi \sum_{n \in V_{\IR}} {\rm Res}_{v=n}[C_X(v) z^v]=\sum c_n z^n \label{invC} \, ,
\ee
where $V_{\IR}$ denotes a set of non-negative poles of $C_X(v)$.
Using Eq.~\eqref{massglu}, $C_X(v)$ is explicitly calculated 
in the massive gluon scheme as
\begin{align}
C_X(v)
&=\int_0^{\infty} \frac{d x}{2 \pi} w_X(x) \int_0^{\infty} \frac{d z}{2 \pi}  \frac{z^{-v-1}}{x-z-i0} \ \non
&=- \frac{1}{2} \frac{e^{-i \pi v}}{\sin(\pi v)} \int_0^{\infty} \frac{d x}{2 \pi} w_X(x) x^{-u-1}  \non
&=-\frac{1}{2} \frac{e^{-i \pi v}}{\sin{(\pi v)}} B_X(v) =-\frac{1}{2} \frac{\cos(\pi v)}{\sin{(\pi v)}} B_X(v)+\frac{i}{2} B_X(v)
\label{Ccal} \, ,
\end{align}
where we used Eq.~\eqref{Borel}.  
(The same equation was derived in Ref.~\cite{Ball:1995ni} in a different context.)
By taking the imaginary part of $C_X(v)$, 
we can check that the usual Borel transformation is obtained 
consistently with Eq.~\eqref{rel}.
The factor $\{\sin{(\pi v)}\} ^{-1}$ in the real part of Eq.~\eqref{Ccal}
generates additional integer poles, 
that is, $U_{\IR} \subset V_{\IR}$. 
In particular, the first few terms of the expansion of $W^{(m)}_X(z)$ 
stem from this factor and reduce to real coefficients: 
\be   
W^{(m)}_{X}(z)=\sum_{0 \leq n
< u_{\IR}} B_X(n) z^n+\dots  \label{WSBorel}  \, , 
\ee
where we use Eq.~\eqref{invC}.
Therefore, from Eqs.~\eqref{eq_Ws}, \eqref{SUV} and \eqref{WSBorel},
the coefficient of the $(e^{5/3} \LQ^2/Q^2)^n$-term of $X_{\UV}$ 
is revealed to be $4 \pi B_X(n)/\bz$.

Incidentally, we have a similar relation for $W^{(m)}_{X+}$ in the massive gluon scheme as 
\begin{align}
C_{X+}(u) \equiv 
&\int_{0}^{\infty} \frac{d z}{2 \pi} \, W^{(m)}_{X +}(z) z^{-u-1} \non
&=\int_0^{\infty} \frac{d x}{2 \pi} \, w_X(x) \int_0^{\infty} \frac{d z}{2 \pi}  \, \frac{z^{-u-1}}{x+z-i0} \non
&=-\frac{1}{2} \frac{1}{\sin{(\pi u)}} \int_0^{\infty} \frac{d x}{2 \pi} \, w_X(x) x^{-u-1} \non
&=-\frac{1}{2} \frac{1}{\sin{(\pi u)}} B_X(u) , \label{C+}
\end{align}
where we used Eqs.~\eqref{W+} and \eqref{Borel}.

Note that Eq.~\eqref{WSBorel} does {\it not} mean 
that the power corrections included in $X_{\UV}$ are 
related to perturbative ambiguity, 
but it is purely a mathematical relation. 
We explore the origin of the expansion of $W^{(m)}_X(z)$ 
and clarify the meaning in terms of the method of
expansion by regions in the next section.

\section{Power corrections and OPE in light of expansion by regions}
\label{sec.EbR}

In this section we investigate (i) the origin of
the power corrections in $X_\mathrm{UV}$,
and (ii) the relation of $X_\mathrm{UV}$ to Wilson coefficients in OPE,
by means of the method of expansion by regions,
or asymptotic expansion in limits of large momentum 
\cite{Beneke:1997zp, Smirnov:1999bza, Smirnov:2002pj}.
With this method,
we can identify
which momentum region 
contributes to each power correction.
We show that the power corrections
in $X_\mathrm{UV}$ for the Adler function Eq.~\eqref{defDUV}
and the interquark force Eq.~\eqref{defalphaFUV}, respectively, 
are genuine UV contributions.
We also provide an effective field theoretical point of view
of our framework presented
in the previous section.

We first discuss some general aspects (Sec.~\ref{ss:general}) and 
subsequently clarify detailed features in the
examples of the Adler function (Sec.~\ref{ss:adler2}) and
the interquark force or QCD potential (Sec.~\ref{ss:force2}).
We also give a supplementary argument using an explicit cutoff (Sec.~\ref{sec:cutoff}).

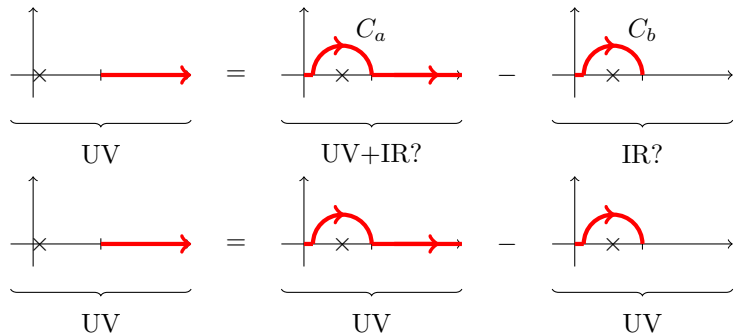
\begin{figure}
\begin{center}
\begin{tikzpicture}[baseline={(0,0.4)},scale=0.3]
\draw [->] (-1-12,0)--(7-12,0);
\draw [->] (0-12,-1)--(0-12,3);
\draw [-](3-12,-0.25)--(3-12,0.25);
\draw [->,ultra thick,red] (3-12,0)--(7-12,0);
\draw (1-12.7,0) node {$\times$};
\draw [decorate,decoration=brace] (7-12,-2)--(-1-12,-2);
\draw (3-12,-3.5) node {UV};
\draw (-3,0) node {$=$};
\draw [->] (-1,0)--(7,0);
\draw [->] (0,-1)--(0,3);
\draw[-](3,-0.25)--(3,0.25);
\draw (1.7,0) node {$\times$};
\draw [->, ultra thick,red] (1.7,1.3)--(1.8,1.3);
\draw [ultra thick,red] (0,0)--(0.4,0);
\draw [ultra thick,red] (3,0) arc (0:180:1.3);
\draw [ultra thick,red] (3,0)--(7,0);
\draw [->,ultra thick,red] (4,0)--(6,0);
\draw (3,2) node {$C_a$};
\draw [decorate,decoration=brace] (7,-2)--(-1,-2);
\draw (3,-3.5) node {UV+IR?};
\draw (9,0) node {$-$};
\draw [->] (-1+12,0)--(7+12,0);
\draw [->] (0+12,-1)--(0+12,3);
\draw[-](3+12,-0.25)--(3+12,0.25);
\draw (1.7+12,0) node {$\times$};
\draw [->, ultra thick,red] (1.7+12,1.3)--(1.8+12,1.3);
\draw [ultra thick,red] (0+12,0)--(0.4+12,0);
\draw [ultra thick,red] (3+12,0) arc (0:180:1.3);
\draw (3+12,2) node {$C_b$};
\draw [decorate,decoration=brace] (7+12,-2)--(-1+12,-2);
\draw (3+12,-3.5) node {IR?};
\end{tikzpicture}
\begin{tikzpicture}[baseline={(0,0.4)},scale=0.3]
\draw [->] (-1-12,0)--(7-12,0);
\draw [->] (0-12,-1)--(0-12,3);
\draw [-](3-12,-0.25)--(3-12,0.25);
\draw [->,ultra thick,red] (3-12,0)--(7-12,0);
\draw (1-12.7,0) node {$\times$};
\draw [decorate,decoration=brace] (7-12,-2)--(-1-12,-2);
\draw (3-12,-3.5) node {UV};
\draw (-3,0) node {$=$};
\draw [->] (-1,0)--(7,0);
\draw [->] (0,-1)--(0,3);
\draw[-](3,-0.25)--(3,0.25);
\draw (1.7,0) node {$\times$};
\draw [->, ultra thick,red] (1.7,1.3)--(1.8,1.3);
\draw [ultra thick,red] (0,0)--(0.4,0);
\draw [ultra thick,red] (3,0) arc (0:180:1.3);
\draw [ultra thick,red] (3,0)--(7,0);
\draw [->,ultra thick,red] (4,0)--(6,0);
\draw [decorate,decoration=brace] (7,-2)--(-1,-2);
\draw (3,-3.5) node {UV};
\draw (9,0) node {$-$};
\draw [->] (-1+12,0)--(7+12,0);
\draw [->] (0+12,-1)--(0+12,3);
\draw[-](3+12,-0.25)--(3+12,0.25);
\draw (1.7+12,0) node {$\times$};
\draw [->, ultra thick,red] (1.7+12,1.3)--(1.8+12,1.3);
\draw [ultra thick,red] (0+12,0)--(0.4+12,0);
\draw [ultra thick,red] (3+12,0) arc (0:180:1.3);
\draw [decorate,decoration=brace] (7+12,-2)--(-1+12,-2);
\draw (3+12,-3.5) node {UV};
\end{tikzpicture}
\caption{\small 
Deformed integral path
introduced in Sec.~\ref{sec.GC}
and different interpretations on relevant
kinematical regions.
[Upper figure]
The pole contribution
is interpreted to be an IR effect.
[Bottom figure]
The pole contribution
is interpreted to be a UV effect,
which is shown to be
legitimate for the
Adler function and the
force between $Q\bar Q$.
}
\label{fig:path}
\end{center}
\end{figure}

\subsection{General aspects}
\label{ss:general}

We discuss two issues
using the method of expansion by regions.
First we answer the question: 
``Which kinematical regions do
the power corrections $(\LQ^2/Q^2)^n$ in $X_{\UV}$
originate from?''
This question can be addressed accurately in the massive gluon scheme.
The motivation to ask this question is as follows.
Since the power corrections 
stem from the contour $C_b$ in Eq.~\eqref{g1} close to the IR pole
at $\tau=e^{5/3}\Lambda _\mathrm{QCD}^2$,
one may suspect that the power corrections originate from
IR contributions, although we claim that $X_{\rm UV}$ consists
of UV contributions. (See Figs.~\ref{fig:path}.)

We can use the method of expansion by regions in the
following manner.
The $1/Q^2$ expansion of $X_{\bz}(Q^2;\mf)$ is determined by
the small-$z$ expansion
of $W^{(m)}_X(z)$ in the integral along $C_b$
[Eq.~\eqref{sum_n_int}], and
the $(\LQ^2/Q^2)^n$-terms are included as a part of it.
Since the pre-weight in the massive gluon scheme
$W^{(m)}_X(\tau/Q^2)$ 
is expressed as
a usual Feynman integral with a massive gluon propagator [Eq.~\eqref{replace}],
we can use the expansion-by-regions technique to
decompose the small-($\tau/Q^2$) expansion of $W^{(m)}_X(\tau/Q^2)$ 
into contributions from different
kinematical regions.

In this analysis, we interpret 
the integral variable $p \in (0,\infty)$ in Eq.~\eqref{replace}
as the gluon loop momentum,
even though it was originally restricted to be higher than the factorization scale $\mf$.
The reason is stated as follows.
$X_{\bz}(Q^2;\mf)$ can be written as
\begin{align}
X_{\bz}(Q^2;\mf)=\int_{p^2>\mf^2} \frac{d (p^2)}{2 \pi p^2} w_X(p^2/Q^2) \alpha_{\bz} (p^2) \, ,
\end{align}
where $p$ denotes the gluon loop momentum.
We can rewrite this integral as
\begin{align}
X_{\bz}(Q^2;\mf)
&=\int_{\mf^2}^{\infty} \frac{d \tau}{2 \pi \tau} \alpha_{\bz}(\tau) 
\int_0^{\infty} d(p^2) w_X(p^2/Q^2) \delta(p^2-\tau) \non
&={\rm Im} \int_{\mf^2}^{\infty} \frac{d \tau}{\pi \tau} \alpha_{\bz}(\tau) \int_0^{\infty} \frac{d (p^2)}{2 \pi} \frac{w_X(p^2/Q^2)}{p^2-\tau-i0} \non
&={\rm Im} \lt( \int_{C_a}-\int_{C_b} \rt) \frac{d \tau}{\pi \tau} W_X^{(m)}(\tau/Q^2) \alpha_{\bz}(\tau)
\, .
\end{align}
Hence, the integral variable $p$ of $W_X^{(m)}$ can be regarded 
as the gluon loop momentum,
whereas $\tau$ can be regarded as an auxiliary parameter. 
Since we discuss the expansion of $W_X^{(m)}(\tau/Q^2)$ 
along $\tau \in C_b$ where $|\tau| \leq \mf^2 \ll Q^2$,
$\tau$ plays the role
of a soft scale, 
whereas $Q^2$ plays the role of a hard scale 
in the analysis by expansion by regions.

If the first few terms of the power corrections are found to
originate from UV region,
we can further deduce that the integral along $C_a$
[$=X_0(Q^2)-4\pi c_0/\bz$]
also originates (dominantly) from UV region
for large $Q^2$.
This is because, $X_{\bz}(Q^2;\mf)$ consists of UV
contributions
(given by an integral over $\tau \ge \mf^2$), and
the $C_a$-integral is given by the difference of $X_{\bz}$ and 
the $C_b$-integral.
(We discuss this issue further in Sec.~\ref{sec:cutoff}.)

Secondly the correspondence between
$X_{\bz}(Q^2;\mf)$ in our formulation
and OPE in
a low-energy effective field theory can be
examined using expansion by regions
of Feynman diagrams \cite{Beneke:1997zp}.\footnote{
This part of the analysis deals with
$X_{\bz}$ at each order of perturbation
and has only minor connection with the separation of $X_{\bz}$
into $X_0$ and
power corrections or with the scheme dependence.
}
Since an early stage of
the development of the 
expansion-by-regions method,
its relation to effective field theory
and OPE has been explored
\cite{Beneke:1997zp, Smirnov:1999bza, Smirnov:2002pj}.
The hard contributions
in the context of expansion by regions
are interpreted as Wilson coefficients
in the effective field theory,
and the soft contributions
are interpreted as
perturbative quantum corrections due to
low-energy degrees of freedom.
In other words,
the low-energy effective Lagrangian
is constructed by including
hard contributions in terms of 
effective vertices
whereas the soft contributions
are left to be evaluated.
This procedure is what is
usually called ``integrating out hard modes.''
In OPE,
the correspondence between
hard contributions and
Wilson coefficients
are unchanged,
while quantum corrections due to
low-energy degrees of freedom
are evaluated as non-perturbative matrix elements.

$X_{\bz}(Q^2;\mf)$ introduced in Eq.~\eqref{start}
can also be interpreted as a Wilson coefficient in OPE,
as we will see in explicit examples below.
However, the way to separate
UV and IR effects is different from
that of expansion by regions
in the following sense:
(i) The Wilson coefficient of our method is
regularized by a cutoff,
whereas the one in the expansion-by-regions method
is formulated in dimensional regularization.
(ii) We separate the UV contribution 
from the IR contribution
by only one measure,
i.e., scale of the gluon momentum.
On the other hand,
the method of expansion by regions
distinguishes momentum regions
with a finer resolution in general.

In the case that the relevant low-energy effective field theory
is known, the expansion-by-regions technique is a standard
tool to systematically compute Wilson coefficients
to high orders.
Detailed connection between the full theory and
the effective field theory can be made, including correspondence
of relevant kinematical regions.
Since $X_{\bz}(Q^2;\mf)$ in our formulation
is well defined in the full theory, using this information
it is possible to establish a firm connection between $X_{\bz}(Q^2;\mf)$ 
and Wilson coefficients, as we have briefly reviewed
in Sec.~\ref{ss:force}.

On the other hand, in 
the case that the relevant effective field theory is unknown,
we can still infer its structure using the expansion by regions,
as well as factorize UV and IR contributions to
the Wilson coefficients and non-perturbative matrix elements,
respectively, in dimensional regularization.
Changing to a cutoff regularization would be less founded,
since consistent treatment is not guaranteed by an
effective field theory framework.
Furthermore,
since the analysis necessarily becomes diagram-based, 
correspondence with operators, gauge symmetry, or
the equation of motion is not transparent.

We can clarify these two issues in explicit computations
for the observables which we studied already,
the Adler function and the QCD force (or the QCD potential).

\subsection{Example 1: Adler function}
\label{ss:adler2}

Using the expansion-by-region method
we first compute the small-$z$ expansion of the pre-weight 
$W^{(m)}_D(z)$ for the Adler function 
in the massive gluon scheme.
In this way we can identify which
kinematical region the power correction 
in Eq.~\eqref{defDUV} originate from.

\begin{figure}[t]
\begin{center}
\begin{tikzpicture}[baseline={(0,-0.1)},scale=0.7]
\draw (-3,0+2.5) node {\large $W_D^\mathrm{(H,H)A}$};
\draw [ultra thick,blue] (0,0+2.5) circle [radius=1];
\draw [ultra thick,blue,gluon] (0,1+2.5)--(0,-1+2.5);
\draw [ultra thick,decorate,photon] (-1.5,0+2.5)--(-1,0+2.5);
\draw [ultra thick,decorate,photon] (1,0+2.5)--(1.5,0+2.5);
\draw (-3+7,0+2.5) node {\large $W_D^\mathrm{(S,H)A}$};
\draw [ultra thick,blue] (0+7,0+2.5) circle [radius=1];
\draw [ultra thick,red,gluon] (0+7,1+2.5)--(0+7,-1+2.5);
\draw [ultra thick,decorate,photon] (-1.5+7,0+2.5)--(-1+7,0+2.5);
\draw [ultra thick,decorate,photon] (1+7,0+2.5)--(1.5+7,0+2.5);
\draw (-3+14,0+2.5) node {\large $W_D^\mathrm{(S,S)A}$};
\draw [ultra thick,red] (1+14,0+2.5) arc [start angle=0, end angle=180, radius=1];
\draw [ultra thick,blue] (-1+14,0+2.5) arc [start angle=180, end angle=360, radius=1];
\draw [ultra thick,red,gluon] (0+14,1+2.5)--(0+14,-1+2.5);
\draw [ultra thick,decorate,photon] (-1.5+14,0+2.5)--(-1+14,0+2.5);
\draw [ultra thick,decorate,photon] (1+14,0+2.5)--(1.5+14,0+2.5);
\draw (-3,0) node {\large $W_D^\mathrm{(H,H)B}$};
\draw [ultra thick,blue] (0,0) circle [radius=1];
\draw [ultra thick,blue,gluon] (0.7,0.7) .. controls (0.3,0.3) and (-0.3,0.3) .. (-0.7,0.7); 
\draw [ultra thick,decorate,photon] (-1.5,0)--(-1,0);
\draw [ultra thick,decorate,photon] (1,0)--(1.5,0);
\draw (-3+7,0) node {\large $W_D^\mathrm{(S,H)B}$};
\draw [ultra thick,blue] (0+7,0) circle [radius=1];
\draw [ultra thick,red,gluon] (0.7+7,0.7) .. controls (0.3+7,0.3) and (-0.3+7,0.3) .. (-0.7+7,0.7); 
\draw [ultra thick,decorate,photon] (-1.5+7,0)--(-1+7,0);
\draw [ultra thick,decorate,photon] (1+7,0)--(1.5+7,0);
\draw (-3+14,0) node {\large $W_D^\mathrm{(S,S)B}$};
\draw [ultra thick,red] (1+14,0) arc [start angle=0, end angle=180, radius=1];
\draw [ultra thick,blue] (-1+14,0) arc [start angle=180, end angle=360, radius=1];
\draw [ultra thick,red,gluon] (0.7+14,0.7) .. controls (0.3+14,0.3) and (-0.3+14,0.3) .. (-0.7+14,0.7); 
\draw [ultra thick,decorate,photon] (-1.5+14,0)--(-1+14,0);
\draw [ultra thick,decorate,photon] (1+14,0)--(1.5+14,0);
\draw (-2.5+13.5,0-2.5) node {\large $W_D^\mathrm{(S,S)C}$};
\draw [ultra thick,blue] (1+14,0-2.5) arc [start angle=0, end angle=180, radius=1];
\draw [ultra thick,red] (-1+14,0-2.5) arc [start angle=180, end angle=360, radius=1];
\draw [ultra thick,red,gluon] (0.7+14,0.7-2.5) .. controls (0.3+14,0.3-2.5) and (-0.3+14,0.3-2.5) .. (-0.7+14,0.7-2.5); 
\draw [ultra thick,decorate,photon] (-1.5+14,0-2.5)--(-1+14,0-2.5);
\draw [ultra thick,decorate,photon] (1+14,0-2.5)--(1.5+14,0-2.5);
\end{tikzpicture}
\vspace*{-1mm}
\caption{\small 
Different kinematical regions contributing
to the Adler function
in light of expansion by regions.
A blue (red) line represents that
a hard (soft) momentum $\sim Q\, (\sim\sqrt{\tau})$
is flowing through the line.
}
\vspace*{-8mm}
\label{fig:adler}
\end{center}
\end{figure}
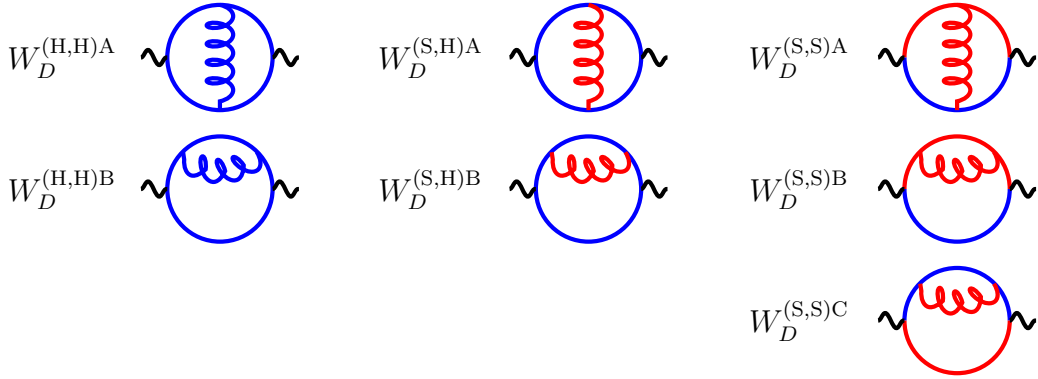

The kinematical regions contributing to the expansion of $W^{(m)}_D(z)$
are shown in Fig.~\ref{fig:adler},
in which blue (red) lines carry hard (soft) momenta.
Note that, since the external momentum is hard,
a hard momentum should flow through between the
two external vertices.
There is no contribution from the
kinematical region, in which the gluon is  hard
and some of the quark lines are soft.\footnote{
This is because in such a region
the soft scale integral becomes 
scaleless, since the soft scale $\tau$ is included
only in the gluon propagator, and if gluon is hard, after expansion in $\tau$
no soft scale remains
in the denominator.
}
Hence, we divide the kinematical regions 
into three regions,
(H,H), (S,H) and (S,S), 
as shown in the figure:
\begin{align}
W^{(m)}_D(z)=W_D^\mathrm{(H,H)}(z)+W_D^\mathrm{(S,H)}(z)+W_D^\mathrm{(S,S)}(z)
\,,
\label{decWD}
\end{align}
where $W_D^{\rm (H,H)}=W_D^{\rm (H,H)A}+W_D^{\rm (H,H)B}$, etc.

For instance, the ``all-hard'' contribution $W_D^{\rm (H,H)}$ is 
computed as follows.
Recall that $W^{(m)}_D$ in the massive gluon scheme
is given as [c.f., Eq.~\eqref{replace}]
\be
W^{(m)}_D(\tau/Q^2)=
\int_0^{\infty} \frac{d (p^2)}{2 \pi} \,\frac{1}{p^2-\tau} \, w_D(p^2/Q^2) \, .
\label{WDint}
\ee
$W_D^{\rm (H,H)}$ is 
obtained by expanding the gluon propagator in
$\tau$
as
\begin{align}
W_D^{\rm (H,H)}(\tau/Q^2)
=\sum _{n=0}^\infty 
\int_0^{\infty} \, \frac{d (p^2)}{2 \pi} \,
\frac{\tau ^n}{(p^2)^{n+1}}\,
w_D(p^2/Q^2) \, ,
\label{RegionNote8}
\end{align}
where it is understood that $w_D$ is regularized by dimensional 
regularization.\footnote{
It means that one 
should not use Eq.~\eqref{wD} for $w_D$.
One expands the integrand before performing any momentum integral while
keeping $\varepsilon\ne 0$.
}
Apart from the gluon propagator, the integrand
does not receive any modification
since the soft-scale parameter $\tau$ is 
contained only in the factor $1/(p^2-\tau )$.
The result of the computation reads
\begin{align}
  W_D^{\rm (H,H)}(z)=N_CC_F
  &\left[
    \frac{1}{4 \pi }
    +\frac{8-6 \zeta _3}{3 \pi }z
    +\left( -\frac{[\epsilon ^{-1}]}{2\pi }
    +\frac{4- 12 \zeta _3}{6 \pi}
     \right)
    z^2\right.\nonumber\\
    &\left. +\left(
    -\frac{[\epsilon ^{-2}]}{6\pi }
    +\frac{5[\epsilon ^{-1}]}{12\pi }
    +\frac{265}{216 \pi }+\frac{\pi }{36}\right) z^3
    +O\left(z^4\right)
  \right]. 
  \label{ebr:wdh}
\end{align}
Divergent terms are denoted as
\begin{align}
[\epsilon ^{-1}]
&=\frac{1}{\varepsilon}-\gamma _\mathrm{E}+\log \left(\frac{4\pi Q^2}{\mu ^2}\right)
+\mathcal{O}(\varepsilon )\,,
  \label{ebr:ep1}
\\
[\epsilon ^{-2}]
&=\frac{1}{\varepsilon^2}
-\frac{2}{\varepsilon}\left[
\gamma _\mathrm{E}-\log \left(\frac{4\pi Q^2}{\mu ^2}\right)\right]
+2\left[\gamma _\mathrm{E}-\log \left(\frac{4\pi Q^2}{\mu ^2}\right)\right]^2
+\mathcal{O}(\varepsilon ),
  \label{ebr:ep2}
\end{align}
where the space-time dimension is denoted as $d=4-2\varepsilon$;
$\gamma_\mathrm{E}=0.57721...$
is the Euler constant
and $\mu$ is the renormalization scale.
If we neglect $\log (Q^2/\mu ^2)$,
these terms correspond to
those which are subtracted in 
the usual $\overline{\mathrm{MS}}$ renormalization.

The results of the other two contributions are given by
\begin{align}
  W_D^{\rm (S,H)}(z)
  =N_CC_F&
  \left[
  \left( \frac{[\epsilon ^{-1}]}{2\pi }
    +\frac{6- 3\log z+3i\pi }{6 \pi}
    \right) z^2\right.
    +
        \left(
        \frac{[\epsilon ^{-2}]}{3\pi }
    -\frac{1+\log (-z) }{3\pi }[\epsilon ^{-1}]
    \right.\nonumber\\
    &\left.\left.
    +\frac{\log ^2(-z)}{6 \pi }+\frac{\log (-z)}{3 \pi }
    -\frac{91}{54 \pi }
    \right)
   z^3
    +O\left(z^4\right)
  \right] \,,
\label{ebr:wdhs}
\\
 W_D^{\rm (S,S)}(z)=
  N_CC_F&
  \left[
    \left(
     -\frac{[\epsilon ^{-2}]}{6\pi }
    +\frac{4\log (-z) -1}{12\pi }[\epsilon ^{-1}]
    \right.\right.\nonumber\\
    &\left.\left.
    -\frac{\log ^2(-z)}{3 \pi }+\frac{\log (-z)}{6 \pi }
     -\frac{\pi}{12}+\frac{35}{216 \pi }
    \right)
   z^3
    +O\left(z^4\right)
  \right],
  \label{ebr:wdss}
\end{align}
where we use a short-hand notation
$\log (-z)\equiv \log z-i\pi$.
Although 
$W_D^\mathrm{(H,H)}$, $W_D^\mathrm{(S,H)}$ and $W_D^\mathrm{(S,S)}$ individually
contain 
the divergent terms \eqref{ebr:ep1}, \eqref{ebr:ep2},
which are $\mu$-dependent,
these terms cancel altogether
in the sum \eqref{decWD}.\footnote{
Cancellation of
divergent terms is a
common feature in the method of expansion by regions
and signifies that the result is independent of the factorization scale
separating the soft and hard regions. 
}

The first two terms of the all-hard contribution [Eq.~\eqref{ebr:wdh}]
are exactly equal to the first two terms ($c_0$ and $c_1$) of
the expansion of $W^{(m)}_D(z)$ [Eq.~\eqref{WDexp}], while
the order $z^0$ and $z^1$ terms are absent in 
$W_D^\mathrm{(S,H)}(z)$ and $W_D^\mathrm{(S,S)}(z)$.
Therefore we conclude that
the $\mu_f$-independent $\LQ^2/Q^2$-term 
of $D_\mathrm{UV} (Q^2)$
belongs to the hard contribution.
Consequently the dominant part of $D_0(Q^2)$ is
also UV origin, according to the argument in the
previous subsection.
Thus, the lower figure of Figs.~\ref{fig:path}
corresponds to the proper interpretation
up to order $1/Q^2$.

In Sec.~\ref{sec.GC} we found that
the imaginary part of the expansion coefficients of $W^{(m)}_X(z)$
results in $\mu _f$-dependent terms,
and the $\mu _f$-dependent terms are related to
IR contributions.
The expansion-by-regions analysis
shows that the imaginary part of the expansion coefficients
stems only from the region where
the gluon has a soft-scale momentum.
This is because the only source of the imaginary part 
is the integral of $1/(p^2-\tau )$.
Indeed $W_D^\mathrm{(S,H)}(z)$ and $W_D^\mathrm{(S,S)}(z)$
include an imaginary part.
Oppositely, the all-hard contribution $W_D^\mathrm{(H,H)}(z)$
in Eq.~\eqref{ebr:wdh}
is explicitly real.
In fact, Eq.~\eqref{RegionNote8} shows that $W_D^\mathrm{(H,H)}(z)$ is
Euclidean and real to all orders in $1/Q^2$ expansion.

In Ref.~\cite{Ball:1995ni}, using the method of massive gluon,
terms which are non-analytic in the gluon mass $\lambda$
are identified as IR contributions,
while terms which are power-like in $\lambda$
as UV contributions.
Written in the form of Eq.~\eqref{RegionNote8},
the source of the imaginary part can be
attributed to the same origin.
For example, a non-analytic term $\log \lambda^2$ 
generates an imaginary part when we substitute $\lambda^2 =-\tau$ 
with $\tau >0$.

\arraycolsep=1.4pt\def\arraystretch{1.5}
\begin{table}
\begin{center}
\begin{tabular}{c||c|c|c|c}
&$\quad c_0\quad $
&$\quad c_1\quad $
&$\quad c_2\quad $
&$\quad c_3\quad $\\\hline
\hline
$W^{(H,H)}_D(z)$
&$\mathbb{R}$
&$\mathbb{R}$
&$\mathbb{R}$
&$\mathbb{R}$\\\hline
$W^{(S,H)}_D(z)$&&
&$\mathbb{C}$
&$\mathbb{C}$\\\hline
$W^{(S,S)}_D(z)$&&&
&$\mathbb{C}$
\end{tabular}
\end{center}
\caption{\small 
Expansion coefficients $c_n$ of the contribution
from each kinematical region to
$W^{(m)}_D(z)$
up to order $z^3$ [Eqs.~\eqref{ebr:wdh},
\eqref{ebr:wdhs}, \eqref{ebr:wdss}].
The symbol ``$\mathbb{R}$"
stands for a non-zero real value, while
``$\mathbb{C}$"
represents a complex value with non-zero imaginary part.
A blank represents that the coefficient is zero.
}
\label{tab:coeffs}
\end{table}
In Tab.~\ref{tab:coeffs} we summarize the
contribution
from each momentum region to the expansion coefficients of $W^{(m)}_D(z)$
up to ${\cal O}(z^3)$
[Eqs.~\eqref{ebr:wdh}, \eqref{ebr:wdhs}, \eqref{ebr:wdss}].
The first two coefficients of $W^{(m)}_D(z)$
originate only from the all-hard region [(H,H) region],
and there is no divergence up to this order.
From the order $z^2$,
the contribution from each region diverges
and only the sum is finite.
In the case that each contribution is
divergent, it is $\mu$-dependent and the separation between
different regions becomes somewhat vague.
The contribution from the soft-gluon and hard-quark region [(S,H) region]
starts at order $z^2$, and
the region where the gluon is soft and some of the quarks
are soft [(S,S) region] contributes from order $z^3$.
As already mentioned above, (S,H) and (S,S) contributions
have a non-zero imaginary part.
Notably these regions also contribute
to the real part, although the values are divergent and 
become definite only after
they are added to the (H,H) contributions.
Namely, from the order $z^2$, the real part of the expansion
coefficients receive mixed contributions
from the hard and soft momentum regions of the gluon.
This is in contrast to the imaginary part, to which only the 
soft-gluon-momentum regions can contribute.

It is worth emphasizing that
the soft contributions are not always pure imaginary,
i.e., not all of the real part of the expansion coefficients 
originate from the hard-gluon region.\footnote{
In particular, the soft contribution may give 
analytic terms (i.e. terms without logarithms).
This point was not correctly addressed in Ref.~~\cite{Ball:1995ni}, Sec.~2.4.
}
Thus, the method of expansion by regions
has a finer resolution than
our analysis given in Sec.~\ref{sec.GC} and
detects soft contributions even in
the real part of $c_n$ for $n\ge u_{\rm IR}$.
From this detailed examination,
we confirm consistency\footnote{
There may occur a contradiction
to the result of Sec.~\ref{sec.GC} in exceptional cases
where the leading soft-gluon
contribution happens to be pure real.
This may happen if the leading IR renormalon 
vanishes accidentally.
} 
of our treatment of $X_{\rm UV}$
in Sec.~\ref{sec.GC},
where we classify as the genuine UV contribution
the $(\LQ^2/Q^2)^n$-terms for $0\le n<u_{\IR}$.

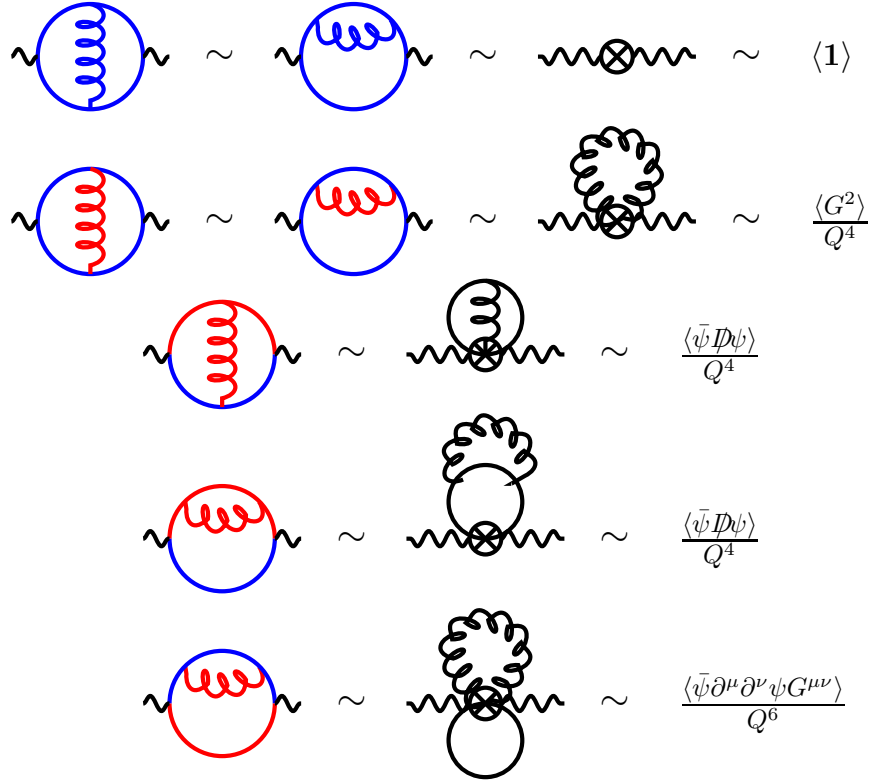
\begin{figure}
\begin{center}
\begin{tikzpicture}[baseline={(0,-0.1)},scale=0.7,text width=1]
\draw [ultra thick,blue] (0-5,0) circle [radius=1];
\draw [ultra thick,blue,gluon] (0-5,1)--(0-5,-1);
\draw [ultra thick,decorate,photon] (-1.5-5,0)--(-1-5,0);
\draw [ultra thick,decorate,photon] (1-5,0)--(1.5-5,0);
\draw (2.2-5,0) node {\Large $\sim$};
\draw [ultra thick,blue] (0,0) circle [radius=1];
\draw [ultra thick,blue,gluon] (0.7,0.7) .. controls (0.3,0.3) and (-0.3,0.3) .. (-0.7,0.7); 
\draw [ultra thick,decorate,photon] (-1.5,0)--(-1,0);
\draw [ultra thick,decorate,photon] (1,0)--(1.5,0);
\draw (2.2,0) node {\Large $\sim$};
\draw [ultra thick,decorate,photon] (0.3+5,0)--(1.5+5,0);
\draw [ultra thick,decorate,photon] (-0.3+5,0)--(-1.5+5,0);
\draw [ultra thick] (0+5,0) circle [radius=0.3];
\draw [ultra thick] (0.3*0.707+5,-0.3*0.707)--(-0.3*0.707+5,0.3*0.707);
\draw [ultra thick] (0.3*0.707+5,0.3*0.707)--(-0.3*0.707+5,-0.3*0.707);
\draw (2.2+5,0) node {\Large $\sim $};
\draw (2.2+6.5,0) node {\large $\langle \mathbf{1} \rangle$};
\end{tikzpicture}
\\
\begin{tikzpicture}[baseline={(0,-0.1)},scale=0.7,text width=1]
\draw [ultra thick,blue] (0-5,0) circle [radius=1];
\draw [ultra thick,red,gluon] (0-5,1)--(0-5,-1);
\draw [ultra thick,decorate,photon] (-1.5-5,0)--(-1-5,0);
\draw [ultra thick,decorate,photon] (1-5,0)--(1.5-5,0);
\draw (2.2-5,0) node {\Large $\sim$};
\draw [ultra thick,red,gluon] (0.7,0.7) .. controls (0.3,0.3) and (-0.3,0.3) .. (-0.7,0.7); 
\draw [ultra thick,blue] (0,0) circle [radius=1];
\draw [ultra thick,decorate,photon] (-1.5,0)--(-1,0);
\draw [ultra thick,decorate,photon] (1,0)--(1.5,0);
\draw (2.2,0) node {\Large $\sim$};
\draw [ultra thick,decorate,photon] (0.3+5,0)--(1.5+5,0);
\draw [ultra thick,decorate,photon] (-0.3+5,0)--(-1.5+5,0);
\draw [ultra thick] (0+5,0) circle [radius=0.3];
\draw [ultra thick] (0.3*0.707+5,-0.3*0.707)--(-0.3*0.707+5,0.3*0.707);
\draw [ultra thick] (0.3*0.707+5,0.3*0.707)--(-0.3*0.707+5,-0.3*0.707);
\draw [ultra thick,decorate,gluon] (0+5,0) .. controls (-2+5,2) and (2+5,2) .. (0+5,0);
\draw (2.2+5,0) node {\Large $\sim $};
\draw (2.2+6.5,0) node {\Large $\frac{\langle G^2 \rangle}{Q^4}$};
\end{tikzpicture}
\\
\begin{tikzpicture}[baseline={(0,-0.1)},scale=0.7,text width=1]
\draw [ultra thick,red] (1,0) arc [start angle=0, end angle=180, radius=1];
\draw [ultra thick,blue] (-1,0) arc [start angle=180, end angle=360, radius=1];
\draw [ultra thick,red,gluon] (0,1)--(0,-1);
\draw [ultra thick,decorate,photon] (-1.5,0)--(-1,0);
\draw [ultra thick,decorate,photon] (1,0)--(1.5,0);
\draw (2.2,0) node {\Large $\sim$};
\draw [ultra thick,decorate,photon] (0.3+5,0)--(1.5+5,0);
\draw [ultra thick,decorate,photon] (-0.3+5,0)--(-1.5+5,0);
  \draw [ultra thick] (0+5,0.7) circle [radius=0.7];
  \draw [ultra thick,decorate,gluon] (0+5,1.4)--(0+5,0);
\draw [ultra thick] (0+5,0) circle [radius=0.3];
\draw [ultra thick] (0.3*0.707+5,-0.3*0.707)--(-0.3*0.707+5,0.3*0.707);
\draw [ultra thick] (0.3*0.707+5,0.3*0.707)--(-0.3*0.707+5,-0.3*0.707);
\draw (2.2+5,0) node {\Large $\sim $};
\draw (2.2+6.5,0) node {\Large $\frac{\langle \bar{\psi} D\!\!\!\! / \psi \rangle}{Q^4}$};
\end{tikzpicture}
\\
\begin{tikzpicture}[baseline={(0,-0.1)},scale=0.7,text width=1]
\draw [ultra thick,red] (1,0) arc [start angle=0, end angle=180, radius=1];
\draw [ultra thick,blue] (-1,0) arc [start angle=180, end angle=360, radius=1];
\draw [ultra thick,red,gluon] (0.7,0.7) .. controls (0.3,0.3) and (-0.3,0.3) .. (-0.7,0.7); 
\draw [ultra thick,decorate,photon] (-1.5,0)--(-1,0);
\draw [ultra thick,decorate,photon] (1,0)--(1.5,0);
\draw (2.2,0) node {\Large $\sim$};
\draw [ultra thick,decorate,photon] (0.3+5,0)--(1.5+5,0);
\draw [ultra thick,decorate,photon] (-0.3+5,0)--(-1.5+5,0);
  \draw [ultra thick] (0+5,0.7) circle [radius=0.7];
  \draw [ultra thick,decorate,gluon] (-0.6*0.707+5,0.7+0.6*0.707) arc [start angle=215,end angle=-45,radius=0.6];
\draw [ultra thick] (0+5,0) circle [radius=0.3];
\draw [ultra thick] (0.3*0.707+5,-0.3*0.707)--(-0.3*0.707+5,0.3*0.707);
\draw [ultra thick] (0.3*0.707+5,0.3*0.707)--(-0.3*0.707+5,-0.3*0.707);
\draw (2.2+5,0) node {\Large $\sim $};
\draw (2.2+6.5,0) node {\Large $\frac{\langle \bar{\psi} D\!\!\!\! / \psi \rangle}{Q^4}$};
\end{tikzpicture}
\\
\begin{tikzpicture}[baseline={(0,-0.1)},scale=0.7,text width=1]
\draw [ultra thick,red,gluon] (0.7,0.7) .. controls (0.3,0.3) and (-0.3,0.3) .. (-0.7,0.7); 
\draw [ultra thick,blue] (1,0) arc [start angle=0, end angle=180, radius=1];
\draw [ultra thick,red] (-1,0) arc [start angle=180, end angle=360, radius=1];
\draw [ultra thick,decorate,photon] (-1.5,0)--(-1,0);
\draw [ultra thick,decorate,photon] (1,0)--(1.5,0);
\draw (2.2,0) node {\Large $\sim$};
\draw [ultra thick,decorate,photon] (0.3+5,0)--(1.5+5,0);
\draw [ultra thick,decorate,photon] (-0.3+5,0)--(-1.5+5,0);
  \draw [ultra thick] (0+5,-0.7) circle [radius=0.7];
\draw [ultra thick] (0+5,0) circle [radius=0.3];
\draw [ultra thick] (0.3*0.707+5,-0.3*0.707)--(-0.3*0.707+5,0.3*0.707);
\draw [ultra thick] (0.3*0.707+5,0.3*0.707)--(-0.3*0.707+5,-0.3*0.707);
\draw [ultra thick,decorate,gluon] (0+5,0) .. controls (-2+5,2) and (2+5,2) .. (0+5,0);
\draw (2.2+5,0) node {\Large $\sim $};
\draw (2.2+6.5,0) node {\Large $\frac{\langle \bar{\psi} \partial^\mu \partial^\nu \psi G^{\mu\nu} \rangle}{Q^6}$};
\end{tikzpicture}
\caption{\small 
Relations between contributions 
to the Adler function evaluated
with the expansion-by-regions method
(colored graphs; c.f.\ Fig.~\ref{fig:adler})
and those with
a (would-be) low-energy effective field theory
(black graphs).
The corresponding terms
in OPE are also shown.
}
\vspace*{-8mm}
\label{fig:adler2}
\end{center}
\end{figure}

Let us turn to examine OPE of the Adler function 
using the method of expansion by regions.
The relevant low-energy effective
field theory is not known.
Applying the expansion-by-regions method to 
the diagrams for the reduced Adler function,
they are decomposed into contributions from different kinematical regions
as shown in Fig.~\ref{fig:adler2}.
For instance, the all-hard region can be
identified with the Wilson coefficient for the identity operator,
as illustrated in the figure.
Similarly, a contribution involving soft gluons/quarks 
can be identified with 
the matrix element of a
higher-dimensional local
operator times its Wilson coefficient.
This includes the local gluon condensate
at order $1/Q^4$.

Using this correspondence,
we argue that $D_{\UV}(Q^2)$ is almost identified with the Wilson coefficient of the identity operator $C_1(Q^2;\mf)$. 
Recall that $D_{\UV}(Q^2)$ is a $\mf$-independent part of $D_{\bz}(Q^2;\mf)$, 
which is diagrammatically given by Fig.~\ref{Fig.Adlerdiagram} 
with an effective coupling $\alpha_{\bz}(\tau)$
and an IR cutoff $\mf$ of the gluon momentum. 
As inferred from the above correspondence,
in general the all-hard region of a diagram,
where all the momenta are larger than the cutoff scale $\mf$, 
contributes to $C_1(Q^2;\mf)$, since the entire loop
integral shrinks to a local vertex.
In contrast a contribution which includes soft modes becomes 
a non-perturbative matrix element times its Wilson coefficient.

$D_{\bz}(Q^2;\mf)$ is slightly different from $C_1$ in that it includes 
both hard and soft quarks.
The leading contribution involving soft quarks reduces to 
the matrix element of the dimension-six operator 
$(\bar{q}\gamma^\mu q) (\bar{q}\gamma_\mu q)$
made only of the quark field.%
\footnote{
Although one may be worried that the cutoff regularization
would break gauge invariance and
generate gauge non-invariant operators,
in fact our regularization method preserves gauge invariance.
} 
(Note that the dimension-three operator $\bar{q} q$ 
is absent since it appears together with the quark mass.) 
Thus, we obtain\footnote{
Contributions from the soft region of the fermion bubble
subgraphs are also suppressed.
}
\be
D_{\bz}(Q^2;\mf)=C_1(Q^2;\mf)+\mathcal{O}(\mf^6/Q^6) \, .
\ee
As a result, the $\mf$-independent part $D_{\UV}$ is identified with $C_1(Q^2;\mf)$ up to
$\mathcal{O}(\mf^4/Q^4)$ via Eq.~\eqref{a20}:
\be
C_1(Q^2;\mf)-D_{\UV}(Q^2)=\mathcal{O}(\mf^4/Q^4) \, .
\ee
In particular, the power correction $\LQ^2/Q^2$ in $D_{\UV}$ is a part of the Wilson coefficient $C_1$.

We note that the matrix element of the dimension-four operator
$\bar{q}\,\sla{\! D}q$ vanishes by the equation of motion
$\sla{\! D}q=0$.
Essentially the same effect is observed in the computation of
the expansion coefficient of $W^{(m)}_D$ in the first part of this subsection.
$W_D^{(S,S)}$ represents contributions in which the gluon and some of
the quarks have soft momenta (see Fig.~\ref{fig:adler}).
By explicit computation we confirm that each of
$W_D^{(S,S)A}$ and $W_D^{(S,S)B}$
is nonzero at $\mathcal{O} (z^2)$,\footnote{
Note that $W_D^{(S,S)C}=0$ by massless quark loop,
see Fig.~\ref{fig:adler2}.
}
while they cancel in the sum,
resulting in the $z^3$ term
as the leading term
in their sum.%
\footnote{
The same cancellation mechanism cannot be seen
explicitly in the computation
of the massless diagrams in Fig.~\ref{fig:adler2}, 
since the soft-scale integrals are scaleless and vanish
for all the diagrams.
On the other hand, in Fig.~\ref{fig:adler}
the gluon mass $\tau$ acts as the soft scale.
}
This property is considered
to be a consequence of gauge invariance and
the equation of motion.
Since each diagram does not respect gauge invariance, 
the soft contribution from each diagram at  order $1/Q^4$
is non-vanishing,
but the sum of all the diagrams should vanish at this order.
The first gauge invariant operator involving soft quarks
is dimension six, as already noted.

\subsection{Example 2: QCD potential}
\label{ss:force2}

The UV contribution to the
force between the static quark and antiquark 
$\alpha_{F,\UV} (1/r^2)$
and its power behavior are analyzed 
in Sec.~\ref{ss:force},
where it is shown that
the $\mu_f$-independent $r^2$-term 
exists in $\alpha_{F,\UV} (1/r^2)$.
This result is obtained from
the one-dimensional integral representation
of the QCD potential \eqref{f1},
and we investigate the QCD potential
in this subsection.

The pre-weight $W^{(m)}_{V} (z)$ of the QCD potential
in the massive gluon scheme is given in Eq.~\eqref{pre_weight_of_V}
and computed from the integration
\begin{align}
  W^{(m)}_{V} (z)=-\frac{2C_F}{\pi}
  \int_{0}^{\infty}
  \frac{\sin (pr) p}{p^2-\tau -i0}dp \,,
  \label{ebr:wv1}
\end{align}
where $z=\tau r^2$.
We investigate the kinematical regions which contribute to
the small-$z$ expansion of $W^{(m)}_V(z)$.
To apply the method of expansion by regions,
we introduce a variant of the dimensional regularization
by replacing 
$dp\to p^{-2\varepsilon}dp$ in Eq.~\eqref{ebr:wv1}.
While in the conventional expansion-by-regions 
method~\cite{Smirnov:2002pj}
only Feynman integrals in momentum space are considered,
Eq.~\eqref{ebr:wv1} contains both
coordinate-space variable ($r$)
and momentum-space variable ($p$).
To our knowledge, there is no systematic argument concerning
validity of the expansion-by-regions method
in such cases.
Nevertheless in the current specific example,
we can show validity of the method
using the argument in Ref.~\cite{Jantzen:2011nz}.

Similarly to Eq.~\eqref{RegionNote8},
the contribution from the hard region $p\sim 1/r$ is given as
\begin{align}
  W_{V} ^{(H)}(z)&=-\frac{2C_F}{\pi}
  \sum_{n=0}^{\infty}
  \int_{0}^{\infty}\frac{\tau^n}{(p^2)^{n+1}}
  \sin (pr) p^{1-2\varepsilon }dp\\
  &=
  -\frac{2C_F}{\pi}
  r^{2\varepsilon}
  \sum_{n=0}^\infty
  z^n
  \frac{\pi \Gamma (-2n-2\varepsilon )}
  {\Gamma (n+\varepsilon +1)\Gamma (-n-\varepsilon)}\\
  &=-C_F\cos (\sqrt{z})
  +{\cal O}(\varepsilon)
  \,,
  \label{ebr:wvh}
\end{align}
while the contribution from the 
soft region $p\sim \sqrt{\tau}$ is given as
\begin{align}
  W_{V} ^{(S)}(z)&=
  -\frac{2C_F}{\pi}
  \sum_{n=0}^{\infty}\frac{(-1)^n}{(2n+1)!}
  \int_{0}^{\infty}\frac{(pr)^{2n+1}}{p^2-\tau -i0}
  p^{1-2\varepsilon}dp\\
  &=-\frac{C_F}{\pi} 
    \sum_{n=0}^\infty
  \frac{(\sqrt{z})^{2n+1}}{(2n+1)!} \tau^{-\varepsilon}
  \Gamma (\varepsilon -\frac{1}{2})
  \Gamma (\frac{3}{2}-\varepsilon )
  e^{i\pi (n-\varepsilon +1/2)}\\
  &=- i C_F \sin (\sqrt{z})
  +{\cal O}(\varepsilon)
  \,.
  \label{ebr:wvs}
\end{align}
Both hard and soft contributions
are finite as $\varepsilon \to 0$
to all orders in the small-$z$ expansion.

The hard and soft contributions separate into
the real and imaginary part, respectively.
There is no mixed contribution
from both regions to each expansion coefficient, 
so that the correspondence is simpler than the
Adler function.
Namely, each coefficient is either real or pure imaginary,
where the former originates from the hard region and the latter
from the soft region.
The real and imaginary coefficients appear alternately.
The order $z^1$ term of $\cos(\sqrt{z})$ gives the
$\LQ^2r^2$ term of $\alpha_{F,\UV}$ [linear potential in $V_{\rm QCD}(r)$],
 which indeed stems from the hard region.

As already explained in Sec.~\ref{ss:force},
the effective field theory for the QCD potential is known as pNRQCD,
and its construction can be understood using the
integration-by-regions method.
According to this understanding, pNRQCD for the
static QCD potential is constructed by integrating out the
so-called ``hard'' and ``soft'' scales.
The remaining  active dynamical degrees
of freedom are those in the ``ultra-soft'' scale and the $\LQ$
scale.\footnote{
There is no contribution from the
``potential region'' in the computation of
the static QCD potential due to the fact that the
static propagator originating from the Wilson line does
not include the kinetic energy term $\sim \vec{p}^{\,2}/(2m)$.
}

Computations in the framework of pNRQCD is systematically organized
using the multipole expansion, which gives an OPE in this
effective field theory.
A number of Wilson coefficients in pNRQCD have been computed
using the method of expansion by regions.
Wilson coefficients are usually regularized by dimensional regularization and
they contain divergences in general.
It is possible to change to another regularization
scheme within pNRQCD framework, and the physical predictions should
not depend on the regularization scheme.
Hence, through such a route, computation of the QCD potential
in our framework can be related to that of pNRQCD or full QCD
without any ambiguity.

In dimensional regularization and in strict expansion
in $\alpha_s$,
the leading Wilson coefficient $V_S(r)$ in Eq.~\eqref{opeQCD}
coincides with $V_{\rm QCD}(r)$ 
to all orders in $\alpha_s$, since 
contributions from the ultra-soft
and $\LQ$ scales (e.g., the second term) 
evaluate to scaleless integrals at each order of $\alpha_s$
in the expansion-by-regions method.
In OPE 
the ultra-soft and $\LQ$ contributions turn into 
non-perturbative matrix elements.
If we adopt the large-$\beta_0$ approximation and
the cutoff in the gluon momentum,
$V_S(r)$ coincides with $V_{\bz}(r;\mf)$ in our formulation.
At the same time,
the leading non-perturbative matrix element 
is estimated as order $\mf^3 r^2$ in this regularization scheme.
By examining the matching between full QCD and pNRQCD in detail,
we can check that 
with this regularization Eq.~\eqref{opeQCD}
also achieves a consistent
separation of the UV (hard+soft)
and IR (ultrasoft+$\LQ$) contributions to the whole static
QCD potential $V_{\rm QCD}(r)$.
Details of the computation can be found, e.g., in 
Ref.~\cite{Sumino:2014qpa}.

\subsection{Relevant kinematical region for $X_{\UV}$: Reconsideration with explicit cutoff}
\label{sec:cutoff}
We revisit the issue:
 ``Which kinematical regions does $X_{\UV}$
originate from?'' 
In this subsection, we address this point 
by introducing an explicit cutoff scale to separate the kinematical regions,
instead of the method of expansion by regions.
The use of the explicit cutoff enables us to understand more directly   
that the contribution from the integral along $C_a$ originates
from UV regions.
In the following,
$X$ is assumed to be the reduced Adler function or the (dimensionless) QCD potential.  

We divide $W_X^{(m)}$ in Eq.~\eqref{replace} into two parts as follows:
\begin{align}
W_X^{(m)}(z) = W_X^\mathrm{H}(z;\mf^2/Q^2)+W_X^\mathrm{S}(z;\mf^2/Q^2)
\end{align}
with
\begin{align}
W_X^\mathrm{H}(\tau/Q^2;\mf^2/Q^2) &\equiv \int_{\mf^2}^{\infty} \frac{d (p^2)}{2 \pi} \frac{w_X(p^2/Q^2)}{p^2-\tau} \label{WXH}\\
W_X^\mathrm{S}(\tau/Q^2;\mf^2/Q^2) &\equiv \int_0^{\mf^2} \frac{d (p^2)}{2 \pi} \frac{w_X(p^2/Q^2)}{p^2-\tau} \, ,
\end{align}
which consist of UV and IR modes, respectively.

We first show that the contribution from 
the integral along $C_a$,
\begin{align}
{\rm Im} \int_{C_a} \frac{d \tau}{\pi \tau}  W^{(m)}_X \lt(\frac{\tau}{Q^2} \rt) \alpha_{\bz}(\tau)
&=\int_0^{\infty} \frac{d \tau}{\pi \tau} W^{(m)}_{X+} \lt(\frac{\tau}{Q^2} \rt) {\rm Im}\, \alpha_{\bz}(-\tau+i0) \non
&\sim -\frac{4 \pi}{\bz} c_0+\frac{4 \pi d_0}{\bz} \frac{1}{\log{(Q^2/\LQ^2)}} ~~\text{as $Q^2 \gg \LQ^2$}\, , \label{CalargeQ}
\end{align}
originates dominantly from $W_X^\mathrm{H}$ [see Eq.~\eqref{asym} for its asymptotic form]. 
For this purpose it is sufficient to show that the part
given through $W_X^\mathrm{S}$ gives subdominant contribution.
Then we investigate
\begin{align}
\int_0^{\infty} \frac{d \tau}{\pi \tau} W_{X+}^\mathrm{S} \lt(\frac{\tau}{Q^2};\frac{\mf^2}{Q^2} \rt) 
{\rm Im}\, \alpha_{\bz}(-\tau+i0) \, ,   \label{Scon}
\end{align}
where $W^\mathrm{S}_{X+}(z;\mf^2/Q^2) \equiv W^\mathrm{S}_X(-z;\mf^2/Q^2)$.
In the computation of $W_{X+}^\mathrm{S}$, there exists some constant $\tilde{c}>0$ such that
\begin{align}
|w_X(p^2/Q^2)| \leq \tilde{c} \, (p^2/Q^2)^{u_{\IR}} \,  \label{upbound}
\end{align}
for sufficiently large $Q^2$.
Using Eq.~\eqref{upbound},
we obtain
\begin{align}
\lt|W_{X+}^\mathrm{S} \lt(\frac{\tau}{Q^2}; \frac{\mf^2}{Q^2} \rt) \rt|
& \leq \int_0^{\mf^2} \frac{d(p^2)}{2 \pi} \frac{|w_X(p^2/Q^2)|}{p^2+\tau}
 \leq \frac{\tilde{c}}{\pi} \frac{\mf^2}{\mf^2+\tau} \lt(\frac{\mf^2}{Q^2} \rt)^{u_{\IR}} \, ,
 \label{evaluation}
\end{align}
where we 
use the fact that the integrand becomes 
a monotonically increasing function of $p$ 
for $u_{\IR} \geq 1/2$.
Therefore, it is shown that Eq.~\eqref{Scon} is suppressed by 
$(1/Q^{2})^{u_\mathrm{IR}}$:
\begin{align}
\int_0^{\infty} \frac{d \tau}{\pi \tau} W_{X+}^\mathrm{S} \lt(\frac{\tau}{Q^2};\frac{\mf^2}{Q^2} \rt)  
{\rm Im}\, \alpha_{\bz}(-\tau+i0)
=\mathcal{O}((1/Q^2)^{u_{\IR}}) \, ,
\end{align}
which is subdominant compared with the asymptotic behavior shown in Eq.~\eqref{CalargeQ}.
Thus, we conclude that the contribution from the integral along $C_a$
originates from UV (large $p$) regions for large $Q$.

We also show that the first few terms of the expansion of $W_X^{(m)}(z)$
is dominantly reproduced by that of $W_X^\mathrm{H}$.  
In this case, 
we use the relation 
$|\tau| \leq \mf^2$
which is satisfied along $C_b$.
The expansion of $W_X^\mathrm{H}$ is obtained by expanding 
the gluon propagator in  
Eq~\eqref{WXH}:
\begin{align}
W^\mathrm{H}_X(z;\mf^2/Q^2)
=\sum_{n=0}^{\infty} c^\mathrm{H}_n z^n \, ,
\end{align}
where $c^\mathrm{H}_n$ is given as a function of $\mf^2/Q^2$:
\begin{align}
c^\mathrm{H}_n=\int_{\mf^2/Q^2}^{\infty} \frac{d x}{2  \pi} w_X(x) x^{-n-1} \, .
\end{align}
In the following, we focus on the case $n<u_{\IR}$, 
which gives the power correction $\sim (\LQ^2/Q^2)^n$ in $X_{\UV}$.
We rewrite $c_n^\mathrm{H}$ as
\begin{align}
c_n^\mathrm{H}=c_n-\int_0^{\mf^2/Q^2} \frac{d x}{2 \pi} w_X(x) x^{-n-1} \, .
\end{align}
Here we use $c_n=\int_0^{\infty} (d x/2 \pi) w_X(x) x^{-n-1}$
as shown in Eqs.~\eqref{WSBorel} and \eqref{Borel}.
Then a similar evaluation as Eq.~\eqref{evaluation} leads 
that the difference between $c_n$ and $c^\mathrm{H}_n$ satisfies
\begin{align}
|c_n-c_n^\mathrm{H}| \leq \frac{\tilde{c}}{2 \pi(u_{\IR}-n)} (\mf^2/Q^2)^{u_{\IR}-n} \, ,
\end{align}
thereby,
 $c_n^\mathrm{H} \sim c_n$ for $n<u_{\IR}$ for large $Q$.
Namely, we have confirmed that
the power corrections in $X_{\UV}$ 
originate from UV (large $p$) regions.

From the above discussion we conclude that each term in $X_{\UV}$
(the integral along $C_a$ and the first few terms of the integral along $C_b$)
originates from UV regions.

We finally note that the method presented in this subsection has a drawback
that it is difficult to discuss $c^\mathrm{H}_n$ 
for $n \geq u_{\IR}$ systematically
in contrast to the method of expansion by regions.
The difference of these two methods is clear in the case of the QCD potential. 
In the method of expansion by regions, 
the hard (soft) contributions are identified with
the real (imaginary) part of the pre-weight 
to all orders in $z$ [Eqs.~\eqref{ebr:wvh} and \eqref{ebr:wvs}].
However,
it is difficult to reach the same result within the method with the explicit cutoff.   
The method presented here has an advantage to examine 
the contribution from the integral along $C_a$.

\section{Relation between $X_{\UV}$ and perturbative series at large orders}
\label{sec.RB}

In this section we show that
$X_{\UV}(Q^2)$ derived in 
Sec.~\ref{sec.GC} is reproduced 
from the perturbative series in the large-$\bz$ approximation
at large orders.
Based on this observation we confirm that our result $X_{\UV}$ 
is consistent with the renormalon uncertainty 
and the OPE framework.
 
The smallest term of the perturbative series of Eq.~\eqref{ExLB}
is given at around $n_*=4 \pi u_{\IR}/(\beta_0 \alpha_s)$, 
hence, it is natural to regard
the truncated series at this order $X_{n_*}(Q^2)$ 
as an optimal prediction within perturbation theory. 
The uncertainty of the prediction $X_{n_*}(Q^2)$ is
of the order of $(\LQ^2/Q^2)^{u_{\IR}}$.
By taking a small $\alpha_s(\mu )$, 
we can examine the large order perturbative series 
keeping the perturbative series finite,
since $n_*$ becomes large in this case. 
The truncated series $X_{n_*}$ is written as
\begin{align}
X_{n_*}(Q^2)
&=\sum_{n=0}^{n_*-1} \int_{0}^{\infty} 
\frac{d \tau}{2 \pi \tau}\, w_X \! \lt(\frac{\tau}{Q^2} \rt) 
\alpha_s(\mu ) 
\lt[ \frac{\bz \alpha_s(\mu )}{4 \pi} \log \lt(\frac{\mu^2 e^{5/3}}{\tau} \rt) \rt]^{n}  
\non
&=\int_{0}^{\infty} \frac{d \tau}{2 \pi \tau} \,
w_X \! \lt(\frac{\tau}{Q^2} \rt) 
\alpha_s(\mu ) \frac{1-L^{n_*}}{1-L}  
\label{ntrun}\\
&={\rm Im} \int_{0}^{\infty} \frac{d \tau}{\pi \tau} \,
W_X \! \lt(\frac{\tau}{Q^2} \rt) 
\alpha_s(\mu ) 
\frac{1-L^{n_*}}{1-L} \, ,
\label{ntrunW}
\end{align}
where we define 
\be
L=\frac{\bz \alpha_s(\mu )}{4 \pi}
 \log{ \lt( \frac{\mu^2 e^{5/3}}{\tau} \rt)} \label{L} \, .
\ee
Since the integrand of Eq.~\eqref{ntrunW} is regular along the
integral path (positive real axis),
we can deform the path into $C_a$:
\begin{align}
X_{n_*}(Q^2)
&={\rm Im} \int_{C_a} \frac{d \tau}{\pi \tau} \,
W_X \! \lt(\frac{\tau}{Q^2} \rt) 
\alpha_s(\mu ) \frac{1-L^{n_*}}{1-L} \non
&={\rm Im} \int_{C_a} \frac{d \tau}{\pi \tau} \,
W_X \! \lt(\frac{\tau}{Q^2} \rt) 
\alpha_s(\mu ) \frac{1}{1-L}
+{\rm Im} \int_{C_a} \frac{d \tau}{\pi \tau} \,
W_X \! \lt(\frac{\tau}{Q^2} \rt) 
\alpha_s(\mu ) \frac{-L^{n_*}}{1-L}  \label{fst} \, .
\end{align}
The first term is a part of $X_0(Q^2)$ since
 \be
 \alpha_s(\mu ) \frac{1}{1-L}=\alpha_{\bz}(\tau) \, .
 \ee
In the second term of Eq.~\eqref{fst},
the integrand has a pole at $\tau=e^{5/3} \LQ^2$. 
We decompose the integral 
into the principle value part and the delta function part,
after taking the integral path again on the positive axis:
\be
\alpha_{\bz}(\tau)
={\rm Pr.} \, \alpha_{\bz}(\tau)
-\frac{4 \pi}{\bz}\, \pi \tau i \, \delta(\tau-e^{5/3} \LQ^2) \, .
\ee 
Thus, we obtain
\begin{align}
X_{n_*}(Q^2)
=&{\rm Im} \int_{C_a} \frac{d \tau}{\pi \tau} \,
W_X \! \lt(\frac{\tau}{Q^2} \rt) \alpha_{\bz}(\tau) 
+\frac{4 \pi}{\bz} {\rm Re} \, 
W_X\! \lt(\frac{e^{5/3} \LQ^2}{Q^2} \rt) 
\non
&+{\rm Pr.} \int_{0}^{\infty} 
\frac{d \tau}{2 \pi \tau} \,
w_X \! \lt(\frac{\tau}{Q^2} \rt) 
\alpha_{\bz}(\tau) (-L^{n_*}) \, ,
\label{eqXnstar}
\end{align}
where we used Eq.~\eqref{rel} for the third term.
By expanding ${\rm Re}\, W_X$ in $\LQ^2/Q^2$, 
we can see that $X_{n_*}(Q^2)$ indeed includes 
$X_{\UV}(Q^2)$; 
see Eqs.~\eqref{exprel} and \eqref{SUV}. 

\begin{figure}
\centering
 \begin{tabular}{ll}
 \begin{minipage}{0.5\hsize}
 \centering
\includegraphics[width=7cm]{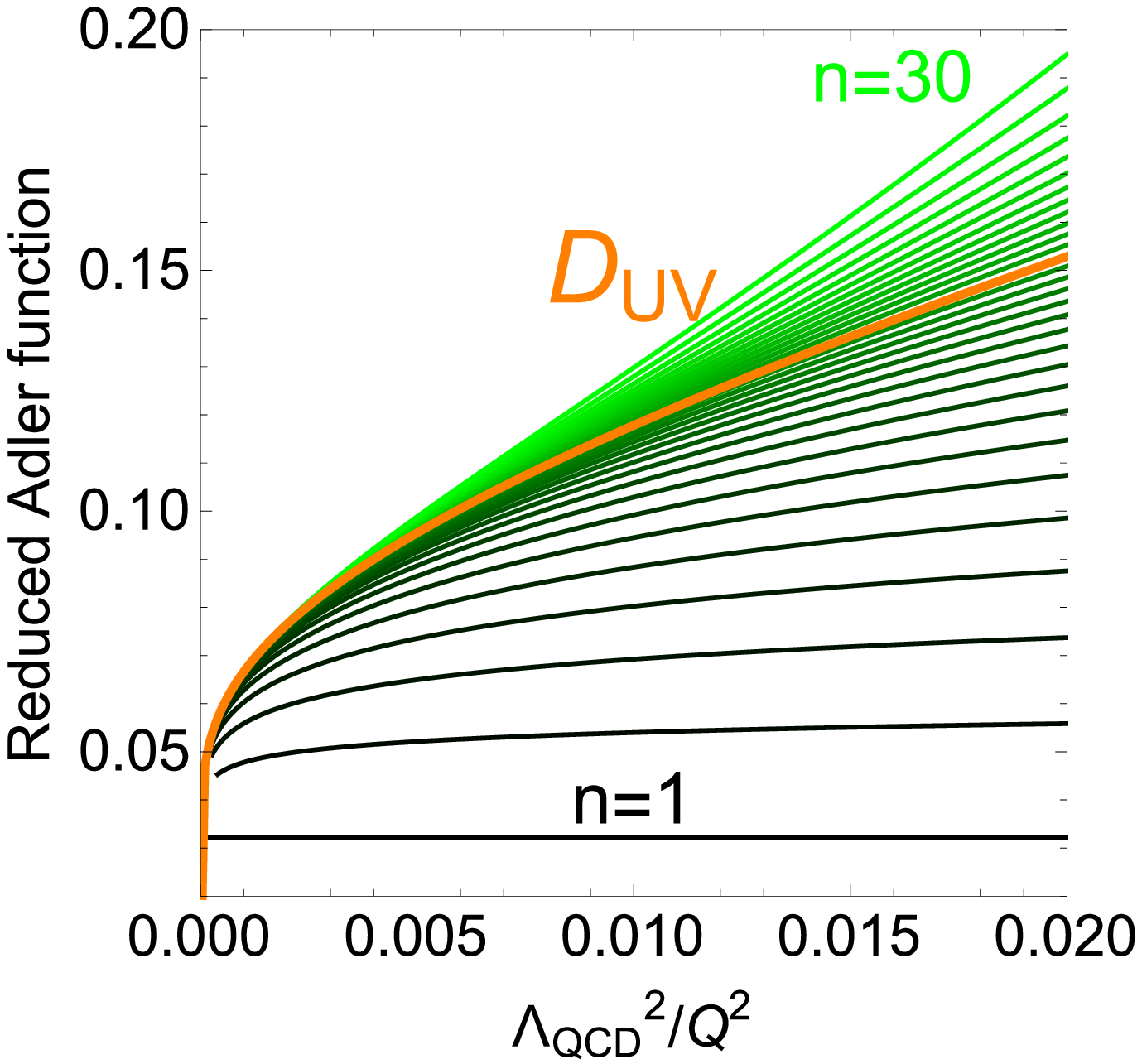}
 \end{minipage}
 \begin{minipage}{0.5\hsize}
 \centering
\includegraphics[width=7cm]{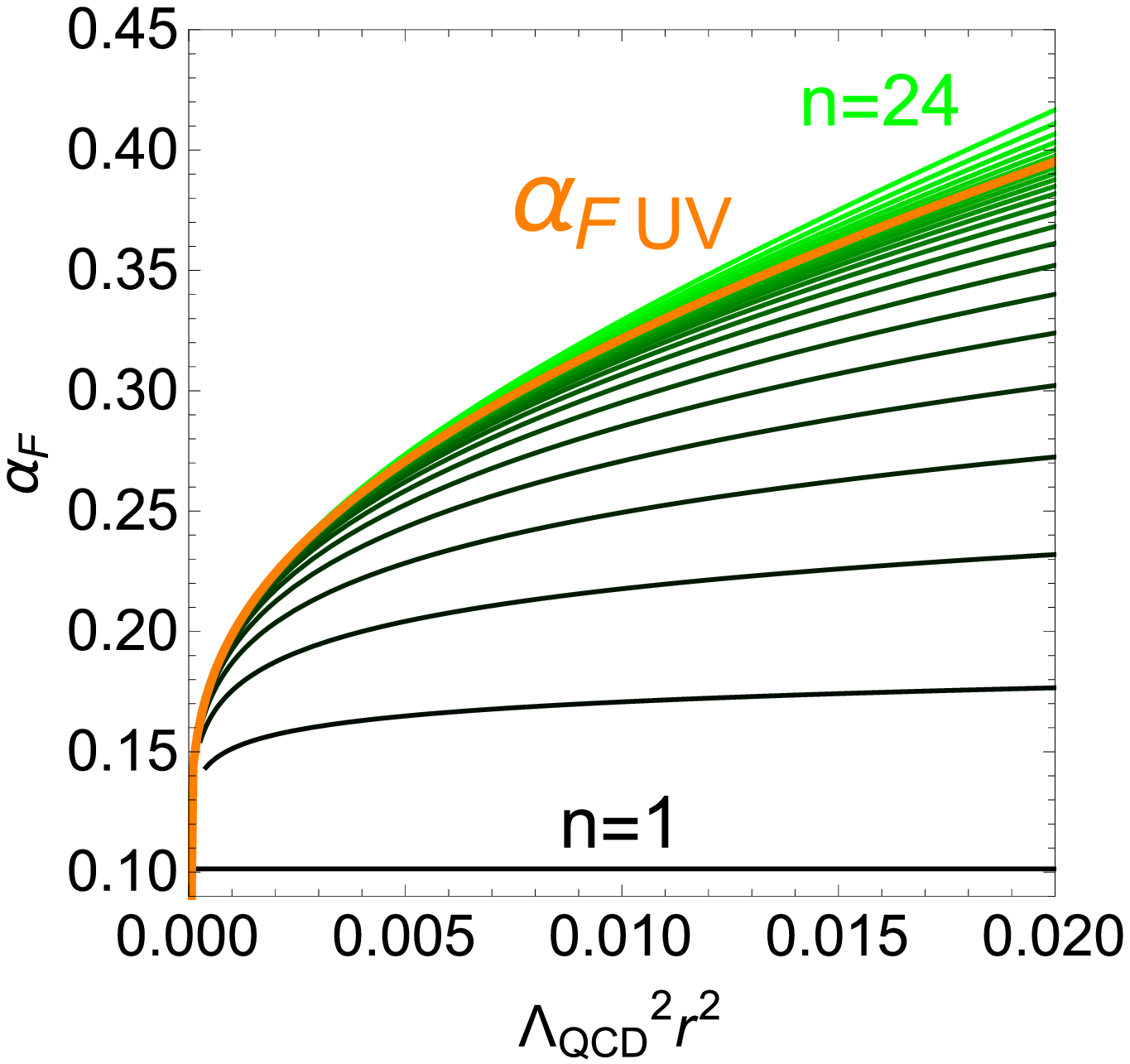}
 \end{minipage}
 \end{tabular}
\caption{\small Truncated perturbative series 
up to $\mathcal{O}(\alpha^n_s(\mu ))$ 
in the large-$\bz$ approximation
and $X_{\UV}$, 
for the reduced Adler function $X=D$ (left) and 
the $F$-scheme coupling $X=\alpha_F$ (right).
We choose $\alpha_s(\mu )=0.0243$ 
corresponding to $n_*=24$ ($n_*=18$) for $X=D$ ($X=\alpha_F$)
and $u_{\IR}=2$ ($u_{\IR}=3/2$).  
Optimal perturbative prediction $X_{n_*}$ lies close
to $X_{\UV}$ in each figure.
We set $n_f=1$.}
\label{fig:green}
\end{figure}

In Fig.~\ref{fig:green},
we show $X_{\UV}$ and perturbative series 
truncated at various orders 
for $X=D$ and $\alpha _F$.
(The truncated order is denoted as $n$.)
One can see that the truncated
perturbative series gradually approaches to $X_{\UV}$ for $n \lesssim n_*$
as we raise $n$. 
For $n\gtrsim n_*$ it starts to deviate from $X_{\UV}$. 

\begin{figure}
\centering
\includegraphics[width=13cm]{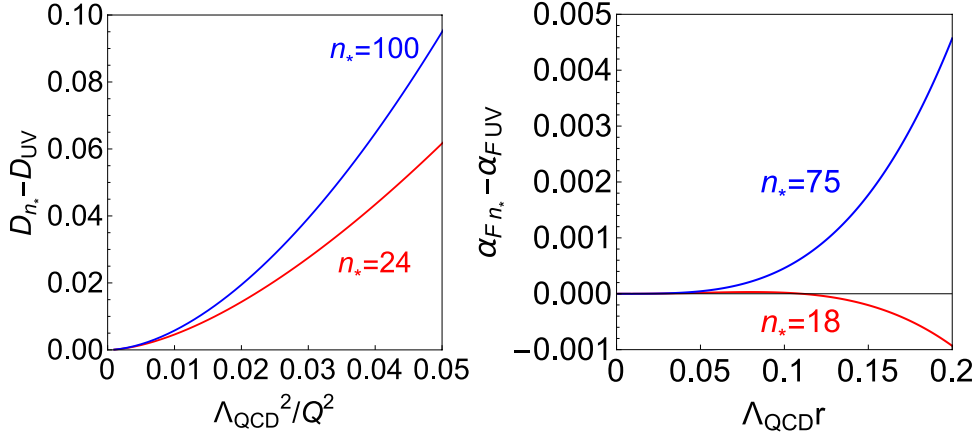}
\caption{\small Difference between 
truncated perturbative series $X_{n_*}$ 
and $X_{\UV}$ for the reduced Adler function (left)
and $\alpha_F$ (right). 
The red (blue) lines correspond to the input $\alpha_s(\mu)=0.1013\,(0.0243)$.
The truncation orders $n_*$ are shown in the plots.
We set $n_f=1$.}
\label{fig:diff}
\end{figure}

The difference between 
$X_{n_*}$ and $X_{\UV}$ is given by 
\begin{align}
X_{n_*}(Q^2)-X_{\UV}(Q^2)
&=\frac{4 \pi}{\bz} \lt[ {\rm Re} \, 
W_X\! \lt(\frac{e^{5/3} \LQ^2}{Q^2} \rt)
-\sum_{0 \leq n <u_{\IR}} c_n 
\lt( \frac{e^{5/3} \LQ^2}{Q^2} \rt)^n \rt] 
\non
&+{\rm Pr.} \int_{0}^{\infty} 
\frac{d \tau}{2 \pi \tau} \,
w_X \! \lt(\frac{\tau}{Q^2} \rt) 
\alpha_{\bz}(\tau) (-L^{n_*})  
\label{Sndif} \, .
\end{align}
For large $n_*$, this difference has a power behavior
$\LQ^2/Q^2$ whose order is the same as the renormalon uncertainty:%
\footnote{
\label{fn:order}
The difference Eq.~\eqref{Sndif} 
can contain a polynomial of 
$\log{(Q^2/\LQ^2)}$, $\log \log{(Q^2/\LQ^2)},\,\cdots$,
as a factor in front of the $(\LQ^2/Q^2)^{u_{\IR}}$-term. 
Therefore, strictly speaking, 
the difference is $o(\LQ^2/Q^2)^{u_{\IR}-\delta}$ 
with $0< \forall \delta \leq 1$.}
\be
X_{n_*}(Q^2)-X_{\UV}(Q^2) 
\sim \mathcal{O}((\LQ^2/Q^2)^{u_{\IR}})  \, . \label{sa}
\ee
More precisely, 
we can detect the $n_*$-dependence of Eq.~\eqref{Sndif} analytically as 
 \be
 X_{n_*}(Q^2)-X_{\UV}(Q^2) 
 \sim \log{n_*} \times \frac{b_{u_{\IR}}}{\bz} 
 \lt(\frac{e^{5/3} \LQ^2}{Q^2} \rt)^{u_{\IR}} 
 \label{lognstar} \, ,
\ee
where $b_{u_{\IR}}$ is an expansion coefficient 
of $w_X$; c.f., Eq.~\eqref{invBexp}. 
(We give a derivation 
in App.~\ref{App.dif}.) 
In Fig.~\ref{fig:diff} we check Eqs.~\eqref{sa} and \eqref{lognstar} 
numerically for $X=D$ and $\alpha_F$.  
We confirm the predicted behavior,
although $n_*=18$ for $\alpha_F$ is not large enough 
to reach the asymptotic forms. 

We can draw some conclusions  
from Eq.~\eqref{sa}. 
First, it shows that 
the power behaviors in $X_{\UV}$ 
are not a new contribution 
which should be added to the perturbative prediction,
rather they are already contained in the perturbative series.
Secondly, using Eq.~\eqref{sa} and the assumption 
of the renormalon uncertainty,
we can extract the following relation 
between $X_{\UV}$ and the true value $X(Q^2)$:
\begin{align}
X(Q^2)-X_{\UV}(Q^2)
&=\left[ X(Q^2)-X_{n_*}(Q^2)\right]
+\left[ X_{n_*}(Q^2)-X_{\UV}(Q^2)\right]
\non &
 \sim \mathcal{O}((\LQ^2/Q^2)^{u_{\IR}}) \, .
\end{align}
This result is consistent with the interpretation
that $X_{\UV}(Q^2)$ is the leading order contribution to $X(Q^2)$ in the OPE framework
and the deviation from $X(Q^2)$ starts 
from the next-to-leading order in
OPE and has the same order of magnitude as
the non-perturbative matrix element of the order of $(\LQ^2/Q^2)^{u_{\IR}}$.

\begin{figure}
\centering
\includegraphics[width=12cm]{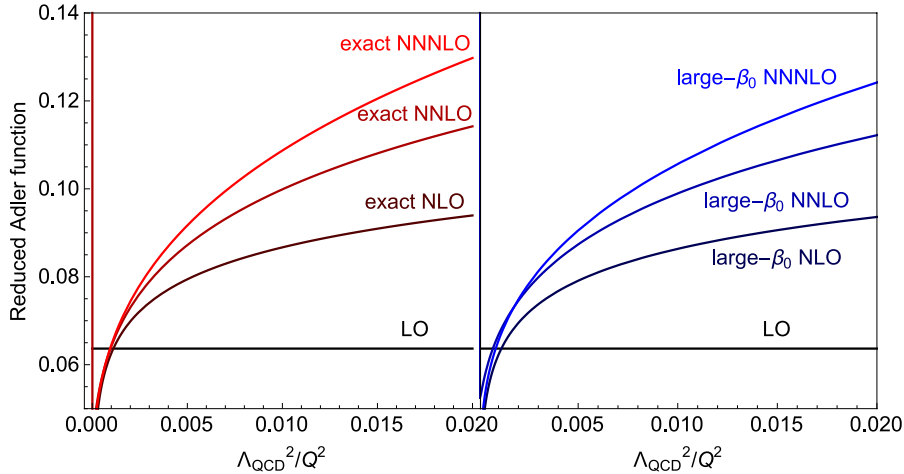}
 \caption{\small{Perturbative series of $D(Q^2)$: 
exact result for the non-singlet component (left) and large-$\beta_0$
approximation (right).
N$^k$LO line represents the sum of the series up to ${\cal O}(\alpha_s^{k+1})$. 
The input is taken as $\alpha_s(\mu)=0.2$, 
and we set $n_f=1$.
}}
\label{fig:compare_adler}
\end{figure}
\begin{figure}
\centering
\includegraphics[width=12cm]{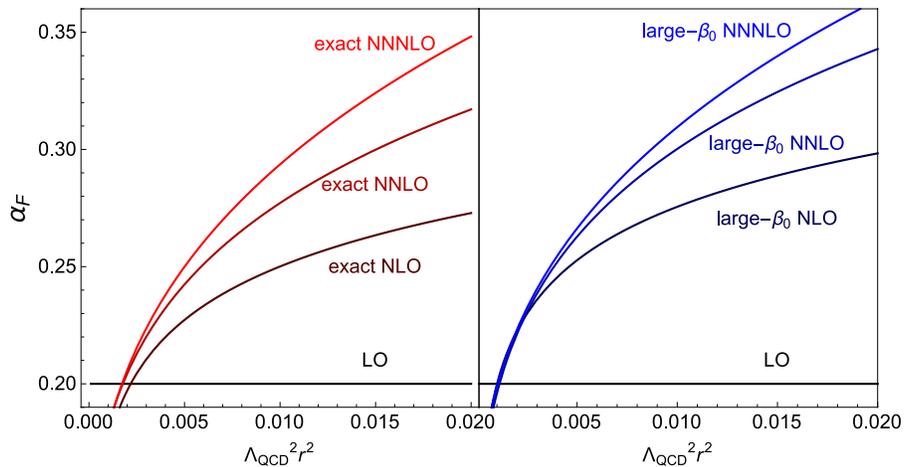}
 \caption{\small{Perturbative series of $\alpha_F(1/r^2)$.
 The parameters are the same as 
 those of Fig.~\ref{fig:compare_adler}.
}} 
\label{fig:compare_force}
\end{figure}

We end this section with comparisons
between the known exact perturbative series
and those obtained under the large-$\beta_0$ approximation
to make sure how far we can trust the result
based on the large-$\beta_0$ approximation.
Figs.~\ref{fig:compare_adler} and \ref{fig:compare_force}
show that the large-$\beta_0$ approximation
reproduces qualitatively the same behavior of the exact series
of $D(Q^2)$ \cite{Baikov:2010je,Baikov:2012zn}
and $\alpha_F(1/r^2)$ \cite{Anzai:2009tm,Smirnov:2009fh,Lee:2016cgz}
\footnote{
There is an IR divergence in 
the exact series of 
$\alpha_F$ from the three-loop order,
and the divergence cancels with contributions from 
the ultra-soft scale.
We do not include the contribution of the ultra-soft scale
because this contribution cannot be regarded as
a part of the Wilson coefficient.
Instead, we simply subtract the term proportional to 
$(1/\epsilon +4(2\log (\mu /p)+\log (4\pi )-\gamma_\mathrm{E}))$
in momentum space
in dimensional regularization.
}.
Therefore we expect that the
results in this paper (especially Fig.~\ref{fig:green}) 
grasp an essential feature of QCD.

\section{Example of timelike quantity: $R$-ratio in $e^{+} e^{-}$ collision}
\label{sec.ET}
So far we have considered Euclidean quantities. 
In this section we investigate how our method can 
be extended to the case of the $R$-ratio in $e^{+} e^{-}$ collision 
as an example of a timelike quantity. 
We obtain a result which can be regarded 
as an extension of the massive gluon scheme. 

In calculating the $R$-ratio, we set
$Q^2<0$ (i.e. $q^2>0$) and take the imaginary part of $\Pi(Q^2)$
according to the optical theorem. 
The difference from Euclidean quantities is that 
we do not have a one-dimensional integral representation of the $R$-ratio 
in the form of Eq.~\eqref{ReLB}. 
Thus, we cannot directly apply the method 
developed for Euclidean quantities in Sec.~\ref{sec.BM},
and a reconsideration is needed.  
Our strategy is to start from a Euclidean quantity and 
to use the analytic continuation
to derive the result for the $R$-ratio.

We consider the reduced vacuum polarization and reduced $R$-ratio, 
in which $\alpha_s$-independent terms are subtracted.  
Let us start from the reduced vacuum polarization\footnote{
Note that we define the reduced vacuum polarization \eqref{Pimu} such that its perturbative expansion does not contain 
the $\alpha_s^0$-part $\frac{N_C}{12 \pi^2} ( \log{(Q^2/\mu^2)+C)}$, which is included in 
the renormalized $\Pi(Q^2)$ of Eq.~\eqref{defPi}. 
Correspondingly, the reduced $R$-ratio is different by $N_C e_q^2$ 
for each quark flavor compared with the $R$-ratio.}
in the Euclid region $Q^2>0$ with an IR cutoff scale,
\be
\Pi_{\bz} (Q^2;\mf)
=\int_{\mf^2}^{\infty} \lt[ \frac{d \tau}{2 \pi \tau} \rt]_r 
w_{\rm \Pi} \lt(\frac{\tau}{Q^2} \rt) 
\alpha_{\bz}(\tau)  \label{Pimu} \, ,
\ee
where we denote by $[d \tau/(2 \pi \tau)]_r$ 
a regularized integral measure which makes the integral UV finite. 
We do not need to specify a way of regularization since the $R$-ratio, 
which we are interested in, is finite, and thus the final result should be 
independent of the regularization method.
The weight
is given as~\cite{Neubert:1994vb}
\be
w_\Pi (x)= 
w_1(x)\theta (1-x)
+w_2(x)\theta (x-1)
\label{wPi}\\
\ee
with
\be
w_1(x)=A
\lt[
 2(1-\log{x}) x+(5-3 \log{x}) x^2
+2 (1+x)^2 \{\Li_2(-x)+\log{x} \log{(1+x)}\} 
\rt]\,,
\label{w1Pi}
\ee
\be
w_2(x)=A
\lt[5+3 \log{x}+2 (1+\log{x}) x
+2 (1+x)^2 \{ \Li_2(-x^{-1})-\log{x} \log{(1+x^{-1})} \} \rt] \, ,
\label{w2Pi}
\ee 
where $A=-N_C C_F/(12 \pi^2)$.
The small-$x$ and large-$x$ behaviors 
of $w_\Pi (x)$ are given, respectively, by
\begin{align}
w_\Pi (x)&=A
\lt[ \frac{3}{2}x^2
-\frac{11-6\log x}{9}x^3
+\dots \rt]
&(x \ll1)  \label{small}\,,
\\
w_\Pi (x)&=A
\lt[ \frac{3}{2}
-\frac{11+6\log x}{9}\frac{1}{x}
+\dots \rt]
&(x \gg 1).
\label{large}
\end{align}
We can see that the first IR renormalon is located at $u=2$ 
for the reduced vacuum polarization. 
The constant $c_{\infty} \equiv 3A/2$ in Eq.~\eqref{large}
stems from the UV renormalon at $u=0$,  
which is the  source of the UV divergence of the integral \eqref{Pimu}.
As shown in Eqs.~\eqref{w1Pi} and \eqref{w2Pi}, 
$w_{\Pi}$ has different analytic forms for $x<1$ and $x>1$, 
hence, we separate the integral path at $\tau=Q^2$ 
in order to represent Eq.~\eqref{Pimu} as an analytic function of $Q^2$:   
\begin{align}
\Pi_{\bz}(Q^2;\mf)
&=\int_{\mf^2}^{Q^2}  \frac{d \tau}{2 \pi \tau}  w_1 \lt(\frac{\tau}{Q^2} \rt) \alpha_{\bz}(\tau)+\int_{Q^2}^{\infty}  \frac{d \tau}{2 \pi \tau}  \lt[w_2\lt(\frac{\tau}{Q^2} \rt)-c_{\infty} \rt]  \alpha_{\bz}(\tau) \non
&+\int_{Q^2}^{\infty} \lt[ \frac{d \tau}{2 \pi \tau} \rt]_r c_{\infty} \, \alpha_{\bz}(\tau) \label{EPi} \, .
\end{align}
We also separate the divergent part which needs a regularization.

Now we replace $Q^2 \to |Q^2| e^{i \pi}$ in Eq.~\eqref{EPi}
and derive an expression for the timelike region.  
The integral path of the
first term in Eq.~\eqref{EPi} can be deformed after replacing $Q^2 \to |Q^2| e^{i \pi}$ as
\begin{align}
\int_{\mf^2}^{Q^2} \frac{d \tau}{2 \pi \tau}\, w_1\! \lt(\frac{\tau}{Q^2} \rt) \alpha_{\bz}(\tau) 
\to &
\int_0^{-|Q^2|} \frac{d \tau}{2 \pi \tau}\, w_1\! \lt(\frac{\tau}{|Q^2|} e^{- i \pi} \rt) \alpha_{\bz}(\tau) 
\non & ~~~
-\int_{C_b} \frac{d \tau}{2 \pi \tau}\, w_1\! \lt(\frac{\tau}{|Q^2|} e^{- i \pi} \rt) \alpha_{\bz}(\tau) .
\\
\begin{tikzpicture}[baseline={(0,0.4)},scale=0.5]
\draw [->] (-5,0)--(5,0);
\draw [->] (0,-1)--(0,3);
\draw [ultra thick,red] (2,0)--(4,0);
\draw [->,ultra thick,red] (2,0)--(3,0);
\draw (1.3,0) node {$\times$};
\draw (2.2,0.7) node {$\mu_f^2$};
\draw (4.2,0.75) node {$Q^2$};
\draw (-4.2,0.75) node {$|Q^2| e^{i \pi}$};
\draw[-](2,-0.25)--(2,0.25);
\draw[-,thick,blue](4,-0.25)--(4,0.25);
\draw[-,thick,blue](-4,-0.25)--(-4,0.25);
\draw (1.57,-1) node {$e^{5/3} \Lambda ^2_{\rm QCD}$};
\draw (4,3)--(4,2)--(5,2);
\draw (4.5,2.5) node {$\tau$};
\end{tikzpicture}
&
\quad \to \quad
\begin{tikzpicture}[baseline={(0,0.4)},scale=0.4]
\draw [->] (-5,0)--(2,0);
\draw [->] (0,-1)--(0,3);
\draw [ultra thick,red] (0,0)--(-4,0);
\draw (1.3,0) node {$\times$};
\draw [->,ultra thick,red] (-2,0)--(-3,0);
\draw (-4.2,0.75) node {$|Q^2| e^{i \pi}$};
\draw[-,thick,blue](-4,-0.25)--(-4,0.25);
\end{tikzpicture}
\quad - \quad
\begin{tikzpicture}[baseline={(0,0.4)},scale=0.4]
\draw [->] (-2,0)--(5,0);
\draw [->] (0,-1)--(0,3);
\draw (1.3,0) node {$\times$};
\draw[-,](2,-0.25)--(2,0.25);
\draw[-,thick,blue](4,-0.25)--(4,0.25);
\draw [ultra thick,red] (0,0)--(0.7,0);
\draw [->, ultra thick,red] (1.3,0.67)--(1.4,0.67);
\draw [ultra thick,red] (2,0) arc (0:180:0.7);
\draw (2,1.5) node {$C_b$};
\end{tikzpicture}
\nonumber
\end{align}
\\
The second term in Eq.~\eqref{EPi} changes as
\begin{align}
&\int_{Q^2}^{\infty}  \frac{d \tau}{2 \pi \tau}  \lt[w_2\lt(\frac{\tau}{Q^2} \rt)-c_{\infty} \rt]  \alpha_{\bz}(\tau)
\to
\int_{-|Q^2|}^{-\infty} \frac{d \tau}{2 \pi \tau}  \lt[w_2\lt(\frac{\tau}{|Q^2|} e^{-i \pi} \rt)-c_{\infty} \rt]  \alpha_{\bz}(\tau) ,
\\
&
\begin{tikzpicture}[baseline={(0,0.4)},scale=0.4]
\draw [->] (-7,0)--(7,0);
\draw [->] (0,-1)--(0,3);
\draw [ultra thick,red] (4,0)--(7,0);
\draw [->,ultra thick,red] (5,0)--(6,0);
\draw (1.3,0) node {$\times$};
\draw[-](2,-0.25)--(2,0.25);
\draw[-,thick,blue](4,-0.25)--(4,0.25);
\draw[-,thick,blue](-4,-0.25)--(-4,0.25);
\draw (4,3)--(4,2)--(5,2);
\draw (4.5,2.5) node {$\tau$};
\end{tikzpicture}
\quad \to \quad
\begin{tikzpicture}[baseline={(0,0.4)},scale=0.4]
\draw [->] (-5,0)--(5,0);
\draw [->] (0,-1)--(0,3);
\draw [ultra thick,red] (-4,0)--(-7,0);
\draw (1.3,0) node {$\times$};
\draw [->,ultra thick,red] (-5,0)--(-6,0);
\draw[-,thick,blue](-4,-0.25)--(-4,0.25);
\end{tikzpicture}
\nonumber
\end{align}
\\
where the end point of the integral path is changed from
$\infty$ to $-\infty$ 
using the fact that the contribution from $C_R$ [defined in Eq.~\eqref{con2}]
vanishes as $R \to \infty$.
The third term becomes
\begin{align}
&\int_{Q^2}^{\infty} \lt[ \frac{d \tau}{2 \pi \tau} \rt]_r c_{\infty}\, \alpha_{\bz}(\tau) 
\to
-\int_0^{-|Q^2|} \lt[ \frac{d \tau}{2 \pi \tau} \rt]_r c_{\infty} \alpha_{\bz}(\tau) 
+\int_{C_a} \lt[ \frac{d \tau}{2 \pi \tau} \rt]_r c_{\infty} \alpha_{\bz}(\tau)  .
\\
&
\begin{tikzpicture}[baseline={(0,0.4)},scale=0.4]
\draw [->] (-5,0)--(7,0);
\draw [->] (0,-1)--(0,3);
\draw [ultra thick,red] (4,0)--(7,0);
\draw [->,ultra thick,red] (5,0)--(6,0);
\draw (1.3,0) node {$\times$};
\draw[-](2,-0.25)--(2,0.25);
\draw[-,thick,blue](4,-0.25)--(4,0.25);
\draw[-,thick,blue](-4,-0.25)--(-4,0.25);
\draw (4,3)--(4,2)--(5,2);
\draw (4.5,2.5) node {$\tau$};
\end{tikzpicture}
\quad \to \quad - \quad
\begin{tikzpicture}[baseline={(0,0.4)},scale=0.4]
\draw [->] (-5,0)--(2,0);
\draw [->] (0,-1)--(0,3);
\draw [ultra thick,red] (0,0)--(-4,0);
\draw (1.3,0) node {$\times$};
\draw [->,ultra thick,red] (-2,0)--(-3,0);
\draw[-,thick,blue](-4,-0.25)--(-4,0.25);
\end{tikzpicture}
\quad + \quad
\begin{tikzpicture}[baseline={(0,0.4)},scale=0.4]
\draw [->] (-1,0)--(7,0);
\draw [->] (0,-1)--(0,3);
\draw[-](2,-0.25)--(2,0.25);
\draw[-,thick,blue](4,-0.25)--(4,0.25);
\draw [ultra thick,red] (0,0)--(0.7,0);
\draw [->, ultra thick,red] (1.3,0.67)--(1.4,0.67);
\draw [ultra thick,red] (2,0) arc (0:180:0.7);
\draw [ultra thick,red] (2,0)--(7,0);
\draw (1.3,0) node {$\times$};
\draw [->,ultra thick,red] (5,0)--(6,0);
\draw (2,1.5) node {$C_a$};
\end{tikzpicture}
\nonumber
\end{align}
\\
Collecting these terms, we obtain an expression for the reduced 
vacuum polarization in the timelike region:
\begin{align}
\Pi_{\bz}
(|Q^2| e^{i \pi}; \mf)
=&\int_0^{\infty} \frac{d \tau}{2 \pi  \tau} 
\lt[ w_{\Pi} \lt(\frac{\tau}{q^2} \rt)-c_{\infty} \rt] 
\alpha_{\bz}(-\tau+i0) \non
&+\int_{C_a} \lt[ \frac{d \tau}{2 \pi \tau} \rt]_r 
c_{\infty} \alpha_{\bz}(\tau)
-\int_{C_b} \frac{d \tau}{2 \pi \tau} 
w_1\lt(\frac{\tau}{q^2} e^{- i \pi} \rt) 
\alpha_{\bz}(\tau)  . \label{Mink}
\end{align}
 By taking the imaginary part, we obtain the reduced $R$-ratio:
 \be
 R_{\bz}(q^2; \mf)=12 \pi \lt(\sum_q e_q^2 \rt) {\rm Im} \, \Pi_{\bz}(|Q^2| e^{i \pi};\mf)  . \label{opti}
 \ee
Setting $\sum_q e_q^2=1$ for simplicity, we have
 \begin{align}
R_{\bz}(q^2; \mf)
&=\int_0^{\infty} \frac{d \tau}{\pi  \tau} W_{R+} \lt(\frac{\tau}{q^2} \rt) {\rm Im}\, \alpha_{\bz}(-\tau+i0) 
-{\rm Im} \int_{C_b} \frac{d \tau}{\pi \tau} W_R \lt(\frac{\tau}{q^2} \rt) \alpha_{\bz}(\tau) . \label{Rstart}
 \end{align}
We regard $W_R$ and $W_{R+}$ as pre-weights 
(although we do not have a weight), 
which are defined as
\begin{align}
&
W_R(z)=6 \pi \{w_1(z e^{-i \pi})-c_{\infty} \} ~~~~(|z| < 1)  \label{pre1}\,,
\\ &
 W_{R+}(z)=6 \pi \{ w_{\Pi}(z)-c_{\infty}\} \, .  \label{pre2}
 \end{align}
In taking the imaginary part of the second term of Eq.~\eqref{Mink} to obtain Eq.~\eqref{Rstart},  
we used
 \be
2\, {\rm Im}  \int_{C_a} \lt[ \frac{d \tau}{2 \pi \tau} \rt]_r c_{\infty} \alpha_{\bz}(\tau)
 =\, {\rm Im}  \int_{C_b} \frac{d \tau}{\pi \tau} c_{\infty} \alpha_{\bz}(\tau) \label{Rmu} \, ,
 \ee
since the imaginary part of the integrand is zero on the positive real axis. 
The way to evaluate the second term of Eq.~\eqref{Rstart},
i.e., the integral along $C_b$,
is no longer different from the case of Euclidean quantities
(see the discussion in Sec.~\ref{sec.GC}).
The expansion of $W_R(z)$ in $z$ reads
 \be
 W_R(z)=N_C C_F \lt[ \frac{3}{4 \pi} -\frac{3}{4 \pi} z^2+\frac{11-6 \log{z}+6 i \pi}{18 \pi} z^3+\dots \rt] \, .
 \label{WRexp}
 \ee
As a result, we can separate the $\mf$-dependence of Eq.~\eqref{Rstart} 
and obtain a $\mf$-independent part $R_{\UV}$ as
\be
 R_{\bz}(q^2;\mf)=R_{\UV}(q^2)+\mathcal{O}(\mf^6/q^6) \label{RUV} \, ,
\ee
where
\be
R_{\UV}(q^2)=R_0(q^2)-\frac{3 N_C C_F}{\bz}\, \frac{e^{10/3} \LQ^4}{q^4} \label{RUVdef}\,,
\ee
\be
R_0(q^2)=\int_0^{\infty} \frac{d \tau}{\pi  \tau} \,
W_{R+}\! \lt(\frac{\tau}{q^2} \rt) {\rm Im}\, \alpha_{\bz}(-\tau+i0)+\frac{3 N_C C_F}{\bz}   . \label{R0}
\ee
The $\mf$-dependence appears first at order $1/q^6$.
In fact,
the first IR renormalon of the reduced $R$-ratio
is known to be located at $u_{\IR}=3$.
Therefore the result is consistent with Eq.~\eqref{genres}.
However, note that the absence of $u=2$ renormalon is
considered to be an artifact of the large-$\bz$ approximation,
and there is a possibility that 
the result [Eqs.~\eqref{RUV}--\eqref{R0}] 
is not based on a good approximation of the exact perturbative series.
Hence, we are cautious in applying our
formulation to serious studies of the $R$-ratio
at the current stage.
Even in such a case, nevertheless, we can still
learn some lessons from the above result.
 
\begin{figure}[t]
\begin{center}
\includegraphics[width=0.7\linewidth]{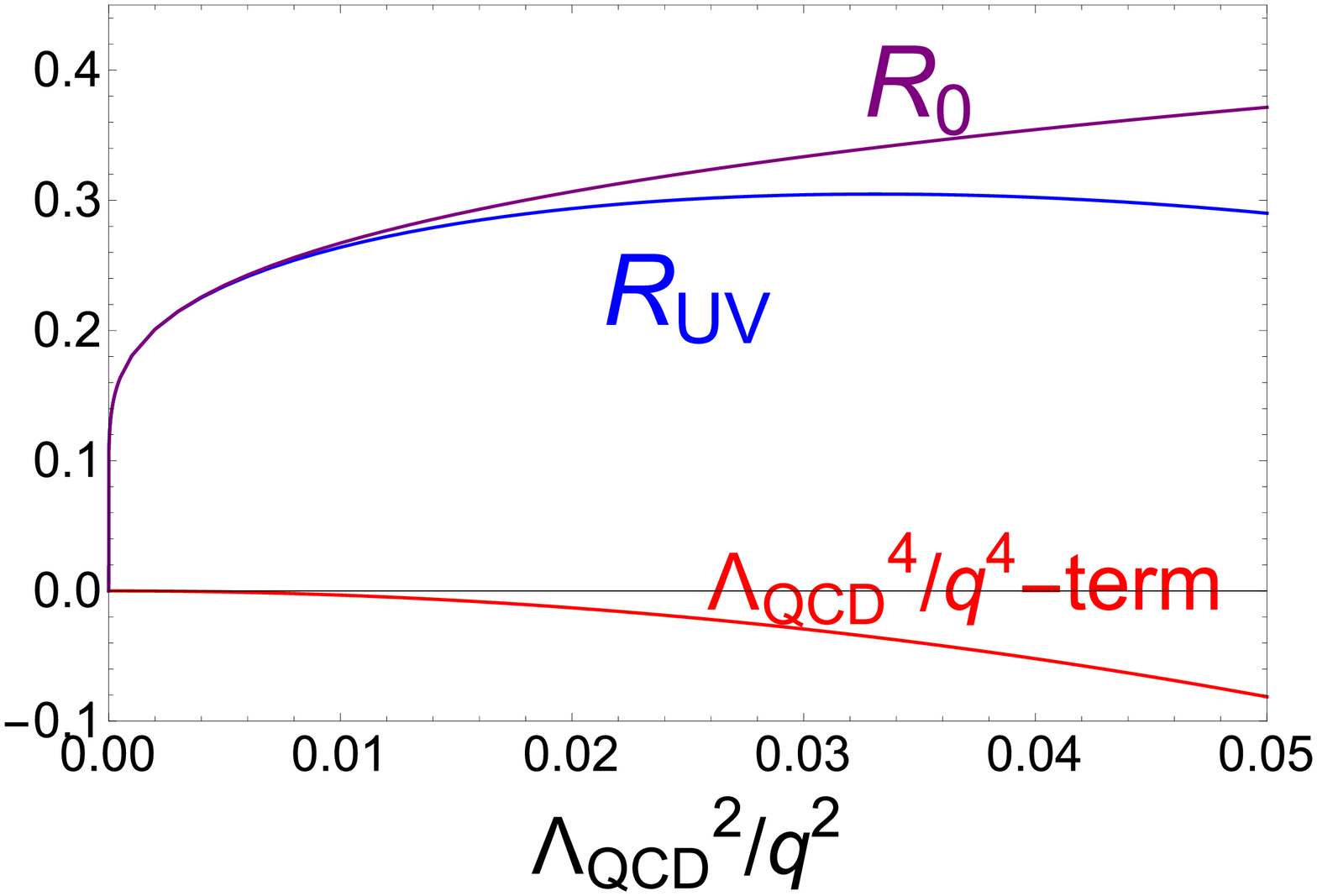}
\caption{\small 
$R_{\UV}$ [Eq.~\eqref{RUVdef}], $R_0$ [Eq.~\eqref{R0}] 
and the $\LQ^4/q^4$-term [Eq.~\eqref{RUVdef}] as functions of $\LQ^2/q^2$.}
\label{Fig:R}
\end{center}
\end{figure}

First, the $1/q^2$-term is absent in Eq.~\eqref{RUVdef}
due to the vanishing $z^1$-term in Eq.~\eqref{WRexp}.%
\footnote{As discussed below Eq.~\eqref{CR},
the absence of $z^1$-term 
is caused by $B_R(1)=0$, 
and it further stems from the absence of a
$u=1$ renormalon in the reduced Adler function.
The absence of the $1/q^2$-term is also understood easily
by reversing the sign of $Q^2$ in Eq.~\eqref{defDUV}.
}
As a result, we obtain a very different behavior of 
the reduced $R$-ratio 
from those of the reduced Adler function and $\alpha_F$,
as seen in Figs.~\ref{Fig.Adler1}, \ref{Fig.Force1} and \ref{Fig:R}.
This fact serves as an evidence that the power corrections indeed 
play an important role in the
determination of the behavior of a physical quantity 
and understanding of it.
     
Secondly, 
our formulation in this section has common features
to those of the Euclidean observables
in the massive gluon scheme. 
Let us clarify this point.
$R_{\UV}$ and $R_0$  [Eq.~\eqref{RUVdef} and \eqref{R0}] 
have the same expressions 
as those of a Euclidean observable obtained in the massive gluon scheme 
[Eqs.~\eqref{SUV2} and \eqref{S0rot}].
In addition, $W_{R+}$ defined in Eq.~\eqref{pre2} 
can be regarded to be ``constructed 
by massive gluon scheme."
To justify this statement,
we can use Eq.~\eqref{C+}, which is satisfied by $W_{X+}$
in the massive gluon scheme.
We regard it as an abstract property of the massive gluon scheme,
since this relation can be checked as long as the observable has
Borel transformation.
The Borel transformation of the (reduced) $R$-ratio is 
known and given in App.~\ref{app:Borel}.
We can show that $W_{R+}$ satisfies the same relation as Eq.~\eqref{C+}: 
\begin{align}
\int_0^{\infty} \frac{d z}{2 \pi} W_{R+}(z) z^{-u-1} 
&=6 \pi \int_0^{\infty} \frac{d z}{2 \pi} ({w}_{\Pi}(z)-c_{\infty}) z^{-u-1} \non
&=6 \pi \int_0^{\infty} \frac{d z}{2 \pi} w_{\Pi}(z) z^{-u-1} \non
&=6 \pi B_{\Pi}(u)|_{Q^2>0} \non
&=-6 \pi \frac{1}{\sin{(\pi u)}} B_{\rm{Im} \Pi} (u)|_{q^2>0} \non
&=-\frac{1}{2\sin(\pi u)} B_{R}(u) ,
\end{align}
where\footnote{
The integral of $({w}_{\Pi}(z)-c_{\infty}) z^{-u-1}$ has the same form as
that of $w_{\Pi}(z) z^{-u-1}$ as a function of $u$
by analytic continuation. 
} 
we use the relation between the Borel transformations 
with opposite signs of $Q^2$ (or $q^2$) \cite{Beneke:1992ch, Beneke:1998ui} and Eq.~\eqref{opti}.
Similarly,  
we confirm that $W_R$ defined in Eq.~\eqref{pre1} 
is consistent with the massive gluon scheme, 
since the expansion of $W_{R}$ is correctly reproduced from the relation
\be
C_R(v)=-\frac{e^{- i \pi v}}{2\sin{(\pi v)}} B_{R}(v) \label{CR}
\ee
and the inverted formula \eqref{WSBorel}, 
which are also obtained in the case of the massive gluon scheme.%
\footnote{
In Ref.~\cite{Ball:1995ni}, the functions $W_{R}$ and $W_{R+}$ were obtained by
the massive gluon method directly. 
Our method can be used to circumvent complicated calculations.}
Namely, our formulation used here can be regarded as
a natural extension of the massive gluon scheme 
developed in Sec.~\ref{sec.BM}
to the timelike quantity.

Thus, our formulation for the $R$-ratio derived by
analytic continuation is
an extension of the massive gluon scheme. 
This is natural if one recalls the discussion in 
Sec.~\ref{sec.FM} that the massive gluon scheme is unique with
respect to the analyticity of an observable.
Namely, if we adopt a formulation which has a good property in terms of 
analyticity, the same result is likely to be obtained.

\section{Conclusions and discussion}
\label{sec.C}

In this paper we proposed a method to extract
a cutoff-independent
UV contribution $X_{\UV}$ from
a general observable $X(Q^2)$
with an explicit IR cutoff,
which is free from IR renormalon ambiguities.
Our method can be applied in the deep Euclidean region
($Q^2\gg\LQ^2$)
and in the large-$\beta_0$ approximation of perturbative series
to all orders.
The UV contribution $X_{\UV}$ consists of the non-powerlike (logarithmic)
term $X_0$ and the power correction terms $\sim (\LQ^2/Q^2)^n$ 
independent of renormalons.

In our method we introduce an analytic function $W_X$,
which we call ``pre-weight,"
for the systematic treatment of
various observables.
General properties of
the pre-weight,
such as its scheme dependence,
were investigated.
Separation of $X_{\UV}$ into $X_0$ and the power corrections
(in particular the coefficients of the power corrections)
depends on the scheme choice.
Among various schemes, the ``massive gluon scheme,''
in which $W_X$ is given by a dispersive integral,
has particularly good analytical properties:
(1) The analyticity of $X_{\UV}$ satisfies physical
requirements within perturbative QCD; 
(2) Origins of the power corrections can be
analyzed accurately using the integration-by-regions method.
We showed that
the feature (1) is satisfied optimally 
in the massive gluon scheme.
We also find that
the analyticity of $X_{\UV}$
and a unique scheme choice follow simultaneously if
the pre-weight satisfies certain good analytical properties
in the upper half-plane.
This discussion establishes natural coefficients of 
the power corrections.

We can use the integration-by-regions method to elucidate
the relation between our formulation and OPE.
Using this relation we showed that
$X_{\UV}$ coincides with the leading Wilson coefficient
in the explicit examples considered.
Thus, we can systematically subtract IR renormalons
from the leading Wilson coefficient in a cutoff-independent way.
Furthermore, we used the integration-by-regions method
to clarify that the 
leading power corrections in $X_{\UV}$ indeed originate from 
UV regions.

As applications of our method,
we investigated the Adler function
and the force between a static quark-antiquark pair.
For each observable,
there is a nontrivial power correction in $X_{\UV}$,
which originates from UV region.
In the context of OPE,
this power correction is a part of the Wilson coefficient of 
the leading identity operator,
and it is consistent with
the structure of OPE.
Comparison with the exact perturbative series indicates
that the large-$\bz$ approximation is
fairly good, hence, it is natural to regard that the
power correction (in the massive gluon scheme)
is inherent in the perturbative series or the UV contribution.

By now there exist ample numerical evidences for
validity of the large-$\bz$ approximation and IR renormalon 
dominance hypothesis.
Apart from these assumptions, we tried to avoid including 
ad hoc assumptions
into our method.
Thus, we believe that we provide a firm connection between
the OPE framework and our
method for subtracting IR renormalons from Wilson coefficients.
Moreover, we consider that our method (in particular in the
massive gluon scheme)
would be an optimal one within the OPE framework,
with respect to systematicity, analyticity, and insensitivity to
the factorization scale (IR cutoff scale).

There remain two directions toward
generalization of our method:
one is to extend it to timelike quantities
and the other is to go beyond the large-$\beta_0$ approximation.
For the former,
we presented an example ($R$-ratio)
but the generalization is left to be done.
We do not have a clear guide to the latter,
since the improvement of the large-$\beta_0$ approximation
in the ordinary perturbation theory 
is still incomplete
and we need a control up to any order in $\alpha_s$.
We speculate that
the method of integration by regions
may play a key role
to achieve the generalization
since the method enables
more complicated scale separation
than the single scale separation which we adopted
in this paper.
In addition, we note that a systematic improvement
beyond the large-$\bz$ approximation has been
achieved
for the static QCD potential 
 and better consistency with OPE has been observed \cite{Sumino:2003yp,Sumino:2005cq},
using the fact that
the pre-weight takes a simple form to all orders in $\alpha_s$.

\section*{Acknowledgments}

The authors are grateful to Y.~Kiyo for fruitful discussion.
The works of G.M.\ and Y.S.\ were supported in part by JSPS KAKENHI 
Grant Number14J10887 and by Grant-in-Aid for
scientific research (No.~26400238) from
MEXT, Japan, respectively.

\newpage

\appendix
\noindent
{\Large \bf Appendices}

\section{Borel transformations}
\label{app:Borel}

We list formulas for the Borel transformations 
in the large-$\bz$ approximation of the
dimensionless observables analyzed in this paper
(reduced Adler function, $F$-scheme coupling defined from the
static QCD force, and $R$-ratio).
\be
B_D(u)
=\frac{8 N_C C_F}{3 \pi} \frac{1}{2-u} \sum_{k=2}^{\infty} \frac{(-1)^k k}{(k^2-(1-u)^2)^2}\,,  ~~~~~~~\text{\cite{Broadhurst:1992si}}
\ee
\be
B_{\alpha_{F}}(u)=\frac{ \sin{(\pi u)}}{\pi u}\Gamma(2-2u)\,,
\ee
\be
B_R(u)=\frac{3 \sin(\pi u)}{\pi u} B_D(u)\,.      ~~~~~~\text{\cite{Beneke:1998ui}}
\ee

\section{Pre-weight
of Adler function}
\label{app:adler}
Another expression of
the pre-weight
of the reduced Adler function in the massive gluon scheme
is given by
\begin{align}
&W^{(m)}_{D+}(z)=W^{(m)}_D(-z)
\nonumber\\
&=
\frac{N_CC_F}{36\pi (z+1)}
\left[
-48 z^3 \text{Li}_2(1-z)
+48 z^3 \text{Li}_2(-z)
+24 z^3 \text{Li}_2\left(\frac{1}{z+1}\right)
-24 z^3\text{Li}_3\left(1-\frac{1}{z}\right)
\right.
\nonumber\\
&
-72 z^3 \text{Li}_3(1-z)
+24 z^3 \text{Li}_3(-z)
-48 z^3 \text{Li}_3\left(\frac{1}{z+1}\right)
+24z^3 \text{Li}_2(1-z) \log (z)
\nonumber\\
&
-48 z^3 \text{Li}_2(1-z) \log (z+1)
-12 z^2 \text{Li}_2\left(1-\frac{1}{z}\right)
+36 z^2\text{Li}_2(1-z)
-48 z^2 \text{Li}_2\left(\frac{1}{z+1}\right)
\nonumber\\
&
-24 z^2 \text{Li}_2\left(1-z^2\right)
-24 z\text{Li}_3\left(1-z^2\right)
+24 z^3 \text{Li}_2\left(1-z^2\right)
+24 z^3 \text{Li}_3\left(1-z^2\right)
\nonumber\\
&
-36 z \text{Li}_2(-z)
+24 z\text{Li}_2\left(\frac{1}{z+1}\right)
+24 z \text{Li}_3\left(1-\frac{1}{z}\right)
+72 z \text{Li}_3(1-z)
-24 z \text{Li}_3(-z)
\nonumber\\
&
+48 z\text{Li}_3\left(\frac{1}{z+1}\right)
+12 \text{Li}_2\left(1-\frac{1}{z}\right)
+12 \text{Li}_2(1-z)
+12 \text{Li}_2(-z)
-24 z\text{Li}_2(1-z) \log (z)
\nonumber\\
&
+48 z \text{Li}_2(1-z) \log (z+1)
+24 z^3 \zeta (3)
+4 \pi ^2 z^3
+4 z^3 \log ^3(z)
+8 z^3 \log ^3(z+1)
\nonumber\\
&
+12 z^3\log ^2(z)
+12 z^3 \log ^2(z+1)
-42 z^3 \log (z)
+24 z^3 \log (z) \log (z+1)
-4 \pi ^2 z^3 \log (z+1)
\nonumber\\
&
+42 z^3 \log (z+1)
+8 \pi ^2 z^2
-66z^2
+6 z^2 \log ^2(z)
-24 z^2 \log ^2(z+1)
-42 z^2 \log (z)
\nonumber\\
&
+66 z^2 \log (z+1)
-24 z \zeta (3)
-4 \pi ^2 z
-57 z
-4 z \log ^3(z)
-8 z \log ^3(z+1)
+12 z \log ^2(z+1)
\nonumber\\
&
+6 \log ^2(z)
-24 z \log (z) \log (z+1)+4 \pi ^2 z \log (z+1)+6 z \log (z+1)-18 \log (z+1)+9
\Biggr]
\label{WDanother} \, .
\end{align}
This expression 
is suited for verifying its
analytical properties, such as,
that $W^{(m)}_D(z)$ has a branch cut
along the positive real axis from $z=0$,
and that $W^{(m)}_{D+}(z)$ takes a real value for $z>0$.
(Note that the polylogarithm $\mathrm{Li}_n(z)$ for $n \geq 2$
has a branch cut along the positive real axis from $z=1$.
In the above expression the 
arguments of $\mathrm{Li}_n$ are less than or equal to 
one for $z \geq 0$.)

\section{Asymptotic expansion of $X_0(Q^2)$}
\label{AppF}

We sketch how to derive the relation \eqref{asymD0}.
Similarly to Eqs.~\eqref{ntrunW}--\eqref{eqXnstar},
we can rewrite the truncated series $X_n$ as follows.
\begin{align}
X_n(Q^2)&=\sum_{k=0}^{n-1}\int_0^\infty \!
 \frac{d \tau}{2 \pi \tau}\, w_X \! \lt(\frac{\tau}{Q^2} \rt) 
\alpha_s^{k+1}\ell^k
\non
&=
\int_0^\infty \!
 \frac{d \tau}{2 \pi \tau}\, w_X \! \lt(\frac{\tau}{Q^2} \rt) 
\alpha_s \frac{1-(\alpha_s \ell)^n}{1-\alpha_s \ell}
\non
&=
{\rm Im}\,\int_0^\infty \!
 \frac{d \tau}{\pi \tau}\, W_X \! \lt(\frac{\tau}{Q^2} \rt) 
\alpha_s \frac{1-(\alpha_s \ell)^n}{1-\alpha_s \ell}
\non
&=
{\rm Im}\,\int_{C_a} \!
 \frac{d \tau}{\pi \tau}\, W_X \! \lt(\frac{\tau}{Q^2} \rt) 
\frac{\alpha_s }{1-\alpha_s \ell}
-
{\rm Im}\,\int_{C_a} \!
 \frac{d \tau}{\pi \tau}\, W_X \! \lt(\frac{\tau}{Q^2} \rt) 
\frac{\alpha_s^{n+1} \ell^n}{1-\alpha_s \ell} \,,
\non
\label{XnIntCa}
\end{align}
where $\alpha_s=\alpha_s(\mu)$, and
\be
\ell=\frac{\bz}{4 \pi} \log{(e^{5/3}\mu^2 /\tau)} \, .
\ee
In the first term of Eq.~\eqref{XnIntCa}, we can rewrite
$\alpha_s/(1-\alpha_s \ell)=\alpha_{\bz}(\tau)$.
In the second term, we can deform the integral path back
to the positive real axis and rewrite
\be
\frac{\alpha_s^{n+1} \ell^n}{1-\alpha_s \ell}~~ \to ~~
{\rm Pr.}\,\frac{\alpha_s^{n+1} \ell^n}{1-\alpha_s \ell} + \frac{4\pi i}{\bz} \pi \tau
\delta (\tau - e^{5/3}\LQ^2 ) \,,
\ee
where Pr.\ denotes the principal value.

Therefore, the difference between $X_0(Q^2)$, given by
Eq.~\eqref{S0}, and $X_n(Q^2)$ 
can be written as
\be
X_0(Q^2)-X_n(Q^2)
=\frac{4 \pi}{\bz} \lt[c_0-{\rm Re} \, W_X \! \lt(\frac{e^{5/3}\LQ^2 }{Q^2} \rt) \rt]
+{\rm Pr.} \int_0^{\infty}\! \frac{d \tau}{2 \pi \tau}\, w_X 
\! \lt(\frac{\tau}{Q^2} \rt) \frac{\alpha_s^{n+1} \ell^n}{1-\alpha_s \ell} \, . \label{F1}
\ee
The first term is ${\cal O}(\LQ^2/Q^2)$.
This follows from $W_X(z)=\sum_n c_n z^n$ and ${\rm Im}\,c_0=w_X(0)=0$.
Hence, the first term is smaller than
$\mathcal{O}(\alpha_s(\mu)^k)$ for an arbitrary positive integer $k$
(or it is zero in expansion in $\alpha_s(\mu)$).
It remains to show that
\be
{\rm Pr.} \int_0^{\infty}\! \frac{d \tau}{2 \pi \tau}\, w_X 
\! \lt(\frac{\tau}{Q^2} \rt) \frac{\alpha_s^{n+1} \ell^n}{1-\alpha_s \ell} =
\mathcal{O}(\alpha_s(\mu)^{n+1}) \, . \label{Prop}
\ee 
It can be shown that the left-hand side is ${\cal O}(\alpha_s(\mu)^{n+1})$
in the case
that $\int^{t}  \frac{dx}{x}\, w_X(x) \times [\text{Polynomial of}~\log x]$ is absolutely convergent as $t \to \infty$, 
and  that the first IR renormalon is a single pole. 
(Although the QCD potential/force does not satisfy the first condition,
we can show Eq.~\eqref{Prop} in another way.)
It is valid for general $\mu$, and in particular 
if we set $\mu=Q$, we obtain the relation \eqref{asymD0}.

\section{Evaluation of $X_{n_*}(Q^2)-X_{\UV}(Q^2)$}
\label{App.dif}
We examine the principal value integral appearing in $X_{n_*}(Q^2)-X_{\UV}(Q^2)$ 
(the second term of Eq.~\eqref{Sndif}) for large-$n_*$.
\be
{\rm Pr.} \int_{0}^{\infty} \frac{d \tau}{2 \pi \tau} w_X \lt(\frac{\tau}{Q^2} \rt) \alpha_{\bz}(\tau) L^{n_*}  \label{Pr}
\ee
Write $L$ of Eq.~\eqref{L} as a function of $n_*$
\be
L(\tL^2/\tau)
=\frac{\bz \alpha_s}{4 \pi} (\log(\mu^2/\LQ^2)+\log(e^{5/3} \LQ^2/\tau)) 
=1+\frac{u_{\IR}}{n_*} \log({\tL}^2/\tau) \, ,
\ee
where $\tL^2 \equiv e^{5/3} \LQ^2$, then we get 
\be
L(\tL^2/\tau)^{n_*} \to \lt( \frac{\tL^2}{\tau} \rt)^{u_{\IR}} ~~~\text{as}~~~ n_* \to  \infty  \label{lim}  \, .
\ee
If we use this form, the integral \eqref{Pr} does not converge around the region $\tau \sim 0$ due to the behavior $w_X(\tau/Q^2)=b_{u_{\IR}} (\tau/Q^2)^{u_{\IR}}+\dots$. Therefore we should calculate keeping $n_*$ finite for this part. It is useful to factorize the integral as follows:
\begin{align}
&{\rm Pr.} \int_{0}^{\infty} \frac{d \tau}{2 \pi \tau} w_X \lt(\frac{\tau}{Q^2} \rt) \alpha_{\bz} (\tau) L(\tL^2/\tau)^{n_*}  \non
&={\rm Pr.} \int_{0}^{\infty} \frac{d x}{2 \pi x} 
\lt[ w_ X \lt(\frac{\tL^2}{Q^2} x \rt)-b_{u_{\IR}} \lt(\frac{\tL^2}{Q^2} x \rt)^{u_{\IR}} \theta(1-x) \rt]
 \frac{4 \pi}{\bz} \frac{1}{\log{x}} L(1/x)^{n_*} \non
&~~+{\rm Pr.} \int_{0}^{\infty} \frac{d x}{2 \pi x} 
b_{u_{\IR}} \lt(\frac{\tL^2}{Q^2} x \rt)^{u_{\IR}} \theta(1-x) 
 \frac{4 \pi}{\bz} \frac{1}{\log{x}}  L(1/x)^{n_*}  \label{sepa}
\end{align}
The second term of Eq.~\eqref{sepa} is separated as
\begin{align}
&{\rm Pr.} \int_{0}^{\infty} \frac{d x}{2 \pi x} 
b_{u_{\IR}} \lt(\frac{\tL^2}{Q^2} x \rt)^{u_{\IR}} \theta(1-x) 
 \frac{4 \pi}{\bz} \frac{1}{\log{x}} \lt\{ L(1/x)^{n_*}-1 \rt\} \non
&+{\rm Pr.} \int_{0}^{\infty} \frac{d x}{2 \pi x} 
b_{u_{\IR}} \lt(\frac{\tL^2}{Q^2} x \rt)^{u_{\IR}} \theta(1-x) 
 \frac{4 \pi}{\bz} \frac{1}{\log{x}} \non
& \equiv \frac{4 \pi}{\bz} b_{u_{\IR}} \lt(\frac{\tL^2}{Q^2} \rt)^{u_{\IR}} F(n_*)
 +{\rm Pr.} \int_{0}^{\infty} \frac{d x}{2 \pi x} 
b_{u_{\IR}} \lt(\frac{\tL^2}{Q^2} x \rt)^{u_{\IR}} \theta(1-x) 
 \frac{4 \pi}{\bz} \frac{1}{\log{x}}
\end{align}
Substituting this into Eq.~\eqref{sepa}, we obtain
\begin{align}
&{\rm Pr.} \int_{0}^{\infty} \frac{d \tau}{2 \pi \tau} w_X \lt(\frac{\tau}{Q^2} \rt) \alpha_{\bz} (\tau) L^{n_*}  \non
&\simeq \frac{4 \pi b_{u_{\IR}}}{\bz} \lt(\frac{\tL^2}{Q^2} \rt)^{u_{\IR}} F(n_*) \non
&+{\rm Pr.} \int_{0}^{\infty} \frac{d x}{2 \pi x} 
\lt\{ w_X \lt(\frac{\tL^2}{Q^2} x \rt)-b_{u_{\IR}} \lt(\frac{\tL^2}{Q^2} x \rt)^{u_{\IR}} (1-x^{u_{\IR}}) \theta(1-x) \rt\}
 \frac{4 \pi}{\bz} \frac{1}{\log{x}} \lt(\frac{1}{x} \rt)^{u_{\IR}}  \, , \label{largenstar}
\end{align}
where we used the limit \eqref{lim} for the second term. 
One can show that the second term of Eq.~\eqref{largenstar} is $o((\LQ^2/Q^2)^{u_{\IR}-\delta})$
(see footnote \ref{fn:order})
although it is fairly complicated.
$F(n_*)$ behaves for large-$n_*$ as
\begin{align}
F(n_*) 
&=\int_{0}^{1} \frac{d x}{2 \pi x} 
x^{u_{\IR}} 
\frac{1}{\log{x}} \lt[ \lt\{1+\frac{u_{\IR}}{n_*} \log \lt(\frac{1}{x} \rt) \rt\}^{n_*}-1 \rt] \non
&=-\int_0^{^\infty} \frac{dt}{2 \pi} \frac{e^{-t}}{t} \lt[\lt(1+\frac{t}{n_*} \rt)^{n_*}-1 \rt] \non
&=-\frac{1}{4 \pi} \lt( \log{n_*}+\log{2}+\gamma_E \rt)+\mathcal{O}\lt(\frac{1}{\sqrt{n_*}} \rt) \, ,
\end{align}  
which gives the result Eq.~\eqref{lognstar}.

\bibliography{fullpaper}

\end{document}